# Towards a compleat theory of ecosystem size spectra

**Ralf Schwamborn**

*Oceanography Department, Federal University of Pernambuco (UFPE), Recife, Brazil*

e-mail: ralf.schwamborn@ufpe.br

## Abstract

The regularity of ecosystem size spectra is one of the most intriguing and relevant phenomena on our planet. Pelagic size spectra generally show a log-linearly downtrending shape, following a power-law distribution. A constant log-linear slope has been reported for many marine pelagic ecosystems, often being approximately b = -1. Conversely, there are variable trophic-level-biomass relationships (trophic pyramids). The contrasting observations of a constant size spectrum and highly variable trophic pyramids may be defined as the *constant size spectrum - variable trophic dynamics paradox*. Here, a mass-specific *predator-prey-efficiency theory of size spectra* (**PETS**) is presented and discussed. A thorough analysis of available data, literature and models formed the basis for the PETS theoretical framework, where pelagic marine ecosystems are controlled by complex trophic processes such as resource-limitation stress and top-down regulation, thus establishing a discrete maximum carrying capacity spectrum. PETS consists of a series of equations and proposed stabilizing mechanisms. The slope b of the biomass-body mass spectrum can be predicted from predator-prey mass ratio (PPMR) and mass-specific trophic efficiency E ($E = d\log_{10}(B) / dTL$), such that $b = E / \log_{10}(PPMR) \approx -1$. The proposed size-specific trophic equilibrium mechanisms stabilize the size spectrum, but not the trophic level - biomass relationship. The complete size spectrum obtained *in situ* (including living organisms and non-living particles) is discussed. This paper is intended as a plea for the integration of modeling approaches, to understand and integrate data and processes across communities including bacteria, plankton, fish and mammals, considering the effects of non-organismic particles.

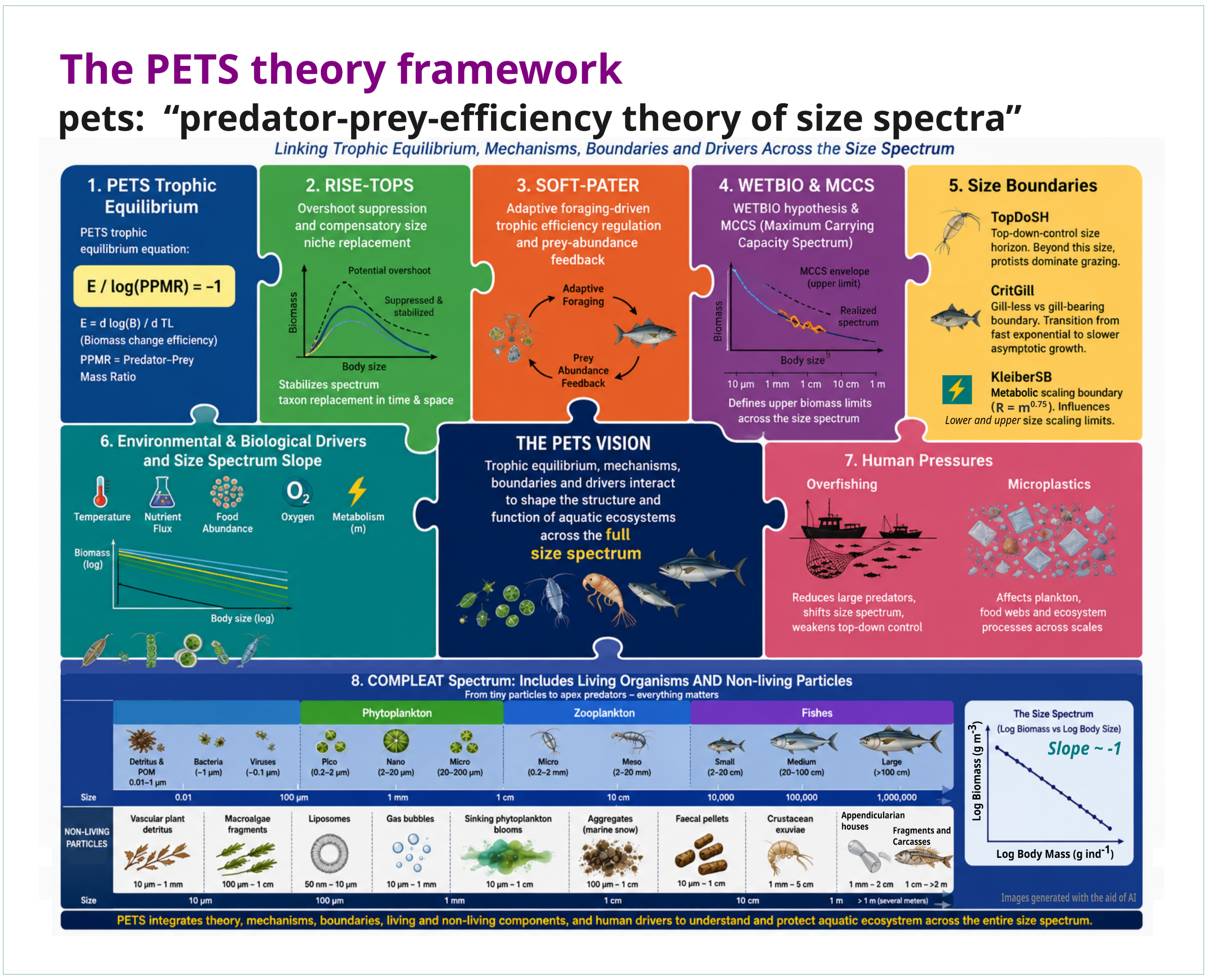

The PETS theory framework
pets: “predator-prey-efficiency theory of size spectra”
Linking Trophic Equilibrium, Mechanisms, Boundaries and Drivers Across the Size Spectrum
1. PETS Trophic Equilibrium
PETS trophic equilibrium equation:
E / log(PPMR) = –1
E = d log(B) / d TL
(Biomass change efficiency)
PPMR = Predator–Prey Mass Ratio
2. RISE-TOPS
Overshoot suppression and compensatory size niche replacement
Potential overshoot
Suppressed & stabilized
Biomass
Body size
Stabilizes spectrum
taxon replacement in time & space
3. SOFT-PATER
Adaptive foraging-driven trophic efficiency regulation and prey-abundance feedback
Adaptive Foraging
Prey Abundance Feedback
4. WETBIO & MCCS
WETBIO hypothesis & MCCS (Maximum Carrying Capacity Spectrum)
MCCS envelope (upper limit)
Realized spectrum
Biomass
Body size
10 µm
1 mm
1 cm
10 cm
1 m
Defines upper biomass limits across the size spectrum
5. Size Boundaries
TopDoSH
Top-down-control size horizon. Beyond this size, protists dominate grazing.
CritGill
Gill-less vs gill-bearing boundary. Transition from fast exponential to slower asymptotic growth.
KleiberSB
Metabolic scaling boundary ($R = m^{0.75}$). Influences
Lower and upper size scaling limits.
6. Environmental & Biological Drivers and Size Spectrum Slope
Temperature
Nutrient Flux
Food Abundance
$O_2$
Oxygen
Metabolism (m)
Biomass (log)
Body size (log)
THE PETS VISION
Trophic equilibrium, mechanisms, boundaries and drivers interact to shape the structure and function of aquatic ecosystems across the full size spectrum
7. Human Pressures
Overfishing
Reduces large predators, shifts size spectrum, weakens top-down control
Microplastics
Affects plankton, food webs and ecosystem processes across scales
8. COMPLEAT Spectrum: Includes Living Organisms AND Non-living Particles
From tiny particles to apex predators – everything matters
Phytoplankton
Zooplankton
Fishes
Detritus & POM 0.01–1 µm
Bacteria (~1 µm)
Viruses (~0.1 µm)
Pico (0.2–2 µm)
Nano (2–20 µm)
Micro (20–200 µm)
Micro (0.2–2 mm)
Meso (2–20 mm)
Small (2–20 cm)
Medium (20–100 cm)
Large (>100 cm)
Size
0.01
100 µm
1 mm
1 cm
10 cm
10,000
100,000
1,000,000
NON-LIVING PARTICLES
Vascular plant detritus
10 µm – 1 mm
Macroalgae fragments
100 µm – 1 cm
Liposomes
50 nm – 10 µm
Gas bubbles
10 µm – 1 mm
Sinking phytoplankton blooms
10 µm – 1 cm
Aggregates (marine snow)
100 µm – 1 cm
Faecal pellets
10 µm – 1 cm
Crustacean exuviae
1 mm – 5 cm
Appendicularian houses
1 mm – 2 cm
Fragments and Carcasses
1 cm – >2 m
Size
10 µm
100 µm
1 mm
1 cm
10 cm
1 m
> 1 m (several meters)
The Size Spectrum
(Log Biomass vs Log Body Size)
Slope ~ -1
Log Biomass ($g\ m^{-3}$)
Log Body Mass ($g\ ind^{-1}$)
Images generated with the aid of AI
PETS integrates theory, mechanisms, boundaries, living and non-living components, and human drivers to understand and protect aquatic ecosystrem across the entire size spectrum.

## Summary

Regular and ubiquitous weight-biomass spectra, with a common slope of b = -1, have been reported within many marine food webs, whether in polar regions or in the tropics. Conversely, a completely different picture, with highly variables shapes, has been observed for trophic level-biomass relationships. Also, predator / prey mass ratios (PPMR) can be enormously variable, ranging across several orders of magnitude. These contrasting observations may be defined as the "*constant size spectrum - variable trophic dynamics paradox*". Here, a mass-specific "*predator-prey-efficiency theory of size spectra*" (**PETS**) is presented and discussed, based on current knowledge in the fields of size spectra analysis and trophic modelling, while utilizing functional responses of predators within size-specific optimal foraging theory. One basic rationale for the choice to prioritize size over trophic level, within PETS ecosystem theory, is that predators, when deciding between feeding strategies directed towards abundant small-sized prey or rare, large sized prey, will consider prey size, but most likely do not care about their prey's trophic level (TL). This also explains why trophic equilibrium mechanisms stabilize the size spectrum, but not the trophic level - biomass relationship. Within PETS, the slope of the size spectrum can be predicted from PPMR (predator-prey mass ratio) and "mass-specific trophic efficiency" E (were: $E = d \log_{10}(B) / d\ TL$). At the core of PETS is the proposed biomass-specific equilibrium constant "c", where $c = E / \log_{10}(PPMR)$, and c = -1. Assuming standard "Kleiber"-scaled metabolic rates R ($R = M^{0.75}$) and standard "Fenchel"-scaled mass-specific growth rates ($g'/M = M^{-0.25}$), as in most eukaryotes: b = c. The ubiquitous b = -1 slope of the weight-biomass spectrum can be used to predict the weight-abundance (-2), size-biomass (-3), size-abundance (-4), weight-production (-1.33) and the weight-turnover (4) spectra, based on simple conversions. Bottom-up, top-down, resource-limit-stress control, and volume mechanics are described and discussed for marine, estuarine, and freshwater pelagic ecosystems.

The "water volume essential to sustain body mass optimum" (WETBIO) hypothesis and its possible relationship to the maximum carrying capacity spectrum are discussed. WETBIO predicts that two-dimensional systems, such as terrestrial and benthic ecosystems, should have a relativity "flat" weight-biomass spectrum slope of approximately b = -2/3.

A discrete "top-down-control size horizon" (TopDoSH) is proposed, that defines the size limit of top-down control by metazoan grazers on phytoplankton. The term "horizon" reflects the concept that there is a prey size boundary, beyond which metazoans cannot reach effectively, i.e., the horizon across which information, energy, and mass disturbances from top-down trophic cascades cannot reach, and where protists take over as the main grazers, defining the phytoplankton size structure. The discrete, critical "gill-less vs gill-bearing size boundary" (CritGill), predicted by GOLT theory, coincides with the transition from fast exponential growth to much slower, asymptotic growth. A perfectly linear size spectrum in natural pelagic ecosystems is expected to occur for organisms that are larger than the TopDoSH size, which coincides approximately with the size range of nanophytoplankton. The observation that large-sized nektonic organisms, such as fish, macroinvertebrates and marine mammals, may follow the global plankton size spectrum (the "CritGill paradox"), is a further proof of the existence of universal laws that shape and stabilize the log-linear size spectrum, regardless of the individual peculiarities of any taxa, such as divergent growth curves and the presence of gills or lungs.

PETS predicts a discrete, fixed, critical size (the “KleiberSB”) between organisms that behave according to Kleiber-scaled size-metabolism relationships ($m_0$ = 3/4), and organisms that have different values of “m“, such as prokaryotes. Metabolic scaling and temperature response of heterotrophic bacteria and picophytoplankton has received little attention in the wider size spectra community, and may be a key parameter within the warming effects on marine ecosystems.

PETS presents a set of concepts, predictions and paradigms regarding carrying capacity in natural ecosystems and the size-spectra effects of overfishing (“Fishing Down the Size Spectrum“), that may modify the complete food web, even beyond the TopDoSH. PETS actually does not predict a slope of b = -1. It predicts that observed mean values of “b” should be between -1 and -0.375, and that the slope of the capacity spectrum is equal to β, within a concept of size spectrum-based carrying capacity (“carrying capacity spectrum hypothesis”). The available data and models suggest that current higher trophic level (HTL) biomass controls and linearizes the biomass of lower trophic level size spectra, down to the TopDoSH, except when there is extreme nutrient limitation stress (then, there is bottom-up control). This could be called the CATCH hypothesis (“Current And Top-down Control by Higher trophic level biomass & carrying-capacity size spectra” - hypothesis).

A thorough analysis of available data, literature and models result in the conclusion that most pelagic marine ecosystems are controlled by trophic processes such as resource-limit stress (bottom-up control) and top-down regulation, with a key role of the carrying capacity of large-sized organisms ($B_{max}$ at $M_{max}$). This has relevant consequences for the prediction and interpretation of size spectra in the context of fisheries, whaling, and the introduction of exotic predators. The complete size spectrum, including living organisms and non-living particles, has received little attention in the current literature. The possible existence of a carrying capacity spectrum for all (living+non-living) particles is discussed. The “carrying capacity spectrum of particles” – hypothesis proposed here has far-reaching consequences for our understanding of ecosystems, for example regarding the effects of the size spectrum of microplastics on living beings.

This paper is intended as a plea for the integration of modeling approaches, to understand and integrate data and processes across communities including bacteria, phytoplankton, fish and mammals, considering also the effect of non-organismic particles such as vascular plant detritus, macroalgae fragments, liposomes, gas bubbles, sinking phytoplankton blooms, aggregates, faecal pellets, crustacean exuviae, appendicularian houses, and animal carcasses. A novel integrated approach promises new insights regarding the effects of warming, acidification, deoxygenation, eutrophication, fisheries, and the inputs of anthropogenic particles such as cellulose fibers and micro- and macroplastics. This size-spectra based approach is intended to enhance the dialogue and exchange of data, concepts, models and predictions between scientific communities, such as theoretical ecologists, microbiologists, microplastics specialists, biogeochemists, planktologists, fisheries scientists, and climate modellers. There is an urgent need for new, integrated size spectrum modelling tools, especially considering the effect of warming on key ecosystem services, such as fisheries and biological carbon pumps.

# 1. Introduction

An immense wealth of theory and data on the relationships of body size, trophic level, abundance, and biomass in natural ecosystems has been published in the past ten decades, since the first quantitative trophic ecology concepts were proposed (Elton,1927, Lindeman, 1942). During all these years, ecology has become a highly fragmented area of research, with countless terminologies, models, and units, without a unifying general theory. There are numerous scientific sub-communities within ecology that have been interacting poorly, such as theoretical ecologists, microbiologists, microplastic laboratories, biogeochemists, primary production laboratories, phytoplankton taxonomists, zooplanktologists, climate modellers, and fisheries scientists, that have completely different different terminologies, models and approaches.

*A problem and a paradox*

The fragmented landscape in ecology is in stark contrast with many other fields of science that are based on well-established central general theories, such as physics, chemistry, plate tectonics, and evolution. One reason for this may be the lack of a clear definition of a single problem or paradox that a general theory should be able to explain and solve, and to derive relevant quantitative predictions from.

The constancy of the marine pelagic size spectrum may qualify as such a fundamental problem. It is one of the most intriguing and relevant phenomena on our planet. Marine pelagic size spectra generally show a log-linearly downtrending shape, following a power-law distribution (Sheldon et al., 1972). Interestingly, a similar distribution shape has been reported from numerous other areas of science (e.g., economy, semiotics, and astronomy), highlighting the importance of the power-law distribution for our understanding of the universe (see, e.g, Vidondo et al., 1997, Convertino et al., 2013).

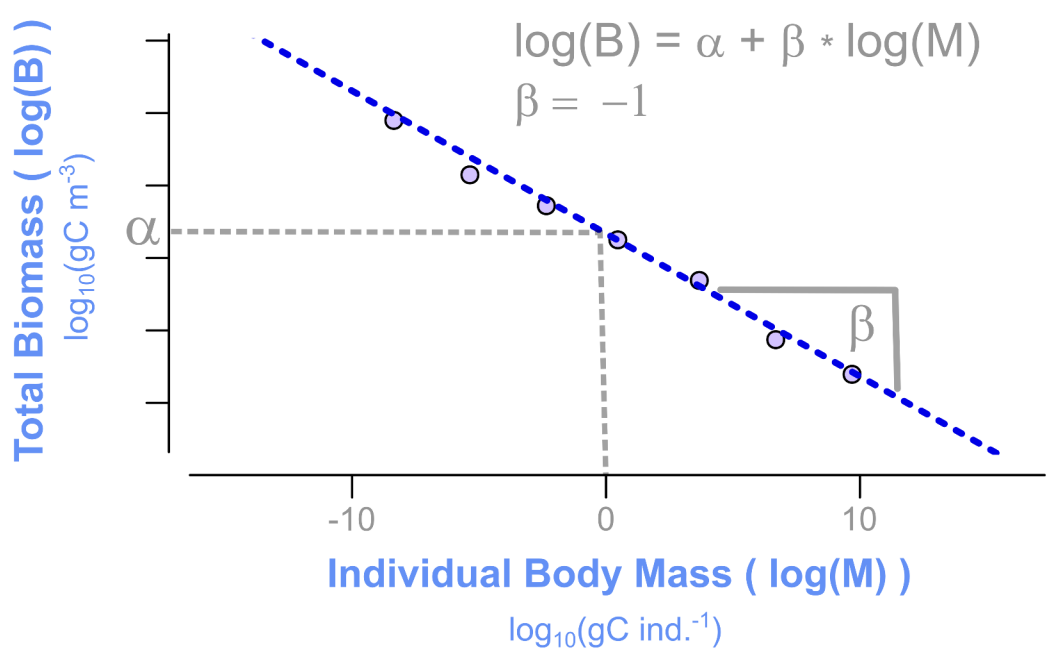

**Figure 1.** The weight-biomass spectrum ( dlog(B) / dlog(M) ) of marine pelagic ecosystems. The slope of the blue dotted line is β = -1.

The intriguing universal ubiquity and constancy of the weight-biomass spectrum slope β (eq. 1), with β = -1, may thus be called the "***size spectra constancy problem***".

A weight-biomass spectrum (Fig. 1) with a slope of β ≈ -1, has been reported within innumerable marine food webs, whether in polar regions or in the tropics (e.g., Dugenne et al., 2024, Soviadan et al., 2024, Ersoy et al., 2025). Conversely, a completely different picture has been observed for trophic level-scaled biomass pyramids. Numerous studies have reported highly variable trophic level-specific biomass pyramids, including the existence of inverted biomass pyramids and top-heavy, hourglass-shaped pyramids (Christensen et al., 2005, Woodson et al., 2018). This has led to a fundamental questioning of the original proposal of a constant trophic efficiency of 10%, as it had been envisaged by early ecologists (Lindeman, 1942, Odum, 1956, Odum, 1957). Another intriguing and unexplained observation is that mean predator / prey mass ratios (PPMR) can be enormously variable, ranging across several orders of magnitude, from approximately 1 to nearly 1 billion (Brown et al., 2004, Barnes et al., 2010, Riede et al., 2011). How is it possible that biomass pyramids, trophic efficiencies, and PPMR are extremely variable, but size spectra are invariant?

In conclusion, we live in a world where *trophic-level-specific* pyramids and trophic efficiencies (whether in units of biomass, production, turnover, or total throughput), and PPMR are highly variable within and across ecosystems, but *weight-biomass* spectra (and their linear transformations, such as abundance-size spectra) are astonishingly consistent and invariable within and across ecosystems, with a constant slope of β ≈ -1. The contrasting observations of a constant size spectrum and highly variable biomass pyramids, trophic efficiency, and PPMR may be defined as the "***constant size spectrum - variable trophic dynamics paradox***".

*The universal slope "β"*

An enormous amount of theory, literature, concepts, models, and data have been produced in the past decades on freshwater and marine size spectra (see comprehensive reviews, e.g., in Andersen et al., 2016, Sprules and Barth, 2016, Dos Santos et al., 2017).

The constant, universal slope "β" of the total biomass / individual mass size spectrum (Fig. 1, eq. 1) can be described as:

$$\beta = \log(B_{i+1}) - \log(B_i) / \log(M_{i+1}) - \log(M_i) = -1 \quad (1)$$

, where Bi = total biomass in any given individual mass (i.e., weight) class Mi)

This equation can be simplified as β = d log(B) / d log(M).

Note that the constancy of β = - 1, means that there is a universally constant, weight-scaled, scale-invariant, weight-specific biomass transfer efficiency β.

*Parameters, descriptors, and theories*

Many different parameters and descriptors have been used in earlier mathematical models to explain the slope of marine ecosystem size spectra, such as predator/ prey mass ratio (PPMR), ecotrophic efficiency (EE), turnover time of body mass, time scale of system energy loss, individual feeding efficiency, community food consumption rate, community assimilation efficiency, body growth, food chain length, population growth, mortality, volumetric search rate, and asymptotic size, with a large diversity of approaches and equations used (e.g., Platt and Denman, 1977, Borgmann, 1982, Borgmann,1987, Zhou and Huntley, 1997, Brown et al., 2004, Andersen and Beyer 2006, Zhou, 2006, Andersen et al., 2009, Barnes et al., 2010, and Dalaut et al., 2025). The large amount of parameters, descriptors and equations in the published literature reflects the lack of consensus and unified general theory for size spectra.

Also, there are several, often complementary theoretical frameworks that have been published in the last few decades, that may be considered important stepstones toward a general theory of size spectra. Among such candidates for a general unified theory are the Allometric Theory (Kleiber, 1932, Peters 1983), and its two most popular extensions, the Metabolic Theory of Ecology (MTE, West et al., 1997; Gillooly et al., 2001; Brown et al., 2004) and the Gill-Oxygen Limitation Theory (GOLT, Pauly, 1981; Pauly et al., 2010; Pauly, 2019; 2021; Cheung and Pauly, 2016) which are both becoming increasingly popular within the scientific community. Recently, a biomass competition theory (Woodson et al., 2018; Fant and Ghedini, 2024) has been proposed, that focuses on inverted biomass pyramids and fisheries effects. A recent size-diversity-trophic-transfer theory may be able to link the size diversity of preys and predators to trophic efficiency, showing another promising connection between biodiversity research and quantitative trophic ecology (García-Comas et al., 2016)

None of these theories, however, has been able to successfully predict and explain the exact value of the size spectra slope on the basis of first principles and simple equations. The theory that would best fit the requirements for a general theory of ecosystems, is undoubtedly the MTE. The core paradigm of Allometric Theory and MTE is that metabolic rate "R" scales with body mass "M" as $R \sim M^{0.75}$ and thus, turnover rate scales as $R / M \sim M^{-0.25}$ (e.g., Kleiber, 1932, Fenchel, 1974, Peters 1983, Brown et al., 2004). MTE lacks the ability to predict the value the slope "β", probably because it is not motivated by explaining the "constant size spectrum - variable trophic efficiency paradox" (by solving the "size spectra constancy problem"). Rather, MTE is a direct extension of Allometric Theory, which is based on assumptions and observations regarding metabolism, temperature, and body size, obtained from laboratory experiments (Kleiber, 1932), including examples of terrestrial plants and mammals.

Furthermore, none of the existing ecosystem size spectra theories has the ability to include phytoplankton (where trophic efficiency may be irrelevant), bacteria, and the size spectra of non-organismic particles (Lins-Silva et al., 2024). While the literature on animal size spectra has focused on trophic efficiency and predator / prey mass ratios, the literature and theory on phytoplankton size distributions has focused on the relationships between nutrient uptake, nutrient flux, surface area / volume ratios, vacuole size, and nutrient storage capacities, i.e., on metabolic processes of phytoplankton population growth, as a function of nutrient dynamics. These theories and models, that are mostly derived from experiments with laboratory cultures, propose a non-linear, unimodal size distribution for phytoplankton (see review in Marañón et al., 2013), that is completely different from the log-linear size spectra observed in nature.

Yet, in spite of different approaches, theories, and equations for phytoplankton and zooplankton, it is an unquestionable fact that these two assemblages are perfectly aligned with each other, with a common slope, within the global size spectra slope of -1 (Hatton et al., 2021; Dugenne et al., 2024, Soviadan et al., 2024; Ersoy et al., 2025; Fock et al., 2026, Schwamborn et al., submitted). Thus, there is an urgent need for a theory and set of equations to explain the "***phyto- zooplankton size spectra paradox***", i.e., the constancy of phyto- and zooplankton size spectra, in spite of apparently completely different processes and theory. Also, there is still no useful generalized ecosystem theory that could be used for global climate change impact models and other applications that focus on biological carbon pumps. A general theory of size spectra should be able to improve and integrate biogeochemical models and other particle-related approaches (e.g. studies on microplastics size spectra, as in Lins-Silva et al., 2024).

*Fisheries, models, and pyramids*

Fisheries science has produced several trophic modelling concepts, approaches and terminologies that are relevant to understand the dynamics of marine size spectra. Among these, the ECOPATH and EwE (Ecopath with Ecosim) theoretical framework, mass-balance approach, set of equations, terminology, and software (Polovina, 1984; Ulanowicz, 1984; Christensen and Pauly, 1992; Pauly et al., 1998; Pauly et al., 2000; Christensen et al., 2005) is especially popular among fisheries biologists. EwE trophic models are based on many simplifying assumptions, such as the existence of discrete, fixed functional compartments. Conversely, the highly complex APECOSM model (Maury, 2010; Dalaut et al., 2025), the OSMOSE model (Shin and Cury, 2001, 2004; Travers et al., 2007), and the "mizer" R package (Scott et al., 2014) are explicitly size- and individual- based models. Yet, all these trophic modeling approaches are inherently hyper-complex, and are generally focused on fish and fisheries. Such trophic models have been extensively utilized as the basis of extremely complex end-to-end (E2E, "from climate to fisheries") modeling efforts (e.g.,Travers et al., 2007).

While trophic modellers and fisheries biologists mostly focus on the highly variable trophic-level-specific pyramids, they usually ignore the ubiquitous size spectrum, which has been observed in numerous zooplankton studies (e.g., Gorsky et al., 2010; Figueiredo et al., 2020;

Dugenne et al., 2024; Soviadan et al., 2024) and in several cross-ecosystem data compilations, from picoplankton to fish (Hatton et al., 2021; Ersoy et al., 2025; Fock et al., 2026; Schwamborn et al., submitted). Indeed, size spectra theory models (e.g., Zhou, 2006; Andersen, 2008; Andersen, 2009; Barnes et al., 2010) and their terminologies and equations have been generally ignored within the trophic modeling and fisheries science community (with notable extensions, e.g., Dalaut et al., 2025).

Similarly, the vast literature on size spectra has generally ignored the existence of variable trophic level-specific pyramids. This has led to a conflicting, fractionated set of concepts and equations between these communities, where size spectra are generally used in studies focusing on zooplankton, while biomass-TL pyramids are mostly used by fisheries biologists. Interestingly, the popularity of such less-quantitative pyramids (it is impossible to read “x” and “y” values on a pyramid), instead of precise two-dimensional plots (as in size spectra) may be an indicative of the use of the pyramid plots as an attempt to equalize (and maybe downplay) the variability and irregularity between trophic level (TL) and biomass (and thus, highly variable trophic efficiencies), within ecosystems.

*Bumps and domes: non-linear spectra*

Another evident challenge to the size spectra theory is the existence of non-linear “bump-shaped” or “dome”-shaped spectra, especially in freshwater systems (e.g., Rossberg et al., 2019). These bumps or domes have been explained mainly through the possible existence of trophic cascades (Rossberg et al., 2019). Rossberg et al. (2019) discuss their proposed trophic cascade mechanisms in the context of lake-manipulation experiments. There is still no consistent theory available that would explain why there are dome-shaped (mostly freshwater systems) versus linear-shaped spectra (e.g., in marine plankton). This apparent contradiction is possibly related to small-scale, low-diversity freshwater *vs* more species-rich and larger-scale marine ecosystems. The troughs or gaps between the “domes” in low-diversity freshwater spectra may be due to the lack of extant taxa that would be capable to fill specific size niches. In marine size spectra, all size classes are seamlessly occupied by numerous taxa and life history stages, that compete for and instantly fill any empty size niche (Schwamborn et al., 2025).

*Towards a “compleat” theory*

Numerous previous studies presented equations to explain ecosystem size spectra with a focus on fish and fisheries, e.g., Kerr & Dickie (2001), Andersen and Beyer (2006), Andersen et al. (2008), and Andersen et al. (2009). Zhou (2006) proposed several complex equations for plankton size spectra, which are widely cited in the zooplankton size spectra literature. Stemmann and Boss (2012) proposed an effort to integrate plankton and particle size spectra research, focusing on the optical properties of particles and plankton for efficient synoptic surveys with *in situ* instruments, such as the UVP (Underwater Vision Profiler).

Lins-Silva et al. (2024) presented quantitative plankton and particle size spectra in marine and estuarine ecosystems, but did not present a theoretical framework for plankton-particle interactions.

A “compleat” (*sensu* Gayanilo et al., 1988) ecosystem size spectra theory should include non-living biogenic particles (which may dominate the particle mass distribution in the oceans), micro- and macroplastics, bacteria, picoplankton, nutrient-limitation-stress effects on phytoplankton size spectra, phyto-zooplankton interactions, functional responses in feeding behaviour, optimal foraging, functional replacement, size-niche competition avoidance, top-down trophic cascades, critical size boundaries and horizons based on allometry, metabolism and Reynolds numbers, fish communities, mammals, macroinvertebrates, and fisheries.

Here, a mass-specific “predator-prey-efficiency theory of size spectra” (**PETS**) is presented and discussed, which is intended, among others, to integrate the well-established ECOPATH mass-balance approach (Polovina 1984, Christensen and Pauly, 1992), with MTE (Brown et al., 2004), and, above all, to provide an updated, simplified, revised and corrected set of trophic equilibrium equations for size spectra, based on the seminal papers of Borgmann (1987), Andersen et al. (2009), and Barnes et al. (2010).

Such a general theory should be able to: 1.) explain the ubiquity and constancy of the marine pelagic size spectrum slope “β”, i.e., solve the “*size spectra constancy problem*”, 2.) resolve several fundamental paradoxes and problems in marine ecology (e.g., the “*constant size spectrum - variable trophic dynamics paradox*” and the “*phyto- zooplankton size spectra paradox*”), 3.) provide a unique, singular solution with β = -1, and 4.) produce a set of quantitative, testable predictions. We start from the observation of an invariant, constant slope of the size spectrum and first principles, towards the establishment of fundamental equations, leading to a discrete set of ten testable predictions.

Most equations presented here were obtained through simple rearrangement, revised and selected based on matching terms and units, and were subsequently tested with extensive simulations with synthetic food webs and example data from literature. Others were obtained directly from simulations with simple synthetic food webs (e.g., equation 4). All logarithmic equations in this study would hold true for any arbitrary log base. For the sake of simplicity, we assume and use base-10 logarithms throughout the text. All simulations and input data are available at *github.com/rschwamborn/pets*.

## 2. Towards a general size-spectra equilibrium model, I: trophic efficiency “E” and biomass-body-mass-transfer efficiency “BBTE”

Any attempt to understand the size spectrum of a given food web will be centered on some measure of trophic efficiency. Several descriptors and units have been suggested for trophic efficiency (Table 1). The most common expression is the “Ecotrophic Efficiency” **EE** (EE = ($P_{exported}$ + $P_{used}$ ) / $P_{total}$ ), the ratio of used (by fisheries and predators) and exported production to the total production in a given ecosystem compartment (Kozlovsky, 1968, Polovina 1984,

Christensen and Pauly, 1992, Gascuel et al., 2011, Eddy et al., 2021). EE is similar (but not identical) to production-specific trophic efficiency **TE** (TE = $P_{TLi+1}$ / $P_{TLi}$ ), the ratio of the production in a trophic level to the production of its prey (i.e., TE is the standard "Lindeman"-type trophic transfer efficiency).

For simplicity and convenience, a log-scaled "biomass transfer efficiency" (**E**), is proposed and used here (eq. 2), where **E = $\log_{10}$(dB) / dTL**):

$$E = \log_{10}(B_{i+1}) - \log_{10}(B_i) / TL_{i+1} - TL_i \qquad (2)$$

, where Bi = total biomass in size class "i", $TL_i$: trophic level in size class "i".

The decline in biomass across trophic levels, E = d log(B) / dTL, defines the mass-specific trophic efficiency "E". Similarly, the decrease in biomass with body mass, b = dlog(B) / dlog(M), i.e., the slope of the spectrum, may be interpreted as another form of trophic efficiency. Thus "β" may also be called the log-scaled, universally constant (β = 1) "*biomass-body-mass-transfer efficiency*" (**BBTE**).

Within PETS, large predator-prey size differences imply larger differences in biomass, within a constant system efficiency, following a strictly **inversely proportional** relationship (Fig. 2). Empirical evidence from numerous ecosystems (e.g., Dugenne et al., 2024; Ersoy et al., 2025) consistently shows BBTE = b = −1, suggesting a constant, scale-invariant trophic efficiency (a universal proportionality between biomass and 1/M), independent of taxonomy, temperature, or metabolism. Simply put, the universally constant BBTE of -1 means that when body mass increases by a given factor (e.g., factor 10), biomass equally decreases by the same factor.

An immense amount of information on ecotrophic efficiency ("EE"), based on ECOPATH trophic models, is available in the literature and within the EcoBase database (Eddy et al. 2021). Conversely, there are few estimates of WPTE, and E available in the literature. These parameters (EE, TE, BBTE, and E) are not identical, but closely related. They may be numerically converted to each other, especially if there is a size-structured food web (i.e. trophic level increases with size, leading to a constant PPMR). Although a size structured-food web has been generally assumed for pelagic ecosystems, there are very few studies that have actually proven the correctness of this paradigm on an ecosystem level (Figueiredo et al., 2020). In spite of otherwise very precise terminology, most previous studies, such as Barnes et al. (2010), kept their definition of trophic efficiency surprisingly vague and ambiguous ("the ratio of the production of a *trophic level or mass category* to that of its prey", Barnes et al., 2010), which highlights the difficulties and challenges associated with this key ecosystem parameter.

Here, several different types of trophic efficiency parameters were tested for their usefulness in predicting β = -1 (Table1), within possible solutions for the "size spectra constancy problem".

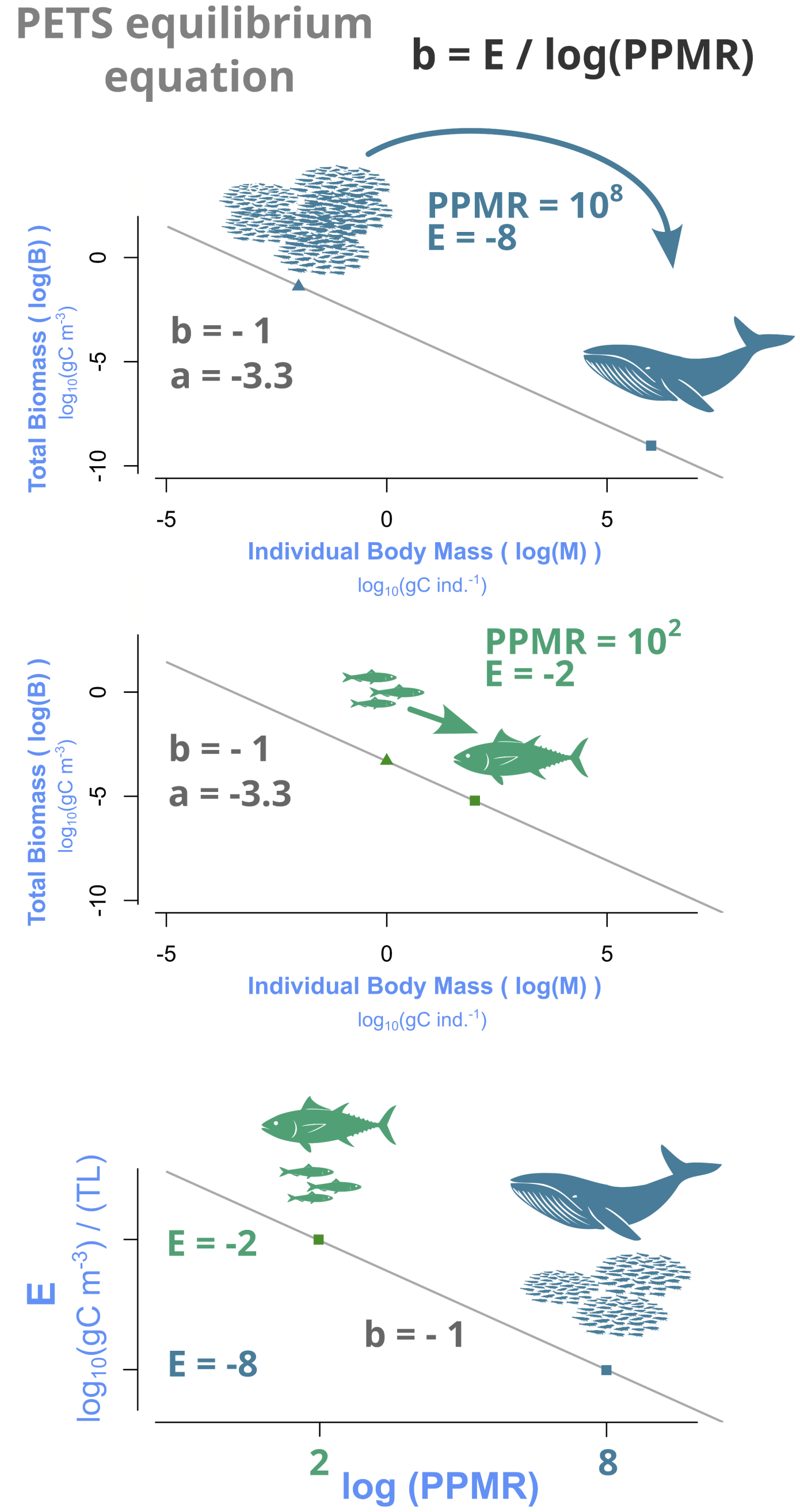

**Figure 2:** Two examples of hypothetical predator – prey pairs and respective "E" (mass-specific trophic efficiency) and "PPMR" (predator - prey mass ratio) equilibrium values. Above: Antarctic krill and baleen whales. Center: sardines and tuna. Below: E vs log(PPMR) plot. Hypothetical examples, for illustration purposes only (not based on actual data). Schematic illustrations of four example organisms were generated with the aid of artificial intelligence, and are not to scale.

**Table 1:** Trophic efficiency definitions and equations.

| Trophic efficiency descriptors tested | Composition and rationale | Selected references |
|---|---|---|
| EE | **Ecotrophic Efficiency**: EE = Losses & Transfers / P or: EE = 1 - $Z_0$ / (P/B) | Polovina (1984), Christensen and Pauly (1992) |
| LE | "Lindeman efficiency, the ratio of total metabolic energy fluxes at trophic level 1 to those at level 0". | Brown et al. (2004) |
| TLPRE | TL-specific production ratio efficiency: $P_i$ / $P_{i-1}$ / $TL_i$ - $TL_{i-1}$ | e.g., Barnes et al. (2010) |
| WPRE | weight-specific production ratio efficiency: $P_i$ / $P_{i-1}$ / $M_i$ - $M_{i-1}$ | this study |
| TLPCE = TE | TL-specific production change efficiency, "Lindeman trophic transfer efficiency", or "**production-specific trophic efficiency**" TE: dP / dTL | e.g., Lindeman (1942), Koslovsky (1968) |
| TLBCE | TL-specific biomass change efficiency: dB / dTL | e.g., Lindeman (1942) |
| WPCE | weight-specific production change efficiency: dP/ dM | e.g., Kozlovsky (1968) |
| WBCE | weight-specific biomass change efficiency: dB / dM, similar to β | e.g., Borgmann (1982) |
| BBTE = β | biomass-body-mass-transfer efficiency BBTE: log-scaled M-specific biomass change efficiency: BBTE = d log(B) / d log(M), identical to the slope (β) of the biomass-weight spectrum | this study |
| MSTE = E | Log-scaled, TL-specific "**mass-specific trophic efficiency**": E = d $\log_{10}$(B) / dTL | this study |

* : TL: Trophic level

## 3. Towards a general size-spectra equilibrium model, II: PPMR, food chain length, and trophic level scaling "S"

Another key parameter for the understanding of size spectra is the predator / prey mass ratio (**PPMR**). Together with food chain length (e.g., Zhou, 2006) and trophic efficiency, PPMR has been among the most often cited parameters to explain variations in size spectra slopes. Barnes *et al.* (2010) demonstrated that PPMR increases with predator body mass M (PPMR = $M^{0.24}$). Thus, under equilibrium assumptions, a log-linear increase in predator mass (individual mass $M_{Predator}$) leads to an increase in trophic level (TL) and to an increase in PPMR, as well as a decrease in weight-specific production change efficiency WPCE (Barnes et al., 2010) and a decline in E ( "E = -log(PPMR)" ). Thus, larger animals should have lower (inversely log-proportional) mass-specific trophic efficiency E ( "E ~ - 0.24 * log(M)" ). This is in line with the well-established fact that larger animals have slower growth rates dM/dt and much lower food assimilation efficiency k1 ( "k1 = dM / ingested food", Pauly, 1986).

If there is a vast range of PPMR values, across many orders of magnitude (Barnes et al., 2010) and β is constant, the only possible conclusion is that there must exist a compensation mechanism in nature, involving a change in trophic efficiency as a function of PPMR, or vice-versa.

Thus, the ubiquity of the invariant ecosystem size spectrum slope β (e.g., Dugenne et al., 2024, Ersoy et al., 2025) implies that there is a constant log("some form of trophic efficiency") / log(PPMR) ratio. Not surprisingly, studies that used a fixed value for PPMR (e.g., PPMR of $10^4$, Mehner et al., 2018), found a linear positive relationship between size spectra slope "b" and trophic efficiency, with steeper slopes representing less efficient ecosystems. Conversely, other studies have assumed an arbitrary, *a priori* fixed value for trophic efficiency (e.g., 70%, Zhou, 2006), and explained variations in size spectra slopes by changes in PPMR and food chain length (Zhou, 2006). Conversely, recent studies (e.g., Hunt et al., 2015, Figueiredo et al., 2020) used the size spectra slope and PPMR values obtained from nitrogen stable isotopes, to estimate trophic efficiency, utilizing the equilibrium equations of Barnes *et al.* (2010).

PPMR describes the Size-TL relationship, and can be directly related to d (TL) / d (M)

$$PPMR = M_{Pred} / M_{Prey} \quad (3)$$

, where $M_{Pred}$ = mean individual mass (weight) of any given prey or predator.

When the weight of the smallest and largest animal in the food web are known, PPMR can be be used to assess the exact food chain length "$F_{exact}$", where

$$F_{exact} = 1 + ( \log( M_{top} / M_{base} ) / \log( PPMR ) ) \quad (4)$$

$M_{base}$: mean body mass of the smallest size bin, at the base of the food chain

$M_{top}$: mean body mass of the largest size bin, i.e., apex predators, in a size-structured food web

When $M_{base}$ and $M_{top}$ are unknown, PPMR may be directly converted into a convenient weight-specific **trophic level scaling coefficient** (S)

$$S = TL_{i+1} - TL_i / \log(M_{i+1}) - \log(M_i) \quad (5)$$

If $TL_{i+1} - TL_i = 1$, and PPMR = $(M_{i+1} / M_i)$, this can be solved as

$$S = 1 / \log(PPMR) \quad (6)$$

S may also be interpreted as a form of intuitive, convenient "*relative food-chain length*".

Typical values in marine pelagic ecosystems may vary from PPMR = 2 to PPMR = 100,000, within realistic limits, as observed for fish and squid (Barnes al., 2010), and for zooplankton (Hunt et al., 2015). When including baleen whales, PPMR values of $10^8$, $10^9$, or even higher may be possible (Fig. 2).

Although PPMR might seem be intuitively simple and apparently easily quantified, at first sight (i.e., by measuring the weight of a predator and its favorite prey), it is actually an immense challenge to determine mean PPMR across an ecosystem, or even, within any given size bin of a size spectrum. That is because predator-prey interactions are highly stochastic and dynamical, and each size bin often contains an immense number of species and life history stages, each with its specific selectivity behavior and diet, leading to an immense number of individual diet compositions within each size bin. This complex, detailed information is not realistically to be obtained *in situ* within any reasonable time limit.

Yet, PPMR can be assessed by bulk analysis of stable nitrogen isotopes in size-fractioned zooplankton samples (Figueiredo et al., 2020), assuming that there is a well-known, constant and reliable relationship between nitrogen stable isotope signature and trophic levels. However, there is considerable uncertainty regarding nitrogen stable isotope fractionation (the relationship between $\delta^{15}N$ and TL). Thus, the conversion from stable isotope data to trophic scaling coefficients S (and thus PPMR) is possible, but it suffers from considerable uncertainty. Up to now, only two studies (Hunt et al., 2015; Figueiredo et al., 2020) have actually attempted to quantify S, and thus PPMR, through nitrogen stable isotopes.

## 4. Towards a general size-spectra equilibrium model, III: the quest for a *universal trophic equilibrium constant* "c"

Based on the observations above, we may start from the premise that trophic efficiency (in forms such as EE, TE, growth efficiency, assimilation efficiency, or log-scaled TL-specific biomass transfer efficiency "E" (proposed herein), and PPMR are two key parameters that interact within an intrinsic compensation mechanism, that stabilizes the size spectrum, leading to a constant slope of $\beta = -1$ (Dickie, 1976, Borgmann, 1987, Boudreau and Dickie, 1989, Brown, et al., 2004, Andersen et al., 2009, Barnes et al., 2010, Woodson et al., 2018).

Numerous equations have been proposed to describe log("efficiency") / log(PPMR) ratios, and to explain size spectra slopes, such as:

$\log(LE) / \log(PPMR) \sim -1/4$ Brown et al. (2004), where

LE: "Lindeman efficiency, the ratio of total metabolic energy fluxes at trophic level 1 to those at level 0" (Brown et al., 2004)

$\beta_{weight\text{-}biomass} = ( z' / g ) + g''$ Zhou (2006), where

g: mean weight-specific growth rate, defined as $g = (1/M)\ (dM/dt)$

z': mean specific rate of net abundance change (i.e., mortality), defined as $(1/N)(dN/dt)$, where N is number of individuals.

g'': a species-specific growth constant.

Trophic equilibrium statement of Zhou (2006): "*the assimilation efficiency and number of trophic levels represent the trophic structure of a community*", with a vaguely defined "*assimilation efficiency*". The "*energy … assimilation or trophic transfer efficiency*", is assumed to be constant at 70% (Zhou, 2006).

$b_{weight\text{-}abundance} = \log(TE'') / \log(PPMR) - 0.75$ Barnes et al. (2018), where

$b_{weight\text{-}abundance}$: slope of the "*numbers spectrum at equilibrium*"

TE'': vaguely defined trophic efficiency ("*ratio of the production of a trophic level or weight category to that of its prey*").

$\beta_{size\text{-}biomass}$ = 0.25 + log(TE''') / log(PPMR)   Woodson et al. (2018)

$\beta_{size\text{-}biomass}$ = change of biomass with linear size, i.e. size-biomass spectrum slope

TE''': vaguely defined trophic efficiency: "*energy … assimilation or trophic transfer efficiency*", between 2.4% and 40% (Woodson et al., 2018).

It is surprising how vaguely "efficiency" is defined in most examples above. This may be due to the fact that none of these equations are able to actually produce a numerically precise size spectrum slope, that is quantitatively identical to the ubiquitous $\beta_{weight\text{-}biomass}$ = -1. Thus, instead of interpreting them as numerically exact predictions, the above equations may rather be considered as relevant conceptual, qualitative models.

The first step towards a numerically exact solution for β will be to verify if a prediction of the β = -1 slope can actually be achieved based on two such parameters (one parameter related to PPMR and one parameter related to trophic efficiency). The next step will be to verify whether, and if so, under which conditions, additional terms may be necessary to explain the size spectrum slope.

Thus, we may preliminarily assume that there is an equilibrium between log("some form of trophic efficiency") / log(PPMR). This equilibrium may be formalized in the form of a trophic equilibrium constant "c" (see below). It seems obvious that the constant "c" should reflect a mechanism where adjustments in PPMR (i.e., in prey choice selectivity of predators) compensate for variations in trophic efficiency (in forms such as "EE", "TE" or "E").

There is a near-infinite list of possible equations for "c", that may qualify as "equilibrium equations". Below are a few possible examples. Here, we tested and considered several possible equations. One important requirement is that the term obtained for "c" should be dimensionless (since "β" is dimensionless, as any ratio of logarithms).

*A few examples of possible equilibrium equations:*

c1 = E / log($M_{Pred}$) / log($M_{Prey}$)

c2 = log(E) / log($B_{Pred}$) / log($M_{Prey}$)

c3 = log(E) / log($M_{Pred}/M_{Prey}$) ; conceptually similar to Brown et al. (2004) and Barnes et al. (2010),

c4 = log(“the ratio of the production of a trophic level or mass category to that of its prey“) / log(PPMR) -m; as in Barnes et al. (2010)

c5 = ( log(E) / S ) - u,           , where u = arbitrary constant (e.g., u = 6.5)

c5 = E / (u’ * S)                    , where u’ = arbitrary constant (e.g., u’ = 229)

c6 = E * S                    ; proposed equation for the trophic equilibrium constant “c”

The last equation in this list, c = E * S, stands out, not only for its simplicity, but also because it is the only equation that leads to perfectly matching terms and units, considering that:

β = d(logB) / d(logM) = -1, (7)

E = d(logB) / dTL, and (8)

S = dTL / d(logM). (9)

Hence, after combining the three equations above (equations 6, 7 and 8) into: E = β / S, we conclude that under equilibrium conditions, the slope of the size spectrum can be predicted from trophic level scaling S, combined with TL-specific biomass transfer efficiency E, into the simple **PETS equilibrium equation**:

**β = E * S** , which is identical to **β = E / log(PPMR)** (10)

A substantial body of prior size-spectrum theory has demonstrated that some form of trophic efficiency and PPMR are functionally coupled (e.g., Brown and Gillooly, 2003; Jennings and Mackinson, 2003; Brown et al. 2004; Andersen et al., 2008; Reuman et al., 2009; Andersen et al. 2009; Barnes et al., 2010, Coghlan et al., 2022). The counterbalance of E and PPMR should not be surprising at all, considering the simple mathematical configuration of the size spectrum slope b and the units and dimensions of E and PPMR, as can be demonstrated and proven very simply and unequivocally: if the slope β = d(logB) / d(logM) = -1, the mass-specific efficiency E = d(logB) / dTL, and PPMR = ($Mi+_1$ / Mi), then β = E / log(PPMR) and E = −log(PPMR).

This simple form of the equilibrium equation (“b = -1”, hence: “E = −log(PPMR)” ) illustrates most clearly that large predator-prey size differences (e.g., baleen whales feeding on krill) correspond to a low biomass-specific trophic efficiency E (e.g., krill has a much higher total biomass than whales, see Fig. 2), maintaining the universal β spectrum slope of approximately b = -1 (in biomass vs mass units), while allowing biomass intercepts to vary.

This PETS equilibrium equation also explains why **trophic pyramids** (i.e., TL *vs* B relationships, i.e., **variations in E** within food webs) can be highly variable, but the size spectrum slope β is invariant (stabilized by the E *vs* PPMR relationship).

Far from being a "universal equation", eq. 10 is based on a series of 9 assumptions:

I.) Equilibrium conditions (i.e., the slope β reflects a mean state, covering many years and many datasets; it may not apply to data from a single sample, day, or season).

II.) Constant metabolic scaling (i.e., "m = 0.75", Kleiber 's law, Kleiber, 1932).

III.) Trophic control (top-down) through density-dependent predation.

IV.) A strictly size-structured food web (TL increases with size).

V.) Constant, complete vulnerability of all prey populations to predation (v = 1).

VI.) The dataset used to estimate β covers many orders of magnitude in body mass, covering many species and populations, where β is determined by d log(B)/d log(M) across many populations (then, the E / log(PPMR) ratio determines the slope). This does not apply to analyses within a single population, or a single species (where the growth / mortality ratio of the population dominates the slope of the size distribution).

VII.) There is no biomass-reducing or biomass-limiting, non-predatory stress (e.g., nutrient limitation stress, oxygen stress, fisheries, sound stress, ... ).

VIII.) The ecosystem is fully three-dimensional (i.e., pelagic), and all processes act in three dimensions.

IX.) The equilibrium mechanism is scale-, and size-invariant (i.e., it acts in the same way at all scales and sizes)

A tangible example of the PETS equilibrium mechanism is the comparison of plankton size spectra in oligotrophic tropical waters vs highly productive upwelling ecosystems (Schwamborn et al., submitted). Highly productive, high-abundance upwelling ecosystems can be expected to be more efficient, due to shorter search times and less search energy necessary for foraging. This could lead to expectations that more productive, more efficient ecosystems should have a flatter size spectra slope, with more large-sized organisms, while oligotrophic systems that are less productive and less efficient should have a steeper ecosystem-wide slope, with less large-sized organisms. Conversely, recent large-scale analyses have shown that there is no such effect, with phyto- and zooplankton NBSS slopes being largely invariant at b ~ -1 (Schwamborn et al., submitted).

Clearly, this is an example of differences in trophic efficiency being compensated by variations in PPMR. Simply put, in more efficient, densely populated, highly productive upwelling systems, higher trophic efficiency is compensated by a higher PPMR (i.e., larger predators such as large-sized copepods and euphausiids feeding on microphytoplankton, with PPMR > $10^4$). Conversely, in less efficient, sparsely populated tropical waters, the low efficiency of the food web is likely compensated by lower PPMR (e.g., small-sized copepods, such as Oithonidae, feeding on large ciliates and giant Rhizaria, with with PPMR < $10^4$).

Numerical example (Fig. 2): If we assume a common PPMR value of $10^4$ (Brown et al., 2004) and β = -1, a realistic example for E could be E = -1 * log10(PPMR) = -4. It is however, important to remember that PPMR can vary across an immense range, e.g., from 10^2 to 10^9, (then, E may vary from -9 to -1). Both PPMR and E are currently near impossible to estimate accurately *in situ* across a complete food web (Hunt et al., 2015, Figueiredo et al., 2020). A simple, intuitive illustration of PPMR and E values and the proposed equilibrium, with two hypothetical predator-prey pairs, may be found in Fig. 2.

It may be possible to transform the PETS equilibrium equation ( eq. 10 ) into units of production (P), instead of biomass (B), if proper conversion factors (i.e. mean P/B ratio for each size class) are available. The commonly used production-specific trophic efficiency (i.e. “ecotrophic efficiency”, “EE”) cannot be constant across different food webs with different PPMRs, considering that this trophic depends on the size ratio, on the metabolism of predators, prey, and food conversion efficiencies, all of which depend on size. This rationale may also contribute to explain why trophic pyramids are highly variable (i.e , the “*constant size spectrum - variable trophic dynamics paradox*”). However, it is still common so see in textbooks that EE should have a specific universal value (for example, 10%), which is incorrect, considering the above. EE is obviously highly variable in nature, with no single value to be expected. Similarly, PETS states that mass-specific “E” is highly variable, but within a stringent relationship with PPMR.

The PETS equilibrium equation (eq. 10) reflects the idea that when E increases by a factor “x”, there is an x-fold decrease in relative food chain length S, a linear increase in -log(PPMR) by x, (i.e., a decrease in PPMR by x orders of magnitude). Similarly, under equilibrium assumptions, an increase in individual mass by “x” orders of magnitude between size classes i and i+1, is followed by a decrease in total biomass in exactly the same order of magnitude. This equilibrium is well represented by the proportional slope (-1) in the pelagic size spectrum.

Simply put, we may conclude, from the initial observation of β = -1, that B and M are inversely proportional ( “B ~ 1/M“ ), that log(M) and log(B) are *linearly (negatively) proportional*, and finally, and that “E” is also *linearly (negatively) proportional* to “log(PPMR)”. All these observations, transformations, and conclusions reflect different aspects of the same equilibrium, i.e., proportionality.

*From describing the mean state towards understanding dynamic spectral stability*

The PETS equilibrium equation does not explain the dynamics maintaining the equilibrium; rather, it is an expression capturing the relationship between mean values of static variables (β, PPMR and E). To mechanistically explain how this equilibrium is maintained, one must consider specific predator-prey, top-down processes that dynamically act as responsive stabilizing forces when deviations from the equilibrium β spectrum occur (e.g., TOPS, RISE, SOFT, PATER, and WETBIO mechanisms, see below).

*The **TOPS** hypothesis (Top-down Overshoot Peak Suppression)*

The most simple and obvious mechanism for maintaining an invariant spectrum is based on the **TOPS** Hypothesis (**T**op-down **O**vershoot **P**eak **S**uppression, Fig 3). Herein, ecosystems self-organize towards a log-log linear biomass spectrum by top-down control. During population growth phases (low mortality and high growth), biomass increases until approaching the maximum carrying capacity spectrum. When population biomass crosses a discrete threshold (i.e. when B overshoots the maximum carrying capacity spectrum **MCCS**), top-down controls (predation, behavioral regulation, antagonistic interactions, trophic feedbacks) kick in, actively suppressing the transient overshooting biomass peak. The proposed TOPS mechanism is probably the main process that maintains a near-steady, scale-invariant spectrum.

However, natural ecosystems are far more complex than this simplified conceptual model. TOPS responses of consumers to ephemerous food peaks likely involve Lotka–Volterra-type dynamics, complex bloom dynamics, prey defenses, and variations in palatability and morphology (e.g., spines), all of which are aspects that challenge simplistic descriptions of top-down control in ecosystems. Consumers are not simple “lawnmowers” (Fig. 3) and trophic interactions are strongly context-dependent. Given the fast and transient nature of the TOPS response to ephemeral perturbations of the spectrum, this mechanism is probably very difficult to observe in nature. Testing the TOPS hypothesis will thus most likely require controlled experiments (e.g., in the laboratory and *in situ*).

*Algal blooms, TOPS and the “Loophole” hypothesis*

Ephemeral algal blooms (regularly observed in many pelagic ecosystems), could be, at first sight, potential candidates for testing the TOPS hypothesis. Yet, the immense available literature on algal blooms (including harmful blooms, i.e. “red” tides) has shown a large body of evidence for numerous trophic and biogeochemical processes involved in early bloom formation and subsequent decay (Fig. 3). Nutrients, light, oxygen, ammonia, and turbulence, together with the taxonomic composition of producers and grazers, palatability, and sinking dynamics, all interplay in complex non-linear ways throughout the bloom succession. One example of this complexity, and the importance of top-down control, is the “Loophole” hypothesis (Irigoien et al., 2005), which postulates a top-down control and relevant predation avoidance mechanisms of phytoplankton under early bloom conditions. Combining concepts of TOPS and Loophole hypotheses, we may predict that early blooms occur when the maximum carrying capacity spectrum is overrun by a taxon that is temporarily able to evade TOPS control (e.g., by synthesizing toxic or grazing-deterrent substances, or by fast growth, far above grazing rates).

*Upwelling ecosystems - pushing the biomass spectrum beyond the maximum*

In upwelling ecosystems (e.g., off Peru, California, Senegal, and Namibia), year-round strong nutrient flux leads to high primary production, increased food density, efficient grazing, and exceptionally high zooplankton biomass (Fig. 3). Under such hypereutrophic conditions, the phytoplankton biomass spectrum, and especially the zooplankton biomass spectrum, can exceed the predicted global pelagic maximum carrying capacity spectrum (Fig. 3). In contrast to the ephemeral nature of algal blooms, hypereutrophic systems, such as upwelling ecosystems and coastal bays with extremely high nutrient input (e.g., Guanabara Bay, Brazil; Schwamborn et al., 2004) can maintain persistently high biomass at all trophic levels for extended periods (see review in Ayón et al., 2009). Such “uplifted” biomass spectra, with zooplankton biomass several orders of magnitude above predictions, have recently been described for the Benguela Upwelling system by Fock et al. (2026) and Schwamborn et al. (submitted).

Within the TOPS terminology, we may summarize that in upwelling ecosystems, the maximum carrying capacity spectrum is significantly modified (“lifted up”), since growth rates exceed predation mortality until a new equilibrium spectrum is reached at higher biomass levels. While the intercepts are shifted upwards, the slope of both spectra (phyto- and zooplankton) is kept invariant at b = -1 by the PETS equilibrium processes (E vs PPMR equilibrium) and by the SOFT - PATER regulation mechanisms (see below).

*The **RISE** Hypothesis (Replacement in Size-structured Ecosystems)*

In contrast to the **TOPS** model, which describes the top-down regulation of excess biomass above the maximum carrying capacity spectrum, the **RISE** Hypothesis (**R**eplacement **i**n **S**ize-structured Ecosystems, Fig 4) describes a mechanism for compensating for *negative deviations* from the smooth log-linear spectrum (i.e., filling gaps and troughs). Essentially, the RISE hypothesis is intended to explain the mechanisms through which all size classes are seamlessly occupied by taxa that instantly fill in any empty size niches, as observed in estuarine and marine ecosystems (Schwamborn et al., 2025).

RISE implies that ecosystems maintain a constant size spectrum despite numerous dynamic processes (varying species composition, advection, local extinctions, massive larval release, blooms, mass mortalities, etc.), through functional replacement within well-defined size niches. Empty size niches (i.e., gaps and troughs in the spectrum) are filled by well-described mechanisms such as adaptive radiation, invasion, density-dependent population growth, and speciation. After any possible gap-generating disturbances (e.g., mass mortalities), species will be replaced by functionally similar, similar-sized, organisms (Fig. 4).

Size-niche-specific functional replacement can also occur seamlessly (without any transient gaps or disturbances in the spectrum, see Fig. 4) through antagonistic and predator-prey interactions, thus maintaining the log-linear biomass size spectrum through compensatory ecological turnover.

The RISE hypothesis, based on the “size niche” concept (*sensu* Schwamborn et al., 2025), may also provide an important link between PETS, Niche and Neutral Theory (refs Neutral Theory). The size niche concept is evidently linked to the classic idea of ecological niche (refs classic ecology with niches), i.e,. the occupation of a size class by a specific taxon and life history stage. Neutral Theory and Niche Theory are both relevant to the occupation of specific beliefs and Functional replacement, and in explaining the diversity of coexisting taxa within given size classes.

While classical Niche Theory is extremely deterministic (each taxon occupies a well-defined ecological niche), Neutral Theory suggests that species assemblages can be replaced randomly within any given niche, with some taxa thriving and others declining by mere chance (refs). Thus, species at the same trophic level are functionally equivalent (refs). Stochastic reproduction, recruitment, feeding, death, dispersal, and extinction shape communities within the Neutral Theory. In this context, “neutral” means that many traits and genes are neutral, not being selected by evolutionary processes, allowing for stochastic competition and succession processes to happen as often observed in nature, e.g., in apparently “chaotic” phytoplankton species succession events in the laboratory and *in situ* (refs). Within the RISE framework, the size spectrum shape is deterministic (with well-defined size niches, according to PETS theory), but the actual taxon occupying a specific size niche may indeed stochastically vary in space and time, following Neutral Theory. Thus, RISE, as a key brickstone of the PETS theory, may be summarized as an integration of size-spectrum approaches, the classical ecological Niche concept, and neutral Theory.

With the PETS framework, dynamic spectral stability emerges from *RISE-TOPS coupling* (overshoot suppression and compensatory size niche replacement) and *SOFT-PATER coupling* (adaptive foraging-driven trophic efficiency regulation and prey-abundance feedback, see below), operating within the constraints of the carrying capacity spectrum (*WETBIO hypothesis*, see below).

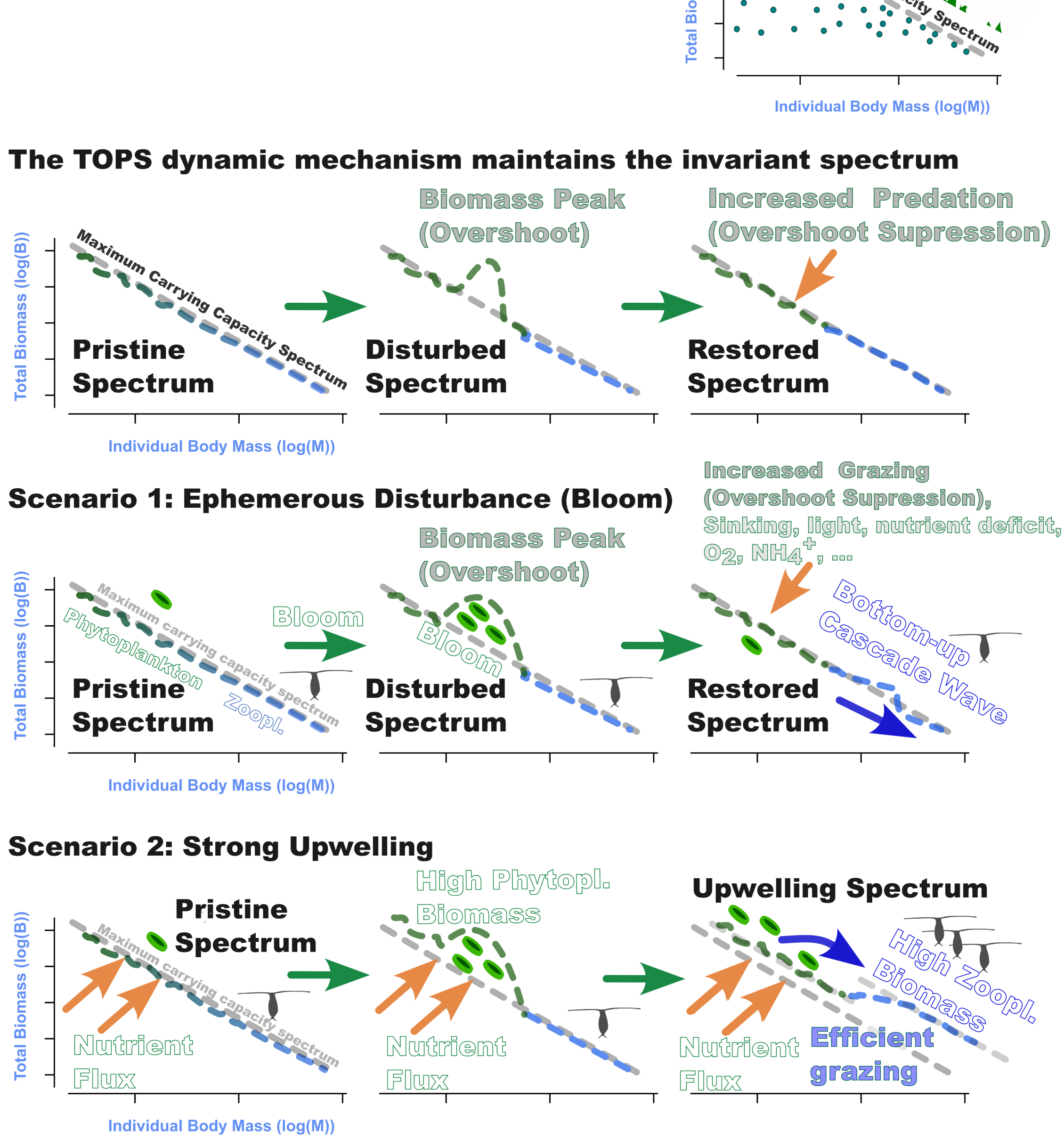

**Figure 3.** Schematic representation of the ***TOPS*** Hypothesis (**T**op-down **O**vershoot **P**eak **S**uppression). Herein, ecosystems self-organize towards a log-log linear biomass spectrum by top-down control. During population growth phases (low mortality), biomass increases until approaching the maximum carrying capacity spectrum. Above: basic TOPS control of the biomass spectrum. Center: Algal bloom dynamics under TOPS. Below: Strong, continuous nutrient flux in upwelling ecosystems leads to high primary production, surpassing the TOPS regulation, leading to uplifted spectra for phyto -and zooplankton.

**RISE: Replacement in Size-Structured Ecosystems**

**Scenario 1: Trough filling (Recovery)**

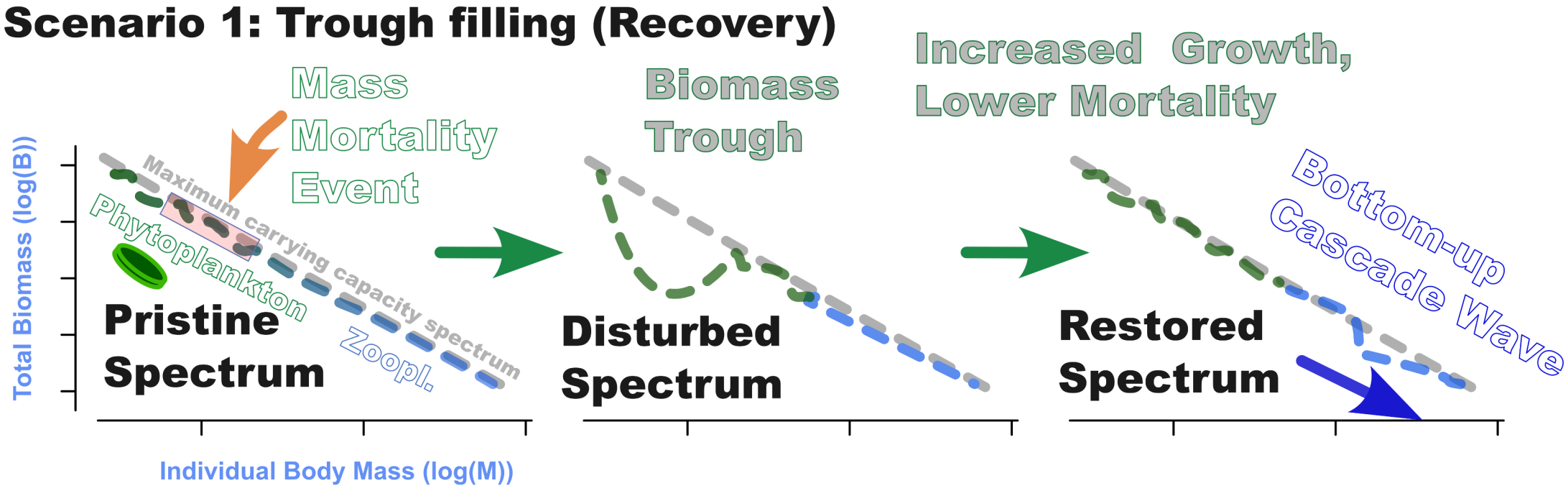

**Scenario 2: Gap filling (Functional Replacement)**

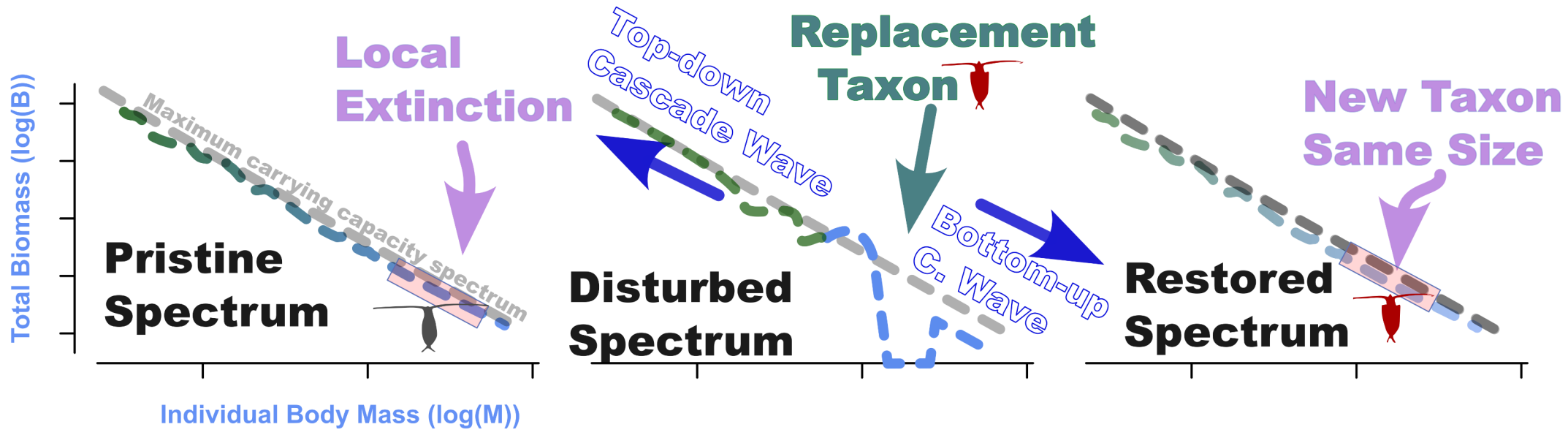

**Scenario 3: Invasive Functional Replacement (Gap carved and filled)**

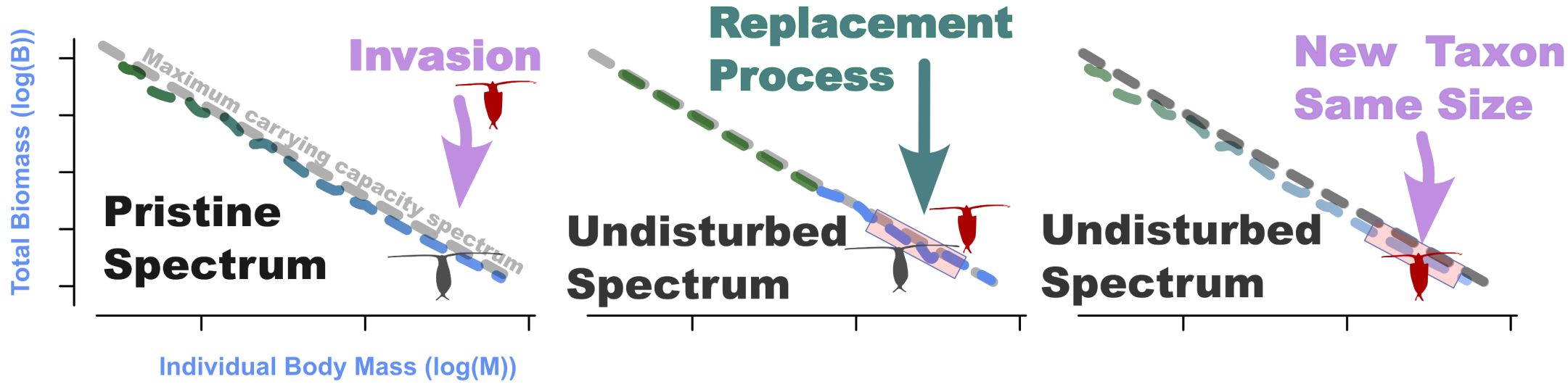

**Figure 4. Schematic representation of the *RISE* Hypothesis (Replacement in Size-structured Ecosystems).** RISE posits that ecosystems maintain a constant size spectrum despite variable species composition and turnover, through the functional replacement of species within well-defined size niches. Empty size niches (i.e., gaps in the spectrum) are filled by well-described mechanics, such as adaptive radiation, invasion, and speciation. Following perturbations, species are replaced by functionally similar organisms via antagonistic and predator-prey interactions, restoring the log-linear biomass size spectrum through compensatory ecological turnover. Above: recovery after perturbation, by the same species. Center: functional replacement after perturbation, Below: continuous invasive functional replacement.

# 5. Towards a general size-spectra equilibrium model, IV: Compensation and equilibrium mechanisms, based on "size-spectra-specific optimal foraging theory" (SOFT) and top-down trophic cascades

There are many mechanistic and dynamic models, and theoretical frameworks, that can plausibly explain the trophic equilibrium equation suggested in this study. Here, we may focus on the most straightforward and simple mechanism: the well-described adaptations in feeding selectivity behaviour of predators, as a function of prey availability, also known as optimal foraging theory, OFT (e.g, MacArthur and Pianka, 1966; Werner and Hall, 1974; Krebs, 1977).

Size-specific OFT is based on two fundamental paradigms I.) when a predator has the choice between different prey, large-sized prey is generally preferred; II.) the choice of whether to capture and ingest small-sized prey (less attractive, but more abundant) or not, is dependent on the abundance of its preferred large-sized prey.

Optimal foraging theory, as applied to size selectivity within a power-law-shaped size spectrum ("**S**ize-spectra-specific **O**ptimal **F**oraging **T**heory" **SOFT**), may be summarized as follows: In times of overall **food scarcity**, predators feed less selectively (i.e., they consume whatever they can find), leading to the consumption of **smaller prey items**, which are several orders of magnitude more abundant than larger prey (power-law distribution). Conversely, **when food is abundant, predators can be more "picky", selecting their preferred large-sized prey**. The effect of SOFT behaviour on the size spectrum equilibrium may be illustrated in two extreme perturbation scenarios. In both scenarios, a perturbation in the system occurs, which is then compensated through the proposed SOFT mechanism (Fig. 5). Such a perturbation may be a sudden decrease or increase in predator biomass, due to fisheries or due to the appearance of exotic predator species, respectively.

In these scenarios, a hypothetical perturbation of the continuous size spectra slope happens initially due to a change in apparent trophic efficiency. This change in "E" is then compensated by changes in feeding behavior, PPMR, and S, leading to a restoration of the original continuous size spectrum slope, but with a change in overall food web biomass. From this simple mechanism it is possible to predict that food webs with higher apparent E will have a higher biomass, but invariant size spectra slope ($\beta$ = -1). These dynamic stabilization mechanisms for the size spectrum are Illustrated schematically in Fig. 2 and briefly explained:

The first scenario (Scenario 1 in Fig. 5) is that of a sudden decrease in apparent E, where the first step (perturbation) leads to a decrease in predator biomass in the system. Subsequently, predators find themselves in an environment with high relative abundance of large prey, and will become more selective, consuming mostly their preferred large prey. This leads to a decrease in biomass of large prey, re-establishing the original size spectrum slope between predator ad prey, but with lower total biomass within this food web.

Conversely, under a scenario of a sudden increase in E (Scenario 2 in Fig. 5), there would be an initial increase in the numbers of predators. In this scenario, predators will find themselves in an environment with food scarcity. This leads to a change in behavior, i.e. predators will switch to smaller prey, which is much more frequently encountered (power-law distribution). This leads to an increase in PPMR. Large prey items then experience decreasing predation mortality, which leads to an increase in their biomass, until the original size spectrum slope is restored, with higher total biomass within this part of the food web.

If we consider the unquestionable observation that immense, hitherto unexplained variations in PPMR (by many orders of magnitude, e.g., Brown et al., 2004) do occur (" We do not yet have a mechanistic theory to explain…" Brown et al., 2004), and β is constant and invariable, then we may have to search for equations that help us to predict E, as a function of metabolism and ambient conditions, and then estimate PPMR as a function of E, utilising equation 10.

**The dynamic mass-specific trophic equilibrium within SOFT & PETS**

$\beta = E * S$

Scenario 1: decrease in mass transfer efficiency "E"

Starting Scenario → Decrease in "E" → Predators encounter more prey → Predators Switch to larger prey → Decrease in PPMR = Increase in dTL / dM = Increase in S → Restored equilibrium

Total Biomass (log(B)); Individual Body Mass (log(M)); $M_{i-2}$, $M_{i-1}$, $M_i$; TL = 1, TL = 1.8, TL = 2.6, TL = ?, TL = 2.8; $\beta = -1$

reduced Biomass at $M_{i-1}$

Scenario 2: increase in mass transfer efficiency "E"

Starting Scenario → Increase in "E" → Predators encounter less prey → Predators Switch to smaller prey → Increase in PPMR = Decrease in dTL / dM = Decrease in S → Restored equilibrium

Total Biomass (log(B)); Individual Body Mass (log(M)); $M_{i-2}$, $M_{i-1}$, $M_i$; TL = 1, TL = 1.8, TL = 2.6, TL = ?, TL = 2; $\beta = -1$

increased Biomass at $M_{i-1}$

**TL: Trophic Level**, TLpred = 1+ TLprey
**E: TL-specific mass transfer eficiency**, E = d(logB) / dTL
**PPMR: predator / prey mass ratio**, PPMR = $M_{pred} / M_{prey}$
**S: weight-specific trophic level scaling parameter**, S = dTL / d(logM), S = 1 / log(PPMR)
**SOFT: Size-Spectra-Specific Optimal Foraging Theory**
**PETS: Predator-Prey-Efficiency Theory of Size Spectra**

**Figure 5.** Schematic illustration of the proposed mass-specific trophic equilibrium (""β = E * S"), that leads to a constant, size-invariant size spectrum slope (β) of "β = -1", within the proposed "Size-spectra-specific Optimal Foraging Theory" (SOFT). Two potential scenarios of perturbations in TL-

specific mass transfer efficiency "E", are shown and their respective compensation response through changes in predator behavior, that lead to changes in PPMR (predator / prey mass ratio) and restore the original size spectrum slope (β). The mean trophic level ($TL_i$) is given for each weight bin "i". Scenario 1 shows a hypothetical example with an initial decrease in E. Scenario 2 shows the ecosystem response to an abrupt increase in E. Note that the size spectrum slope is maintained constant at "β = -1", but the mean trophic level ($TL_i$) of each weight bin is variable, depending on the E / PPMR ratio in each scenario. Total ecosystem biomass and biomass of weight bins Mi and $Mi_{-1}$ increase with increasing "E".

*Relationship of ecotrophic efficiency EE and mass-specific trophic efficiency E*

If we want to utilize the large amount of literature on ecotrophic efficiency EE, it may be useful to find a way to transform EE into E, which can be achieved through multiplication with the turnover rate $\tau = (P/B)$: $EE = P_{exported} / P_{total} = 1 - Z_0 / \tau$

, where $Z_0$ = "other" mortality, due to age and disease (not related to predation or fisheries).

First, we may analyze the relationship of EE to the TL-specific production change efficiency TLPCE = d log(P) / dTL, where TLPCE = log(EE).

Then, we just will have to convert P to Biomass:

$$E = \log(EE) - ( d \log(P/B) / d\, TL ) \qquad (11)$$

Based on this equation, we may conclude that

1.) Mass specific trophic efficiency "E" is positively related to "log (EE)"
2.) "E" is negatively related to d log(P/B) / d TL, the rate of change in P/B with TL.
3.) The log scale - relationship between log(EE) and E means that minor changes in EE will only slightly, possibly unnoticeably, affect E, and that E is probably near-constant, while EE is highly variable. This reflects the constancy of the size spectrum (double logarithmic scale) vs trophic pyramids (often not log scale, but linear, or cubic, or "schematic", e.g., Woodson et al., 2018).

Based on current literature on EE, we can preliminarily assume that:

4.) EE (and thus "E") tends to decline with trophic level: higher-level, larger animals have fewer predators, and larger animals have a less efficient food conversion (i.e., physiological assimilation efficiency, k1).

5.) EE (and thus "E") increases with prey abundance (higher prey encounter rates, smaller search volumes, and shorter search times, less energy used for foraging, … )

6.) EE declines with temperature (but E? … an important question mark…): Since current knowledge indicates that EE decreases with increasing temperature, and "d log(P/B) / d TL" is probably invariant with temperature, one may preliminary assume that "EE" and "E" both decrease with increasing temperature. However, it is also hypothetically imaginable that "d log(P/B) / d TL" changes with temperature in a way that it compensates for changes in EE. Yet, most observations hint at a decrease in E with decreasing temperature (e.g. huge PPMR (e.g., baleen whales feeding on krill), in polar regions, which will lead to high E values). This highlights the urgent need for more in-depth studies on temperature effects on poorly investigated key ecosystem parameters, such as the change in turnover with trophic level ("d log(P/B) / d TL").

*Trophic equilibrium equation in units of E, EE and TE*

If we assume that mass-specific growth rate scales as $\mu = -0.25$, mass-specific trophic efficiency E (E = d (logB) / dTL) can be transformed in production (P) units, and thus into EE ( Production consumed by predators or exported / Total Production, for each TL) and TE (Production-specific trophic efficiency TE = dP/ dTL). Individual biomass production (i.e., instantaneous body growth $g' = dM/dt$) may be described as: $g' \sim M^{0.75}$ and mass-specific growth ($g = dM/dt\ M^{-1}$) may be described as $P/B = \tau = g = M^{-0.25}$ (Fenchel, 1974, Kjørboe and Hirst, 2014). Thus, we may be able to express the trophic equilibrium equation (eq. 10, i.e., the equation that describes the slope "b" of the biomass-body mass spectrum) in units of EE and TE.

$$E = d(\log B) / dTL;\quad b = E / \log PPMR$$

$$PPMR = dM / dTL$$

$$P/B = M^{-0.25}$$ (Fenchel, 1974; Kjørboe and Hirst, 2014)

$$TE = dP / dTL = (dB / dTL) * (PPMR^{-0.25})$$

$$\log(TE) = E - 0.25 * \log(PPMR)$$

$$b = -1 = \log(TE) / \log(PPMR) + 0.25$$

If variations in EE are mainly driven by predation (i.e., exports and fisheries terms are small), then EE scales similarly to TE (EE ~ TE).

Barnes et al. (2010, citing Borgmann, 1987 and Andersen et al., 2008) stated that the "*numbers spectrum at equilibrium can be approximated as* $b_{numbers}$ = *(log TE / log PPMR) - 0.75), where 0.75 is the assumed scaling of consumption (assumed to be driven by metabolic rate) with body mass.*" Their equation, that subtracts the "Kleiber" mass-metabolism scaling of -0.75, is evidently different form our above equations (that adds +0.25, i.e. "Fenchel"-scaling). This may look, at first sight, as an inconsistency or error. Yet, this difference (+ 0.25 in the above equation *vs* -0.75 in previous studies) may be explained by the different spectra used (numbers spectrum vs biomass spectrum). The main issue for a quantitative comparison of our eq. 10 and the Barnes et al. central equation is their ill-defined efficiency, which if we consider it to be production-specific trophic efficiency TE (TE = dP/ dTL, will match our PETS equilibrium equation (eq. 10) with astonishing precision (see below).

The slope of the numbers spectrum (or body-mass-abundance spectrum) is actually not -1, but $b_{numbers}$ = -2 (Figueiredo et al., 2025; Landreth et al., 2025). The difference in slopes between the weight-numbers spectrum ($b_{numbers}$ = -2) and weight-biomass spectrum (b = -1) is "+1": -2 + 1 = -1. Interestingly, the difference in the terms used by Borgmann (1987), Andersen et al. (2008), Andersen et al. (2009), and Barnes et al. (2010) *vs* this study is also "+1": -0.75 + 1 = + 0.25. Thus, both above equations for a trophic equilibrium equation in units of production, that have been formulated independently and considering completely different processes (scaling of consumption and numbers *vs* biomass and body mass scaling), are precisely congruent. This is a further indication of the consistency and correctness of modern size spectra equilibrium theories.

Interestingly, the hugely popular and widely cited "metabolic theory of ecology" (MTE, Brown et al., 2004) is not able to predict the correct size spectrum slope. Instead, MTE predicts a different, wrongful (completely flat) size spectrum slope, with $b_{weight\text{-}biomass}$ = 0 and $b_{weight\text{-}numbers}$ = -1 (see, e.g., Fig. 10 on page 1785 in Brown et al., 2004). This shows that MTE is not suitable to predict or explain natural ecosystem size spectra.

The "-0.75" term in the $b_{numbers}$ , TE, PPMR equation ("$b_{numbers}$ = (log TE / log PPMR) – 0.75)") had been previously interpreted as the metabolic scaling of food consumption.(e.g., by Andersen et al. (2008), Andersen et al. (2009), and Barnes et. al. 2010). Yet, Barnes et. al. 2010, did not actually define whether efficiency was in production or biomass units, and used the erroneously "flat" non-normalized spectrum of Boudreau and Dickie, (1992) as their reference "b" value. They interpreted the insertion of the Kleiber-scaling term -0.75 in their equation as representative of metabolic scaling : " … *where 0.75 is the assumed scaling of consumption (assumed to be driven by metabolic rate) with body mass*". Instead, we showed (see above) that the term -0.75 is the result of two simple dimensional (units) conversions: i.) subtracting -1 from the "b" value, when converting from biomass to numbers and ii.) converting trophic efficiency from biomass (E) to production units (TE) by adding "+ 0.25" ("Fenchel-scaling") in log space, i.e., adding the scaling of mass-specific growth (production/biomass) with body mass.

Far from being a mere coincidence, the precise congruence between earlier size spectra equilibrium equations (Andersen et al. 2008, 2009, Barnes et al., 2010) and the proposed

PETS equilibrium equation (eq. 10) shows that in spite of different standpoints, units, processes, dimensions, and interpretations (e.g., consumption vs production, Kleiber- *vs* Fenchel-scaling), a correct, congruent grasp of ecosystem structure and equilibrium phenomena can be obtained.

## 6. Towards a general size-spectra equilibrium model, IV: Bottom-up regulation and equilibrium through the intrinsic prey abundance - PPMR relationship: the "Prey Abundance - Trophic Efficiency Regulation (PATER).

The mechanism described above, that exemplifies top-down regulation and PETS equilibrium through optimal foraging (SOFT), is by itself completely sufficient to explain the ubiquitous size spectra slope $\beta = -1$ and to solve the initially stated problems and paradoxes. However, PETS does not require an exclusively top-down regulated food web. Instead, the PETS framework assumes the simultaneous co-occurrence of top-down regulation (see above) and bottom-up regulation (see below) mechanisms (i.e. top-down and bottom-up-traveling equilibrium waves, Scott et al., 2014), both of which unavoidably lead to the size spectrum slope of -1, within a dynamic equilibrium between E and PPMR.

Within the PETS framework, bottom-up regulated equilibrium is straightforwardly obtained through the intrinsic prey abundance - trophic efficiency - PPMR relationship. Numerous authors have suggested that pelagic ecosystems are regulated through both top-down and bottom-up mechanisms, such as the effect of prey abundance on the growth of different fish species (Scott et al., 2014). Similarly, within PETS, the existence of a simple bottom-up mechanism can be assumed, through the well-described immediate effect of variations in prey abundance $N_{prey}$ on the trophic efficiency E of predators (Novak et al., 2024).

Before investigating the relationship of $N_{prey}$ and E, let us have a look at the terms and processes that may affect E. Any tentative equation to understand the main components that determine E, should consider that the most likely components are the physiological assimilation efficiency k1 (which depends on Kleiber-scaled metabolic rates, food composition, food quality, and growth rate), foraging energy use fe (energy used for foraging, and predation, which is Kleiber-scaled), and foraging energy efficiency "kf" (foraging energy predator body mass$^{-1}$ prey mass$^{-1}$, which is Kleiber-scaled, as it depends on predator mass, were foraging energy needed to locate, successfully capture or filter, handle and ingest each prey energy his time $t_s$ is proportional to spent per prey and thus has proportional effect on efficacy). Several studies have demonstrated that fe and kf are proportional to foraging search time $t_{s,}$ which is proportional to prey abundance $N_{prey}$ (Murdoch and Oaten, 1975, Novak et al., 2024).

For the sake of simplicity, one may thus assume that the mass-specific foraging energy efficiency is a linear function (Type I functional response *sensu* Novak et al., 2024) of prey abundance $N_{prey,\,j}$ of all prey effectively foraged by predator "j" in weight bin 'i". The effects of these two components ($k1_j$ and $N_{prey,\ j}$ ) on the mass-specific efficiency $E_j$ of a predator may be given as:

$$E_{ji} \sim \log(N_{prey,\ j}) + m + \log(k1_j) \quad (12)$$

Considering that assimilation efficiency (k1) cannot vary within several orders of magnitude and thus, log(k1) is nearly constant, but $N_{prey}$ varies by several orders of magnitude, it is safe to assume that variations in E are mainly determined by variations in prey abundance (and not by m and log(k1) which are near-constant), i.e. :

$$E_j \sim \log(N_{prey,\ j}) \quad (13)$$

This explains why highly productive upwelling ecosystems such as the Benguela and Humboldt current systems, where there is an enormous supply of nutrients, and organisms are crammed together in extremely particle-rich, shallow oxygenated surface layers, are generally reported with extremely energy-efficient food webs. This study also shows that PETS predicts that such efficient systems will have extremely high food web biomass, but with a constant β slope of -1, as observed in recent studies (Fock et al., 2026, Schwamborn et al., submitted).

*The "prey abundance - trophic efficiency regulation" (**PATER**)*

Within a power-law-distributed ecosystem, $N_{prey,\ j}$ is inversely proportional to $PPMR_j$. This may be illustrated by a theoretical example of a predator behavior switch event as a functional response to decreasing prey abundance, such as: 1.) sudden decrease in prey abundance (e.g., in a seasonal period of food scarcity 2.) functional response (SOFT): predator behavior switches towards smaller prey (i.e., predators switch from low PPMR to higher PPMR), 2.) predators prey upon more abundant prey, and 3.) predators switch from low $N_{prey}$ to high $N_{prey}$). Thus, it is clear that:

$$\log(N_{prey,\ j}) \sim 1/\log(PPMR_j) \quad (14)$$

, which, by combining eq. 13 and eq. 14, can be rearranged in a form that is similar to the original PETS equation (eq. 10):

$$E_{ij} \sim \log(PPMR_j) \quad (15)$$

This simple equation exemplifies the bottom-up PETS equilibrium mechanism, mediated through "**p**rey **a**bundance **- t**rophic **e**fficiency **r**egulation" (**PATER**). In spite of its simplicity, the proposed mechanism exemplifies how disturbances in prey abundance may propagate from lower trophic levels towards apex predators, i.e., how variations in prey density propagate upward within the size spectrum, within a strictly bottom-up-regulation mechanism.

Thus, PETS predicts that both processes (**SOFT and PATER**, i.e., top-down and bottom-up cascades) act simultaneously in stabilizing the size spectrum in natural ecosystems. In summary, **PETS** consists of a key equilibrium equation (eq. 10) and four key dynamic stabilizing mechanisms: **SOFT** and **PATER** both stabilize the *slope* of the spectrum through cascade waves, **TOPS** maintains the biomass below the maximum carrying capacity spectrum (i.e., it stabilizes the *intercept*), and **RISE** keeps a continuous (gapless) spectrum. This theoretical framework is intended as a contribution to our understanding of the intriguing hyper-stability of pelagic size spectra.

## 7. Towards a general size-spectra equilibrium model, V: the "carrying capacity spectrum" β, variations in resource abundance and in metabolic scaling (m).

All previous considerations focused on regulation through size-specific predator-prey interactions. These can explain and predict the existence of a constant, universal slope of the spectrum, which makes them interesting candidates as key components for a "universal ecosystem size spectrum theory", such as PETS. Yet, these mechanisms *per se,* are most likely not sufficient to predict the exact value of the intercept and to construct a precise, predictive linear model (i.e. to produce the exact values of $\alpha$ and Bi). These predator-prey equilibrium mechanisms (although useful to explain a constant $\beta$, with variable PPMR and E), will most likely, in the near future, not be practically useful to predict the exact numerical value of $\beta = -1$, given the immense uncertainties that still hamper precise *in situ* assessments of E and PPMR, with current methods.

*The maximum carrying capacity spectrum (**MCCS**)*

In contrast to the complex trophic dynamic mechanics described above, there is a very simple rationale that explains and predicts surprisingly well the exact numerical value of $\beta = -1$. It starts with the observation that there is a constant size spectrum, that is superposed by wave-, or dome-shaped deviations, and also by situations where ecosystems may present much lower biomass than expected from the β spectrum. Herein, we may preliminarily consider the ubiquitous ideal β spectrum to represent the ***maximum*** attainable biomass within a given weight (or size) bin, under ideal, standard conditions.

According to the proposed PETS theory, this ideal, universal ***maximum carrying capacity spectrum β*** is not characteristic for a specific ecosystem or taxon, but it is a ***universal spectrum*** that follows a *universal law* that applies to all three-dimensional (i.e. pelagic) ecosystems. Accordingly, the ***universal* β spectrum** is the base on which variations through resource limit stress (collapse in Bi), excessive nutrients (increase in Bi) or trophic cascade waves (domes and waves) impose temporary or constant alterations. This rationale utilizes the common concept of carrying capacity, i.e., the population size that an environment can support over a long term, being constrained by multiple biotic and abiotic **limiting factors**, including **resource availability** (e.g., nutrients, food), environmental conditions (e.g., oxygen, temperature), and species interactions, such as competition and **predation**.

The concept of ***carrying capacity*** (Verhulst, 1838, Pearl & Reed, 1920) was immensely popularized by early systems-ecology works. Odum (1953) defined it as the **asymptotic constant "K"** in the logistic population growth equation, as an equilibrium concept tied to energy flow and ecosystem stability. Yet, since the second half of the 20th century, it has been often criticized as a vague, "one-fits all" concept, being applicable to numerous conflicting scenarios, such as bottom-up and top-down limited systems, as can be seen in statements such as that the "… utility of the concept is questionable" (Hixon, 2008).

Nevertheless, carrying capacity may be a useful concept for understanding the "maximum size spectrum." It helps interpret how observed size spectra can collapse rapidly and fall far below the β spectrum under resource-limited conditions (e.g., strong nutrient stress), while generally not exceeding the β spectrum in offshore pelagic ecosystems (Dugenne et al., 2024), except under unique circumstances (e.g, persistent extremely high E; persistent extremely high nutrient flux). The collapse of the biomass may be very quick (e.g., for most phytoplankton) and may lead to a decrease of several orders of magnitude, with no obvious "bottom". Conversely, biomass does not generally reach far above the β spectrum in offshore pelagic ecosystems (Dugenne et al., 2024), except under the most extreme, unique circumstances (e.g., in estuaries). Thus, for most pelagic ecosystems, the β spectrum can be viewed as an upper "ceiling," particularly in resource-limited (i.e., hyper-oligotrophic) systems.

Also, the concept of a *maximum carrying capacity spectrum* (MCCS) helps us understand the link between size spectra theory and population dynamics theory (i.e., the logistic population growth equation, with population growth "r", and the asymptotic constant "K" = *carrying capacity*). This is especially important considering that in each size class, there is an equilibrium of size-scaled positive (e.g., population growth, body growth, and recruitment) and negative (mortality) processes that act in each size class of the spectrum, within an equilibrium biomass.

The invariant, perfectly log-linear shape of the spectrum means that these equilibrium processes are size-scaled (negative and positive processes are similarly scaled relative to size) and scale-invariant (acting in the same, universal way, throughout all size scales).

*The weight-specific carrying capacity constant $\kappa$*

To understand the meaning of an exact, universal slope of $\beta$ = -1, of the MCCS, we may consider that, simply put, this means that there is a *universal, scale-invariant law* (Capitán & Delius, 2010), that leads to a universal***, inversely proportional relationship between B and M*** in all three-dimensional (i.e., pelagic) ecosystems of our planet.

This universal law means that there is a *universal weight-biomass constant $\kappa$, that* can be obtained from eq. 1, as follows:

$$\beta = d(\log B) / d(\log M) = -1 \Rightarrow d(\log B) = -d(\log M) \Rightarrow \log(B) = -\log(M) + \alpha$$

$$\Rightarrow \log(B) = \log(\kappa) - \log(M) \Rightarrow B = \kappa / M \Rightarrow \kappa = B * M \qquad (16)$$

, where $\alpha$ = universal log-biomass intercept. Interestingly, $\alpha$ encodes a relevant scaling factor, $\alpha = \log(\kappa)$. $\kappa$ may be simply called the ***weight-specific carrying capacity constant****.* Although visually intuitive (Fig. 1), interpreting the intercept of a power-law distribution is far from trivial. Its arbitrary numerical value depends on units and logarithms used for B, and more importantly, on the logarithm, unit and reference standard for "M", which defines the position of the exact log(M) = 0 value, and thus, the arbitrary value of "M" used to obtain the intercept $\alpha$, and subsequently $\kappa$.

From these considerations, it is clear that defining the exact units, dimensions and numerical value of $\kappa$ is far from trivial, and there is no single universal value of $\kappa$ to be found in the near future, until there is a consensus, standardized model and method for size spectra with standardized x and y axis units and $M_0$ reference standard (i.e., the definition of the $\log(M_0)$ = 0 value, see Fig. 1). For example, using "$M_0$ = 1 g carbon $ind.^{-1}$" as a universal size spectrum standard reference weight ("$\log(M_0)$ = 0 = log10(1 g carbon)"), would take the intercept definition to the smallest-sized nekton (e.g., small anchovies and small myctophids). The (arbitrary) definition of $M_0$ = 1 g C $ind.^{-1}$ leads to an estimate of the intercept a = -3.3 log10 (gC $m^{-3}$) (Schwamborn et al., submitted) and thus $\kappa = 10^{-3.3} = 0.0005\ g^2 * m^{-3}$.

Alternatively, " $M_0$ = 1 mg carbon $ind.^{-1}$ " could be the universal standard, with "$\log(M_0)$ = 0 = $\log_{10}$(1 mg carbon)", which would have us looking at the numerical value of the intercept obtained at size bin that is approximately within the large-sized zooplankton and small-sized micronekton, close to the relevant, critical **CritGill** size boundary (the boundary between mostly gill-less zooplankton and gill-bearing nekton). The definition of an appropriate $M_0$ value, leading to the placement of the "$\log(M_0)$ = 0" value in the center of the spectrum (ideally at the exact median of the body mass M data used for analysis), is most important, to avoid auto-correlation artifacts between estimations of of $\alpha$ and $\beta$. This **auto-correlation impedes the simultaneous assessment of both values** (since the estimate of $\alpha$ affects the estimate of $\beta$, and vice-versa), when $M_0$ is defined in such a way that "log(M) = 0" is located far from the center of the spectrum (i.e. far from the median of Mi).

A standard (consensus) size spectra methodology and terminology is urgently needed for any relevant progress in size spectra research. This is especially important considering numerical artifacts, and confusing, wrongful terminologies, misconceptions, dimensions, and units (e.g., NBSS with biomass values of B that are usually given in intriguing "pseudo-abundance" units) that are nowadays used in most size spectra studies (see Schwamborn, submitted).

The existence of an invariant carrying capacity spectrum and a universal constant $\kappa$ shows that there is a maximum biomass concentration within a given volume of water, that depends on the individual weight (i.e., body mass) M. Since individual body mass and individual body volume are approximately proportional, this perfectly inversely proportional volume vs biomass-per-volume relationship is to be expected, and *per se* fully explains and predicts the $\beta$ = -1 value of the slope. This rationale may be formulated as a simplistic, but convincing "minimum water volume necessary to sustain a given body volume", or **"water volume essential to sustain body mass optimum" (WETBIO)** hypothesis. The WETBIO hypothesis predicts that biovolume-based size spectra studies in pelagic ecosystems should yield slopes that are steeper (closer to -1) than size spectrum studies based on carbon (or nitrogen, or calories) only. Independently whether in carbon or in volume units, the volume-volume or weight-volume (i.e., WETBIO) rationale is probably the most promising rationale that leads invariably to the exact and unique numerical value of $\beta$ = -1.

*The minimum water volume necessary per unit body mass*

The "minimum water volume necessary per g body mass", $V_{min,M,i}$ / Mi in each body mass bin "$M_i$" may be estimated as:

$$B_i = M_i / V_{min,M,i} \Rightarrow V_{min,M,i} = M_i / B_i \; ; \; B = \kappa / M \Rightarrow V_{min,M,i} = M_i^2 * \kappa^{-1}$$

$$\Rightarrow V_{min,M,i} / M_i = M_i \kappa^{-1} \qquad (17)$$

This gives a numerical estimate for $\kappa$ as $\kappa = M_i^2 / V_{min,w,i}$, and a preliminary unit and dimension for $\kappa$ ($g^2 * m^{-3}$). Intuitively, this unit and dimension for $\kappa$ makes sense, since biomass is given in $g/m^3$, and within a power-law size spectrum, biomass is scaled by weight (g), thus $g^2 m^{-3}$.

Thus, in standard $\beta$ ecosystems, $V_{min,M}$ / M scales proportionally with M (and not with $M^{0.75}$, as predicted by MTE, see Brown et al., 2004). Interestingly, the observation that $V_{min}$ / M scales in perfect linear proportionality with M, is coherent with the perfect proportionality of the $\beta$ spectrum (slope of $\beta$ = -1), indicating the correctness of the equations above.

*The minimum water volume necessary per individual*

Now, let's calculate the "minimum water volume necessary per individual" $V_{min,ind.}/ N$, as a function of body mass M:

$$B = Abund * M \; ; \; Abund = N * V^{-1} \Rightarrow B = N * V^{-1} * M \Rightarrow V_{min,ind.} = N * M * B_{max}^{-1} \quad (18)$$

, where V is the water volume around each organism, $V_{min,ind.}$ is minimum the water volume around each organism, necessary to sustain its existence, and Abund is volume-specific abundance (Abund = N / V).

Combining eq. 16 and eq. 18:

$$V_{min,ind.,i} = (N_i * Mi) \; / \; (\kappa / Mi) \Rightarrow V_{min,ind,i} = N_i * M^2_i * \kappa^{-1} \Rightarrow V_{min,ind,i} / N_i = M^2_i * \kappa^{-1} \quad (18)$$

Considering variations in "m" (e.g., in prokaryotes), we may also formulate equation 18 as:

$$V_{min,ind.,i} \; / \; N_i = M^{(m + 1.25)}{}_i * \kappa^{-1} \quad (19)$$

This means that minimum volume per individual ($V_{min,ind.}$ / N) is inversely correlated to the weight-specific carrying capacity constant $\kappa$, which is unsurprising, and indicates the correctness of the above calculations.

A recent meta-analysis of zooplankton feeding (Kiørboe, 2011) found that clearance rates (mL $day^{-1}$) scale linearly with body mass on a log–log scale (slope ≈ 1). This implies that clearance per unit body mass is constant, meaning that the volume of water processed per unit biomass scales proportionally with body mass. Consequently, the required space, or effectively the utilized volume per unit biomass, is independent of size. This alone may already provide a sufficient, simple and straightforward mechanistic explanation for the biomass-body mass spectrum slope of b = - 1 predicted by the **WETBIO** hypothesis. It is likely that both processes, ecophysiological constraints on the utilized water volume (WETBIO), and top-down trophic interactions (e.g., the TOPS-RISE and SOFT-PATER mechanisms described above) jointly contribute to the observed invariant biomass spectrum (slope ≈ 1) in nature.

*Minimum mean spherical radius and minimum mean spherical contact surface area*

If $V_{min\ per\ indiv.} \sim M^2$, then we can predict the minimum mean radius and the minimum mean contact surface area of the spherical (3D) water volume surrounding each organism, with $r_{min,3D} \sim M^{2/3}$ and $A_{min,contact,3D} \sim M^{4/3}$.

Since body size "L" is more intuitively related to the minimum spherical radius surrounding each individual ("L" and "r" have the same linear dimension), we can estimate $r_{min,3D} \sim M^{2/3} \Rightarrow r_{min,3D} \sim (L^3)^{2/3} \Rightarrow r_{min,3D} \sim L^2$.

The result that both volume and radius scale in a non-linearly increasing (square) relationship with body mass and size ($\sim M^2$ and $\sim L^2$, respectively), may explain the ubiquitous observation that very large marine animals (such as whales) seem to be near-infinitely apart from each other, while small-sized organisms such as phytoplankton, seem to be extremely densely packed, with thousands of cells per milliliter.

While we, as terrestrial beings, intuitively think of the ocean as being near-infinite, the limited volume available and the need to adapt and adjust to a given volume (the minimum volume needed) for each individual may have been a key evolutionary driver for the establishment of a universal size spectrum $\beta$ in the oceans, along many millions of years, which is the rationale at the core of the WETBIO hypothesis.

Accordingly, too small (or too large) water volumes (i.e. deviant, non-$\beta$ biomass Bi) will trigger trophic regulation mechanisms, such as TOPS-RISE and SOFT-PATER (see above), or an increase in population growth, until the ideal biomass $B_\beta$ is reestablished.

*Slope, minimum area and radius in 2D systems (e.g., benthic or terrestrial ecosystems)*

All equations above assume three-dimensional, volume-related processes, that are typical of pelagic ecosystems. Following this rationale, size spectra in two-dimensional (benthic or terrestrial) ecosystems should be expected to behave differently from pelagic ecosystems. Within a weight-area relationship (instead of weight-volume), the size spectra slope of two-dimensional ecosystems (assuming m = 0.75 and 100% vulnerability) should be considerably flatter than $\beta$, with:

$$b_{two\text{-}dimensional} = -\ 2/3 \qquad (20)$$

For two-dimensional systems (e.g. benthic or terrestrial ecosystems), WETBIO (derived from $\beta$ = -1, and subsequently with $V_{min} \sim M^2$) may predict (assuming $V_{min} \sim M^2$) the minimum mean radius "r" and the minimum mean surface area "A" of a flat (disc-shaped, as a disc with constant thickness and variable radius) volume surrounding each organism:

$$V_{Disc} \sim A_{Disc}\ ;\ V_{min} \sim M^2 \Rightarrow A_{min,2D,\ Disc} \sim w^2\ ;\ r_{min,2D,\ Disc} \sim M \Rightarrow r_{min,2D,\ Disc} \sim L^3 \qquad (21)$$

Although there are fewer published weight-biomass spectra (e.g., NBSS, normalized biomass size spectrum) studies for two-dimensional systems, than for marine plankton, the few available datasets indicate that they have flat, temperature-invariant, productivity-invariant (the intercept being productivity-related), and consistent slopes across and between ecosystems, which are in accordance with PETS and WETBIO predictions. For example, two-dimensional terrestrial invertebrate communities, such as soil invertebrates, tend to have flat, invariant, NBSS, with slopes between -0.3 and -0.6 (Mulder et al., 2008). Benthic NBSS in polar regions also tend to have invariant, flat slopes, from - 0.46 to - 0.57, with no differences among localities (Quiroga et al., 2014, Mazurkiewicz et al., 2020). In the East China Sea, Hua et al. (2013) reported benthic NBSS slopes that ranged from -0.596 to -0.953, with most samples showing NBSS slopes between approximately -0.6 and -0.7 (approx. $b_{two\text{-}dimensional} = -2/3$), as predicted by PETS and WETBIO. These observations strongly support the idea that weight-area and weight-volume considerations (such as the WETBIO hypothesis) are relevant to explain the ubiquitous $\beta = -1$ carrying capacity spectrum in pelagic (i.e., three-dimensional) ecosystems. As such, the existence of a ***universal carrying capacity spectrum*** $\beta$ (within the WETBIO rationale) is, together with regulation by trophic cascade waves (within size-specific predator-prey interactions, such as SOFT and PATER), the backbone of the PETS theoretical framework.

For very small-sized organisms, such as nano- and picoplankton, the existence (or not) of a TopDoSH, will be a key determinant to check whether there is a predominance of minimum-volume-related (WETBIO) processes. If there is actually no discernible TopDoSH (as in Dugenne et al., 2024), the likely explanation is that top-down control by metazoans AND by protozoans (as in Taniguchi et al., 2023) reaches deep into the smallest-sized picoplankton, and it simply does not matter whether the top-down control is exerted by metazoans or by protozoans (within the “E *vs* PPMR” equilibrium mechanism described in chapters 2 to 6). Simply put, a global ocean with a clear TopDoSH is mostly regulated by top-down control, while an ocean with no TopDoSH can be also explained by a predominance of top-down control (see above) or, alternatively (and most likely), by control thorough a combination of multiple (e.g, minimum volume, bottom-up, and top-down) processes.

The existence of a universal carrying capacity spectrum implies that not only $\beta$ but also $\alpha$, are both invariant in standard pelagic ecosystems. Accordingly, Dugenne et al. (2024) found an invariant value of the intercept $\alpha$ across global ocean phyto- and zooplankton size spectra, with a similar coefficient of variation (approx. 20-30 %) as for the invariant slope $\beta$. Actuality the coefficient of variation for $\alpha$ was consistently narrower than for $\beta$ in Dugenne et al. (2024), indicating that the intercept (i.e. carrying capacity) is even more invariant than $\beta$ in marine phyto- and zooplankton. This indicates that there is a universal carrying capacity spectrum, with a well-defined biomass Bi for each weight bin i.

This means that not only the slope, regulated by predator-prey-relationships, is universally constant (as generally acknowledged), but also that there exists a universal law, that determines the absolute numerical value of the biomass (and abundance) in each size bin. Similarly, Fock et al. (2026) and Schwamborn et al. (submitted) found near-invariant values of $\alpha$ and $\beta$ across the Atlantic Ocean, except for the most extreme situations such as for the hyper-oligotrophic Tropical Southwest Atlantic off Brazil, under extreme nutrient limitation stress (phytoplankton biomass far below predictions), and for the highly productive Benguela upwelling system (crustacean zooplankton biomass above $B\alpha$, probably due to high efficiency "E", very high $\alpha$, with $\beta = -1$). These results (near-invariant $\alpha$ and $\beta$ across the global ocean, as in Dugenne et al., 2024) support the existence of a deterministic, exact numerical value of $V_{min}$ within each size bin, as predicted by the WETBIO hypothesis (see above), which is a key component of PETS theory.

In the following, we will consider some common variations from the universal carrying capacity spectrum $\beta$, as observed in several ecosystems and communities, such as effects of bottom-up control through variations in resource abundance and variations in metabolic scaling. In natural systems, the realized mean carrying capacity size spectrum may be defined by the combination of factors such as bottom-up and top-down cascade waves (see Chapter 6). This perfectly linear carrying capacity spectrum is disturbed by trophic cascade waves, nutrient stress in hyperoligotrophic scenarios, excess nutrient inputs in hypereutrophic systems (e.g., estuaries and lagoons), variable vulnerably (as in shallow lakes and streams with abundant shelter from predation), and divergent metabolic scaling (e.g., in bacteria).

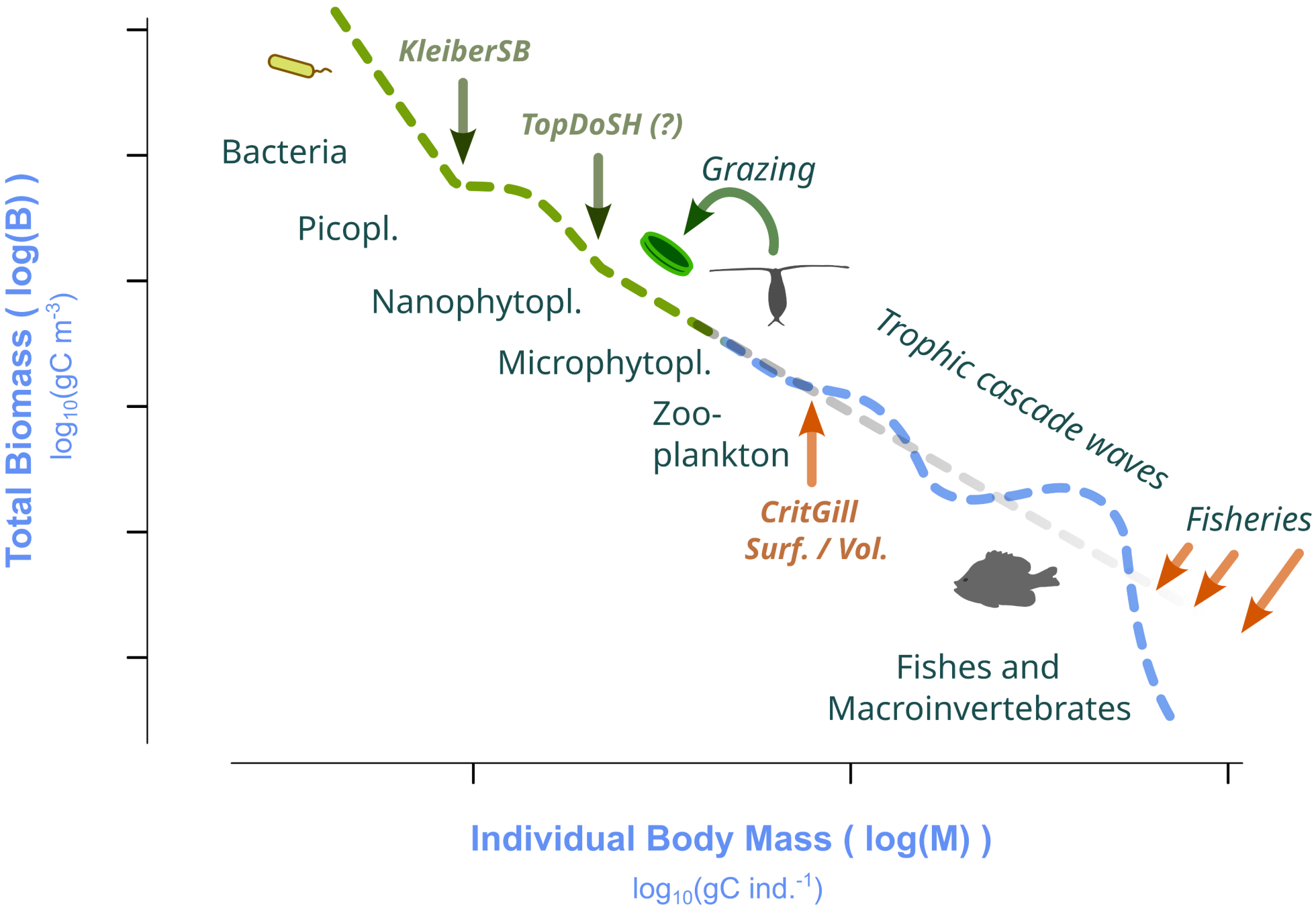

**Figure 6.** Hypothetical example of a weight-biomass spectrum, indicating trophic cascade waves, the extent of the linear spectrum, between TopDoSH and CritGill. KleiberSB: Kleiber's "m" metabolic scaling size boundary, i.e., the size were an abrupt change in the size spectra slope occurs, at the size limit between eukaryotes (b = -1) and prokaryotes. TopDoSH: size limit of top-down control by size-structured food webs, such as grazing by copepods. CritGill: critical size of switching from exponential to asymptotic growth, where gills become necessary.

*Resource-stress-regulated size spectra*

The trophic equilibrium equations proposed above assume a control of populations through trophic interactions (top-down or bottom-up regulation, or both). Alternatively, it may be possible to imagine a resource-limit-stress-control scenario, where all organisms are so sparsely distributed that there are no relevant prey-predator interactions at all, and all organisms are limited by one density-independent “stressor”, or limiting resource (Fig. 7). For example, such a scenario may be imaginable for sparsely distributed plants, such as gramineans in a desert (limited by rainfall, not by grazing), or for thinly abundant phytoplankton cells in hyper-oligotrophic ecosystems under extreme nutrient limitation stress (limited by nutrients, not by grazing).

The size spectrum of a non-trophically interacting ecosystem may be simply estimated by the relationship between individual size (or weight) and the spacing between individuals, i.e. the space (and resources in that space) used by each individual, which scales with metabolism (i.e., with “m”). Starting from the assumption that clearance rate “C” (e.g., L $ind^{-1}$ $d^{-1}$) scales with 0.79 (Huntley and Lopez, 1992), which is similar to the “Kleiber- scaling” of 0.75, it is possible to calculate the cleared volume (or foraged volume, or nutrient-depleted volume, depending on the taxon) used by each individual in a given unit of time. Henceforth, it is possible to calculate the number of individuals per unit volume “N”, for each weight bin M, where N scales with $N \sim M^{-0.79}$. Thus, a theoretical “non-trophic” abundance-weight spectrum has a slope of b = -0.79 (approximately Kleiber–scaled), which is much flatter than the predicted abundance-weight spectrum slope of approximately -2 (Figueiredo et al., 2025). Similarly, Brown et al. (2004) also predicted that, within the “Energetic Equivalence Hypothesis” (Damuth, 1981), the weight-dependence of population density (and thus, the abundance-weight spectrum) scales with approximately -0.75, before being corrected for trophic losses, within their “Energetic Equivalence Hypothesis with Trophic Transfer Correction” , where b = -0.75 + (log(TE) / log(PPMR)). One interesting conclusion from this thought exercise is that for a completely predation-free, resource-stress-limited system, that is Kleiber–scaled by metabolism, one may expect an extremely flat weight-biomass spectrum slope “b”, with, a slope of

$$b_{resource\text{-}stress\text{-}limited} = -1 * (-0.75 / -2) = -0.375 \quad (22)$$

Thus, we may predict that pelagic ecosystems will have a mean weight-biomass spectrum slope “b” between -1 (regulated by density-dependent trophic interactions and $V_{min}$), and -0.375 (resource-stress-regulated, and scaled by metabolism).

*The size spectrum slope in microbial (“non-Kleiber-scaled”) food webs*

Initially, it is important to emphasize that the general mass-specific trophic equilibrium equation ("β = E * S") is already scaled by 0.75, i.e., that "E" is already "Kleiber-scaled" (i.e., proportional to $M^{0.75}$, see above). Thus, there is probably no need to include a specific term for "m" into the trophic equilibrium equation for β (as done by Barnes et al., 2010).

However, there may be deviations from m = 0.75 in specific taxa (e.g., in bacteria), or under specific conditions, which will affect the weight-biomass spectrum slope "b".

$$b_i = c - (m_i - m_0) \quad (23)$$

, where

$$c = E * S \quad (24)$$

$b_i$ : weight-biomass spectrum slope within size class "i"

c: trophic equilibrium constant ( c = -1 ), or c = β m: metabolic scaling, i.e., the relationship between metabolic rate R and body mass M, expressed as $R = a * M^m$

$m_0$: predicted value of "m", $m_0 = 0.75$ (Kleiber, 1932)

$m_i$: *in situ* metabolic scaling within size class "i"

For microbial taxa, the deviation ($dm = (m_i - m_0)$ ) may be used for corrections and to explain possible departures form the β = -1 slope (e.g., for bacteria, where "m" may be much higher than Kleiber's $m_0$, up to $m_i = 2$, DeLong et al., 2010). The "exotic", non-Kleiberian scaling of these organisms may also explain why bacteria and picoplankton are generally ignored in the trophic modeling community. In spite of being dominant in secondary production in all ecosystems, bacteria are generally not included in trophic models (e.g., EwE, APECOSM).

The term "$(m_i-m_0)$" is only relevant in a scenario where a given group of organisms has a metabolic scaling $m_i$ that is considerably different from $m_0$ ($m_0 = 0.75$), otherwise, the trophic equilibrium equation reverts to its most simple form: β = c. Yet, if for example, $m_i$ is considerably greater than $m_0$, e.g., $m_i = 1.75$ (as reported for bacteria and picoplankton), then β = -1 - (1.75 - 0.75) = -2. A higher $m_0$ value, as for bacteria and picoplankton, will thus lead to steeper size spectra slopes (much steeper than -1) for those organisms. A steep NBSS slope is to be expected for these extremely small-sized taxa, since they have much higher total biomass and secondary production than expected from global size spectra models (see Fock et al., 2026, Schwamborn et al., submitted).

Thus, PETS predicts a discrete, fixed boundary (i.e., a critical size) between organisms that behave according to Kleiber-scaled size-metabolism relationships ($m_0$ = 3/4), and organisms that have completely different values of “$m_i$“ such as prokaryotes. This critical size (or critical mass, or volume), may be called the ***KleiberSB*** size (“Kleiber’s “m” metabolic scaling size boundary”), i.e., the size were an abrupt change in the size spectra slope occurs, at the size limit between eukaryotes ($\beta$ = -1) and prokaryotes (b < -1, steep slope, extremely high biomass and production, Fig. 6).

However, it is important to note the assumption of a size-structured food web (TL increasing with body size throughout the food web), an essential prerequisite of the equilibrium equation presented above, has not yet been tested for microbial food webs. The fact that these requirements and assumption as are not valid for all types of food webs, but mostly for marine pelagic food webs, and the need for large numbers of organisms in the datasets, prior to binning, makes it clear why size spectra theory has evolved mainly in marine plankton studies (Zhou, 2006), and not, for example, in terrestrial ecosystems, where several assumptions are often not fulfilled, with notable exceptions (Mulder et al., 2008).

Below the KleiberSB size (i.e., in the “microbial loop”, involving DOM, bacteria, picoplankton, flagellates, ciliates, and phagocytosis), the size spectrum is uncoupled from top-down control exerted by metazoan zooplankton and higher trophic levels, and has higher metabolism / weight ratios (i.e., a higher “m”) value, leading to nonlinear- shaped size spectra. Organisms smaller than KleiberSB (prokaryotes) will thus have higher biomass, above predictions from the $\beta$ spectrum.

A discrete size limit could be expected where there is a dramatic change in this size spectrum, which is the “top-down regulation size horizon” (***TopDoSH***). This discrete horizon may be discernible at the minimum size limit, where top-down control by size-structured food webs can reach, i.e., where grazing by metazoans such as copepods, lead to a top-down induced linear shape of the phytoplankton, within the microphytoplankton and parts of the nanoplankton (Schwamborn et al., submitted), towards smaller sizes (nano– and picoplankton), where filter feeding through meshes of setae and other filter-feeding apparatuses is no longer viable, due to the increasingly unfavorable viscosity - size relationships (considering the low Reynolds numbers of picoplankton). Simply put, the hypothetical TopDoSH would be the boundary between effective top-down control (i.e., filter--feeding grazing) by metazoans on the larger side, and phagocytosis by protists on the smaller side of this horizon. Also, it would be the boundary between rigorously size-structured food webs and the microbial loop (see Azam et al., 1983, Fenchel, 2008), which is defined by a critical value of prey Reynolds numbers.

The term size “horizon” proposed here, reflects the idea that it is the prey size boundary, beyond which metazoans cannot reach effectively, i.e., the horizon across which information, energy and mass disturbances from top-down cascades cannot reach, and where protists take over as the main grazers, defining the phytoplankton size structure. Considering that viscosity is temperature- dependent, the TopDoSH size should be at a larger size in colder waters, which would predict larger prey organisms in metazoan grazing food webs, in polar regions. PETS predicts the phytoplankton size spectrum in nature to be shaped by such discrete size boundaries (i.e., TopDoSH and KLeiberSB), while the literature on phytoplankton size distributions has focused mainly on nutrient dynamics. The possible existence of a discrete TopDoSH, and a linear spectrum at sizes larger than TopDoSH, would indicate that top-down regulation mechanisms are the main drivers for a linear size spectrum in nature.

Another discrete boundary is to be expected between the plankton and nekton, due to the different metabolism and growth dynamics. Dramatic changes occur at the critical mass, where the volume / surface ratio increases (within allometric theory and GOLT) to a point where the body surface is not large enough to supply the organism with oxygen, and thus, gills become necessary (Pauly, 2021). This discrete boundary, where growth decreases, switching from exponential to asymptotic growth, and gills become necessary, may be called the "critical gill-bearing size" (***CritGill***), which approximately corresponds to the boundary between mostly gill-less zooplankton and gill-bearing nekton (although some large-sized zooplankton, such as euphausiids, actually do have gills). In summary, PETS predicts a linear "$\beta$" spectrum to occur within a discrete size range, between the TopDoSH to CritGill size limits (i.e., from nano- to macroplankton).

The observation that large-sized nektonic organisms, such as mesopelagic fish and macroinvertebrates (Fock et al., 2026, Schwamborn et al., submitted), and large marine mammals, actually follow the global plankton size spectrum (Hatton et al., 2021), in spite of completely different growth dynamics, physiology etc., may be called the "**CritGill paradox**", which still requires more data and theory. Alternatively, we may regard this apparent paradox as a further proof of the existence of a universal law that shapes and stabilizes the log-linear size spectrum, regardless of the individual peculiarities of any taxa, such as divergent growth curves and the presence of gills or lungs.

It is still unclear whether the other terms (other than $m_i$) in the equilibrium equation (e.g., E and S) can be temperature-dependent, certainly a key topic for future studies. PPMR is evidently body-size dependent, increasing with predator size (Barnes et al., 2010), but its possible dependence on temperature is unknown. If E should be dependent on temperature (which is still unknown), then probably PPMR would also vary, compensating for such variations (eq. 10). Within the preliminary knowledge that is nowadays available, it seems likely that any kind of trophic efficiency has a strong component of assimilation efficiency "k1". There is already a vast literature on the temperature dependency of k1, especially from aquaculture studies. For example, in many fish and invertebrate (e.g., shrimp) species, assimilation efficiency increases with temperature, until reaching a species-specific thermal optimum (e.g., Jobling 1997, Clarke, and Johnston, 1999).

Conversely, previous modeling efforts using EwE indicated that within any given food web, warming may not have a consistent effect on trophic efficiency "Energy transfer efficiency between trophic levels … was not affected by either warming or acidification." (Ullah et al., 2018). Yet, if we preliminarily assume that E varies with temperature and body size, and there really is a dynamic and effective mechanism between E and S, that means that any temperature-dependent and body-size-dependent change in E will be compensated by a change in S (i.e., a change in PPMR and food chain length), leading to consistent, temperature-independent slope of $\beta = -1$, and constant body size-independent slope throughout the size spectrum, as observed in numerous ecosystems under various temperature settings, from polar to tropical ecosystems Dugenne et al., 2024, Soviadan et al., 2024).

## 8. From the maximum carrying capacity spectrum β towards the realized spectrum “b” - considering stress and dynamic non-equilibrium effects

According to PETS, the size spectrum slope “b” of natural pelagic ecosystems can be expected to be close to the theoretical equilibrium size spectrum slope β, only when the assumptions listed above are fulfilled (e.g., unlimited nutrients and regulation through trophic interactions, as in productive upwelling ecosystems). In such nutrient-rich pelagic ecosystems, the total biomass $B_i$ in any given weight bin Mi should be close to the Biomass $B_{predicted}$, predicted from the global size spectra model with slope = β (i.e., $B_i = B_{predicted}$).

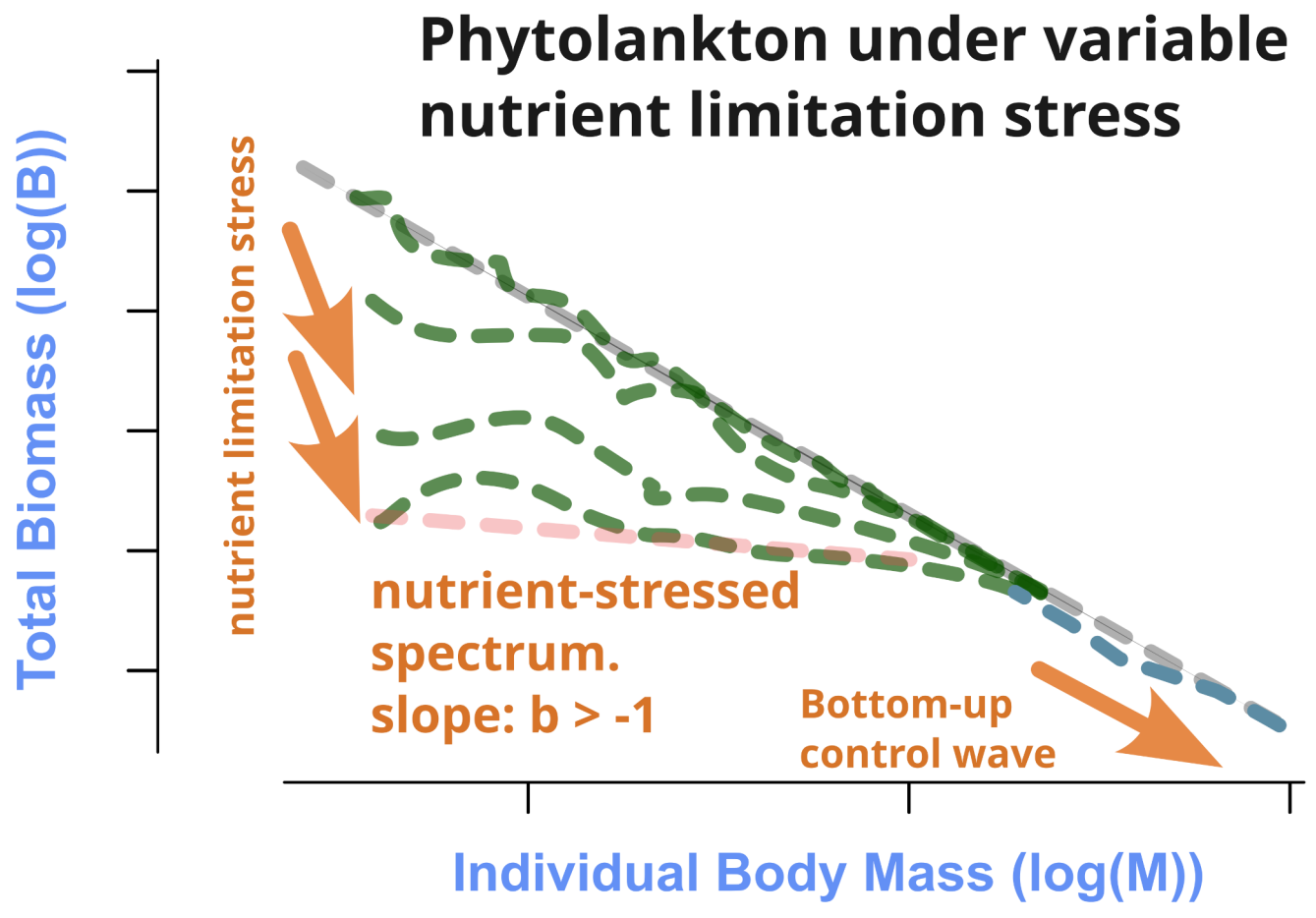

**Figure 7. The carrying capacity spectrum under variable nutrient limitation stress, with a bottom-up-control wave.** In *resource-stress-regulated* systems, the biomass $B_i$ varies from zero to $B_{max}$, which may be defined by the carrying capacity spectrum β, where the slope of the spectrum is: $b = \beta = -1$.

Conversely, in hyper-oligotrophic or mixed (e.g., seasonally or spatially variable ecosystems), the biomass $B_i$ in any given weight bin "i", should exhibit a huge variably, when considering biomass estimates from repeated sampling under varying conditions (as in many large size-spectra datasets). Mean (and median) $B_i$ values in such nutrient-limited or nutrient-poor ecosystems should be well below $B_{predicted}$, expected from the linear model with slope = β. In such non-ideal systems, the maximum observed (or the 99 % quantile) is expected to follow the slope = β model. The maxima in Bi, that follow the ideal linear model, can thus reflect ephemerous situations of resource abundance (e.g., during local blooms), while the biomass minima in each bin reflect situations with extreme resource limitation stress (Fig. 7).

In systems with variable resource availability (i.e, a *resource-stress-regulated* systems that are often nutrient-limited) the biomass maximum in each bin should represent the biomass under resource-rich situations (e.g. seasonal, spatially patchy, or stochastic blooms). The maxima in each bin should be close to the values expected form an ideal equilibrium linear models with slope = β (where there is no resource-stress limitation, and trophic processes, such as grazing, start taking place). Thus, in resource-stressed scenarios and in systems with varying resource variability, the linear model with β = 1 represents the carrying capacity spectrum, not the mean or median biomass, in each size bin, and the minimum in each bin represents the predation-free, resource-stressed biomass. The proposed existence of a carrying capacity spectrum may be called the "carrying capacity spectrum hypothesis", or WETBIO hypothesis (see above), within PETS.

The observed weight-biomass spectrum slope "b", which may be variable (not necessarily b = -1), may be described by:

$$b = c + \sigma + s + \gamma + z/g + v + n - e - (mi-m0) \qquad (25)$$

, where

b: slope of the weight-biomass spectrum b = dlog(B) / dlog(M), where B = total biomass, M = individual body mass i.e., individual weight s: mean of the density-independent stress factors $s_i$ across in all size bins "i".

σ: weight-specific change in stress σ, where $\sigma = d\, s_i / d \log(M_i)$.

The stress factor "s" represents the log-linear effect of extreme, density-independent stressors, that affect biomass and trophic interactions (e.g., nutrient-limitation stress, oxygen-limitation stress, sound pollution stress, or stress due to fisheries). s = 1 means zero trophic interactions, e.g., when all individuals (predators and prey) are near-infinitely apart from each other. s = 0 means full control by trophic interactions (i.e., the "β food web"). The weight-specific stress factor "σ" represents how individuals of different sizes are affected by stress factor s. The inclusion of s in the model permits a transition from trophic regulation (s = 0) to stress-limited scenarios (s > 0). Unitless and non-dimensional. The inclusion of "σ" in the model permits to include the effect of changes in s with weight (e.g., as predicted by the GOLT theory for the weight vs oxygen deficiency tolerance - relationship).

$\gamma$: weight-specific change in growth/mortality ratio

$$\gamma := g\, z^{-1}\, M^{-1} \quad (26)$$

, where

c: trophic equilibrium constant, c = E * S (see above)

z: instantaneous total mortality rate, $z = dN/dt\ N^{-1} = 1/dt$

g: mass-specific body growth rate, $g = dM/dt\ M^{-1} = (1/dt)$

g': instantaneous linear body growth (dM/dt)

However, the term "g/z" may be irrelevant across large size ranges, encompassing many TLs, and size across many orders of magnitude, where b is determined by trophic, not growth-related, processes.

n: non-predatory mortality scaling $n = g/z_0\ dM^{-1}$, where $z_0$ = "other" (i.e., non-predatory and non-fisheries mortality). The term "n" is relevant only when non-predatory mortality (e.g. mortality from disease, parasitism, stress, or old age) dominates and "n" is extremely different from beta.

v: weight-specific vulnerability to predation: $v = d\ Rv\ dM_{-1}$, where Rv is the Ratio vulnerable / invulnerable (e.g. hiding in vegetation or refugia) individuals in each size class. This term is especially important in systems with large non-predatory mortality terms. Similar to the vulnerability term used in Ecosim simulations (Christensen et al., 2005). Note that the basic PETS equation (eq. 10) is only valid under 100% vulnerability (i.e., 100% of the prey population is always vulnerable to predation) and density-dependent trophic regulation through the food web. Many ecosystems (e.g., coral reefs, or freshwater systems with slush vegetation) may exhibit size classes with low vulnerability levels, leading to biomass above predicted, and to "bumps" in the spectrum (as observed by Rossberg et al., 2019). Also, some taxa have zero vulnerability to predation, as adults (e.g., species with long spines and toxins, such as in lionfish, and large-sized apex predators, such as sperm whales).

e: size-specific biomass export (e.g., though migration, sinking, and advection), $( (dB/dt)_i - (dB/dt)_{i+1} / ( M_i - M_{i+1})$

The ecosystem - scale weight–biomass spectrum "b" is composed of numerous individual species, with their species-specific weight-biomass spectra, which may be described by the following equation:

$$\text{mean}(g/z) \sim \text{weighted mean} \left( \text{mean}(b_{T1}), \text{mean}(b_{T2}), \text{mean}(b_{T3}), \ldots, \text{mean}(b_{Ti}) \right) \quad (27)$$

, where $b_{Ti}$ is the slope of the weight-biomass size spectrum of taxon "T", for example one specific species (e.g., "T = species x"). Note that mean(g/z) is not equal to b.

For a discrete sample obtained from nature (e.g., a sample taken with a net) we may consider the following equation:

$$b_{sample} = b + p + \boldsymbol{\varepsilon} \quad (28)$$

, where:

$b_{sample}$ = slope of the weight–biomass spectrum in a discrete sample.

p: weight-specific scaling of the probability of capture, related to gear selectivity and gear avoidance.

**ε**: error term, accounts for stochastic variability.

The fact that the stress factor is a positive addition to the equation may initially seem counter-intuitive considering that stress will lead to a reduction in biomass for a given size bin, but the idea is that the stress will lead to a flatter slope less to an increase in b.

PETS predicts a constant, temperature-invariant (dos Santos et al., 2017) universal value of all terms "c", "m", and "β". The actual size spectrum measured in real ecosystems, may of course show considerable deviations from the expected value of -1, due to changes in "m" (e.g. prokaryotes should have higher "m" and thus, steeper size spectra slopes, especially at higher temperatures) and due to stress (e.g., nutrient-limitation stress) related reductions in biomass (i.e., b = β, is valid only under equilibrium, trophic regulated systems, with m = 0.75), additionally to stochastic and non-equilibrium effects (especially on in small samples and when dome-shaped trophic cascades appear (Rossberg et al., 2019).

Thus, in real marine ecosystems, where oligotrophic situations are common, the actual size spectra slope b may change with tropicalization, that means a transition from resource-rich systems ( b = -1) to resource-limited, flat size spectra (b > -1).

## 9. Predictions (and observations from recent size spectra datasets)

Based on the scenarios above (Fig. 5), it is possible to make a short list with 10 simple, testable predictions from PETS (see below). Below are these straightforward predictions and a few recent observations that support the proposed equations and mechanisms.

### Predictions from trophic equilibrium theory (PETS):

1.) **Constant slope, but variable biomass in food webs and ecosystems**. Assuming that the trophic efficiency is a function of search time and search volume, and thus of prey density, it is possible to predict that more efficient food webs should have higher biomass (and not, flatter size spectra, as stated in many other studies, e.g., Wang et al., 2025). High meso- and macrozooplankton biomass has been reported from several highly productive and extremely efficient upwelling systems, where the food web is compacted into a shallow oxygenated layer, as in scenario 2 (Fig. 5). Inefficient systems, such as sparse, mesopelagic food webs, should have lower biomass, below predicted from the overall size spectrum, but with a perfect $\beta = -1$ slope, such as in scenario 1 (Fig. 5). A universal $\beta = -1$ is to be expected for all three-dimensional (i.e., pelagic) ecosystems, and $b = -0.67$ for two-dimensional (shallow-water, benthic, or terrestrial) ecosystems (see chapter 7).

2.) **Nutrient limitation stress and excessive nutrient flux**: PETS assumes that size spectra with $\beta = -1$ are shaped by trophic regulation mechanisms (i.e., predator / prey interactions). Nutrient-stress limited (hyper-oligotrophic) systems should not exhibit a -1 slope log-log-linear shape, but rather have much lower biomass than predicted. When under nutrient-limitation-stress, phytoplankton should have exotic ("non-Sheldon") size spectra, such as flat or unimodal distributions (Marañón et al., 2013). Excessive nutrient inputs, as in hyperoligotrophic estuaries and lagoons, may lead to higher Bi in phytoplankton, and thus, steeper ecosystem spectra.

3.) **Discrete change in the spectrum shape at well-defined critical sizes (KleiberSB, TopDoSH and CritGill).** Assuming that the ubiquitous size spectrum β, is due to trophic (e.g., top-down) regulation processes, PETS predicts that in phytoplankton communities that are subject to intensive zooplankton grazing, there should be a horizon (a lower size limit), where top-down processes reach and linearize the phytoplankton spectrum. For extremely small-sized phytoplankton, beyond the TopDoSH horizon, a deviation from the β spectrum is to be expected, with higher biomass than predicted (See Figure 5, scenario 2). From PETS theory, a discrete trophic control horizon (or boundary), from top-down controlled large-sized phytoplankton, within size structured food webs, and a completely different food web, with “microbial lo loop” (see Azam et al., 1983, Fenchel, 2008), small-sized-picoplankton, should be expected between TopDoSH and CritGill. PETS predicts that these critical sizes (KleiberSB, TopDoSH, and CritGill) can and will be detected in natural pelagic size spectra.

4.) **Picoplankton**. Smallest-sized organisms (i.e., small-sized primary producers) should display high biomass, above prediction, as in scenario 1 (Fig. 5). High picoplankton biomass is indeed generally observed in tropical oligotrophic oceans (Schwamborn et al. submitted)

5.) **Lakes vs Oceans**. The top-down regulation mechanisms explained above request a highly vulnerable prey population. It is not applicable to ecosystems where prey may hide and protect themselves from predators, for example in rivers and lakes with slush vegetation, where a considerable fraction of prey populations will spend time hiding. This may explain that β spectra are ubiquitous in marine pelagic food webs but sampling in streams and lakes has yielded numerous exotic, non-linear (domes and bumps), and flat (see chapter 7) size spectra.

6.) **Diversity**. According to PETS, a continuous, rigorously size-structured food web (Figueiredo et al., 2020) is necessary for a perfectly linear size spectra shape. The existence of discrete bumps within the spectrum may be due to the existence of low diversity, with very few species and size classes with gaps, troughs and bumps which may lead to non-continuous trophic cascades (Rossberg et al., 2019).

7.) **Fisheries.** The continuous removal of large-sized organisms (e.g., pelagic fishes, such as tuna, mackerels, etc.) leads to top-down cascades, reducing the total biomass of consumers throughout the food web (as in scenario 1), but without affecting the mean biomass spectrum slope ($\beta$ = -1) of the food web, and finally leading to an increase in the biomass of small-sized primary producers (as in scenario 1, Fig. 5).

8.) **Exotic, invulnerable predators** (e.g., lionfish). The introduction of exotic predators that have low or zero vulnerability to predation (e.g., lionfish *Pterois volitans*) may have a disrupting, non-linear effect in food webs (reducing biomass of consumers, including their preys, their competitors, and of much larger organisms, and increasing the biomass of algae) since the continuous size spectrum assumes full predation vulnerability.

a.) **Polar *vs* tropical ecosystems** In polar ecosystems, lower temperatures may lead to higher E, and thus higher PPMR and larger mean predator size, as observed in many obvious examples (e.g., baleen whales). Polar ecosystems have typically very large-sized zooplankton species (e.g., *Calanus finmarchicus*), that are able to store enough lipids for the long winter (e.g.., Lee, 1975, Clarke and Peck, 1991, Kwasniewski, et al., 2012, Balazy et al., 2018). Still, the weight-biomass relationship of such extremely large-sized animals is expected by PETS to be placed perfectly within an intact "β" spectrum, with β = -1. Indeed, large datasets have shown that zooplankton spectra have a slope of approximately - 1, regardless of whether in tropical or polar seas (Dugenne et al., 2024, Soviadan et al., 2024).

9.) **The effects of warming.** The relationship between temperature and "m" is a central topic of MTE, within Arrhenius-like relationships and equations that describe the relationship between metabolic rate "R" and temperature. Simply put, R increases with increasing temperature, and "m" may increase or not (depending on the organism type, e.g., prokaryotes *vs* eukaryotes). Therefore, several authors have suggested that size spectra should be steeper in warmer waters. Instead, PETS predicts that if increasing temperatures (possibly) lead to lower efficiency "E", this would subsequently be compensated by lower PPMR, thus maintaining the invariant "β" size spectrum slope. Considering that PPMR is directly related to predator size, increasing temperatures are expected to lead to smaller-sized organisms and lower HTL biomass (scenario 1 in Fig. 5), while maintaining a constant size spectrum slope of -1. This pattern may be expected within a warming world, for top-down controlled food webs, based on the PATER equilibrium mechanism described above. Interestingly, GOLT (based on oxygen-limitation-related processes) also predicts smaller-sized organisms and lower fish biomass in a warming ocean. Conversely, for the smallest-sized picoplankton, PETS predicts that an increase in temperature may lead to steeper size spectra slopes and higher biomass, given the strong temperature response of "m" in prokaryotes.

These mechanisms and dynamic processes, predicted from PETS, are yet to be examined and scrutinized by future large datasets of field observations. A recent study based on an extensive data set from picoplankton to mesopelagic fish (Fock et al., 2026, Schwamborn et al., submitted) has presented size spectra data for different communities across the Atlantic Ocean, that support the mass-specific predator-prey-efficiency equilibrium theory of size spectra (PETS) described above, based on the works of Borgmann (1982), Borgmann (1987), Brown et al. (2004), Zhou (2006), Andersen et al. (2009), and Barnes et al. (2010).

One key result of these studies (Fock et al., 2026, Schwamborn et al., submitted) was regarding the size spectrum of net-caught crustacean zooplankton (mostly copepods and euphausiids) in upwelling and oligotrophic systems. These data showed that the size spectrum of these zooplankton was invariant at b = -1, but the biomass was much higher in productive and efficient upwelling systems, than in less productive and less efficient tropical oligotrophic systems. This observation may confirm the prediction that an increase in food web efficiency can produce an invariant size spectra slope, but with higher biomass, as described above (see Fig. 3).

In these recent studies the offset between phytoplankton and zooplankton (linear model "contrast" *sensu* Fock et al., 2026), was a particularly large in the highly productive Benguela upwelling system, i.e., zooplankton biomass was much higher than predicted from the overall size Spectrum model, but with a constant-1 size spectrum slope within the zooplankton community food web ( Fock et al., 2026, Schwamborn et al., submitted)

## 10. Three Paradoxes solved - what have we learned?

*Paradox A.* ***Constant size spectrum and variable trophic pyramids***

Numerous previous studies have highlighted the prevalence of exotic (inverted, barrel-, or hourglass-shaped) trophic pyramids and offered a series of possible explanations (Jackson et al., 2006, Fath and Killian, 2007, Wang et al., 2009, Trebilco et al., 2013, Sandin & Zgliczynski, 2015, Kolding et al., 2016, Vadeboncoeur & Power, 2017, McCauley et al., 2018, Woodson et al., 2018, Barbier & Loreau, 2019, de Omena et al., 2019, Woodson et al., 2020, Schwamborn et al., in prep.). PETS provides a novel terminology and theoretical framework to establish a concise set of possible factors (i.e., a list of ***9 hypotheses, see below***) contributing to explain the existence of constant size spectra *versus* variable trophic pyramids, based on methodological (Schwamborn, 2026) and theoretical (this study) considerations.

At the beginning, we stated that a general trophic equilibrium theory (such as **PETS**) should be able to solve the "**constant size spectrum - variable trophic dynamics paradox**", that arises when comparing size spectra and trophic pyramids.

Most simply put, the relationship between log(abundance or biomass) and log(size or weight) is regular, but with regard to trophic levels, relationships are highly variable. Clearly, the log-log-transformation plays an important role in this comparison. Size spectra are double-log transformed, while trophic pyramids are usually not. This may already be sufficient to explain why size spectra generally look very constant and trophic pyramids are more variable. However, inverted and hourglass-shaped pyramids *vs* linear size spectra *cannot be explained* just by log-log-transformation or by binning artifacts (see Schwamborn, 2026).

The PETS theoretical framework discussed here, indicates that there is a compensating mechanism between biomass vs trophic level change efficiency E and log(PPMR), leading to a constant biomass vs weight change efficiency $\beta$.

Conversely, the paradox above indicates that apparently, there is no such mechanism in place that stabilizes trophic level-based relationships. Thus, we conclude that there is a strictly weight-scaled compensation between E and log(PPMR), which cannot be simply converted to trophic levels. One straightforward, simple explanation for this difference is that there is a size-based (or weight-based) compensation mechanism regarding the choice of small-sized and large-sized prey within functional responses of predators to varying prey abundance that does not apply to trophic-level specific models. In summary, the SOFT functional response mechanism of predators is intrinsically size-specific, not trophic-level-specific. Simply put, predators adjust their functional feeding strategy according to prey size and abundance within SOFT, and do not care about prey trophic level. This explains why trophic equilibrium mechanisms are size (or weight) - specific, not trophic level specific, leading to highly trophic pyramids and rigorously stable size spectra.

## *9 Factors Contributing to Constant Size Spectra vs. Variable Trophic Pyramids*

The apparent paradox between highly consistent size spectra (slope $\beta \approx -1$) and highly variable trophic pyramids (not linear, often showing hourglass- or inverted shapes) can be explained by several fundamental methodological and ecological differences:

### *I. Logarithmic vs linear plots*

Size spectra are inherently represented on double-logarithmic scales, whereas trophic pyramids are often visualized on linear scales. This distinction is critical: the logarithmic relationship between trophic efficiency (EE) and mass-specific efficiency (E) implies that even large variations in EE translate into relatively small changes in E. As a result, E remains approximately constant, while EE can vary over orders of magnitude.

This explains why size spectra appear stable and invariant, while trophic pyramids appear highly variable.

### *II. Compensation between E and PPMR (PETS equilibrium equation)*

Within the PETS framework, mass-specific trophic efficiency (E) and the predator–prey mass ratio (PPMR) are dynamically coupled, leading to a constant slope (b).

This compensatory mechanism ensures that energy transfer efficiency and feeding structure co-adjust, and that the slope remains invariant despite food web variability.

This equilibrium mechanism applies to the size biomass relationship only, not to TLs. PPMR and E (E being related to trophic pyramid slope, see below) may be variable, b is not.

### *III. E vs TP relationship: E is B-normalized, while TP scales with biomass*

The trophic pyramid slope (TP = dB/dTL) is related to E (E = dlog(B) / dTL ) as:

$$d \log(B) / d TL = 1 / B * d B / d TL,$$

hence:

$$E = TP / B$$

While E is the relative (log-scale) rate of biomass change per trophic level, TP is the absolute biomass change per trophic level. This explains our key insight: large variability in TP (trophic pyramids) can coexist with near-constant E (size spectra), because TP scales with biomass, while E is B-normalized.

*IV. Size-based vs trophic-level-based interactions*

Predators respond to prey size, not to prey trophic levels. The side-spectra stabilizing top-down SOFT mechanism (Size-spectra-specific Optimal Foraging Theory) is inherently size-based and not constrained by discrete trophic levels. Simply put, SOFT will not stabilize trophic pyramids, only size spectra.

*V. Normalization (NBSS vs pyramids)*

Size spectra are typically expressed as Normalized Biomass Size Spectra (NBSS), ensuring that variations in bin width do not affect the spectrum shape and slope. Conversely, trophic pyramids are generally not bin-width-normalized, often based on raw or aggregated biomass values across variable size ranges.

*VI. Size ranges increase with TL - the origin of the “inverted” pyramids artifact?*

Higher trophic levels typically span broader body-size ranges (max. – min. of several meters or tons within a TL) than lower trophic levels (e.g., size ranges of < 1 mm or < 1 g, within each TL). Within trophic pyramids, this distortion of bin widths mathematically leads to overestimation of biomass in higher trophic levels, if not properly size-normalized (as in NBSS).

This may result in a mathematical artifact that creates inflated biomass estimates at higher trophic levels and erroneous “inverted” pyramids. Within the logic of trophic pyramids, this is not actually an error (and it cannot be remedied), rather an example of how to explain the difference between these two approaches. Size spectra avoid this issue by either using continuously linear size bins (independent of trophic level), or using nonlinear binning with subsequent normalization (NBSS).

*VII. Allochthonous inputs into higher TLs*

High trophic level predators often rely on allochthonous (externally derived) resources.

Examples include wide-ranging predators (e.g., sharks) that forage outside the studied ecosystem. This may also result in inflated biomass estimates at higher trophic levels and “inverted” pyramids. Plankton size spectra are less affected, since they integrate biomass locally within size classes.

*VIII. Stable plankton size spectra vs variable fish communities*

While zoo- and phytoplankton size spectra are usually very stable (e.g., Dugenne et al., 2024), trophic pyramids based mostly on fish and other higher trophic levels are highly variable (e.g., Woodson et al. 2018). The apparent paradox between size spectra science, which has focused primarily on plankton, and trophic pyramids obtained from EwE models being largely dedicated to understanding fisheries and higher trophic levels, may help explain the contrast between invariant size spectra and highly variable trophic pyramids, given that larger higher TL are more variable than plankton. For example, top-down propagating dynamic trophic cascade waves are initially generated at high amplitude (for example, due to changes in fisheries or by intrinsic Lotka-Volterra-dynamics) and then are subsequently stabilized (damped) when reaching high-turnover food webs in the plankton (as in Andersen and Pedersen, 2010).

*IX. Scale issues, fishing, non-linear regime change, and gear avoidance in higher trophic levels*

Considering the short turnover times and small spatial scale of plankton interactions, as well as the relatively well-defined water volume cleared per individual, plankton sampling is comparatively simple and quantitative compared to the substantial challenges of sampling fish communities and higher trophic levels. For extremely large-sized organisms, it is nearly impossible to sample across adequate spatial scales and timescales. While mean plankton biomass can be well assessed from a few samples within a reasonable time series, many years or decades are required to capture population recovery and interaction processes in higher trophic levels.

One example of our limited understanding of extreme variations in higher trophic level biomass over decades and centuries is the sudden and unexplained appearance of extremely high numbers of the Galapagos shark (*Carcharhinus galapagensis*) around the remote St. Peter and St. Paul Archipelago (Brazil). Historical accounts from early explorers frequently described sharks there as being extraordinarily abundant, with expressions such as “water swarms with sharks” and “countless” individuals being common in 18th–19th century reports (Luiz & Edwards, 2011). Charles Darwin visited the archipelago in 1832 during the voyage of the Beagle, documenting its extreme isolation, sparse terrestrial life, and astonishingly high shark abundance (“*The sharks and the seamen in the boats maintained a constant struggle which should secure the greater share of the prey caught by the fishing-lines*”, Darwin, 1845). Later records suggest that shark populations declined dramatically during the 20th century (possibly due to fishing pressure), with reports indicating near disappearance from the mid-20$^{th}$ century until the early 21$^{st}$ century (Luiz & Edwards, 2011). For example, the author of this paper visited this remote archipelago during a two-week expedition in 2009–2010 and did not observe any visible shark aggregations. Since 2010, however, there have been more recent reports indicating signs of recovery and renewed extremely high abundances following protection measures. This long-term variability, spanning centuries, highlights how variable and difficult to sample higher trophic level dynamics can be.

This may help explain why plankton-based size spectra are generally more stable than trophic pyramids focused on fish and higher trophic levels. In addition, observational and sampling biases, such as gear avoidance and the difficulty of sampling large, mobile organisms across appropriate spatial and temporal scales, further complicate our understanding of higher trophic level variability with increasing organism size, in fast-swimming organisms (e.g fish and squid).

*X. Data quality, origin, standardization and model uncertainty*

Trophic pyramids - especially those derived from Ecopath with Ecosim (EwE) models - often include data that were not obtained *in situ*. EwE models are often assembled as a complex puzzle, including data from from far away regions and exotic ecosystems. Also, in such models, biomass is often estimated indirectly by the model (assumptions filling gaps where empirical data are lacking). In contrast, size spectra are typically based on standardized quantitative *in situ* data, and simultaneous sampling across size classes. This leads to greater variability and uncertainty in trophic pyramids, and higher consistency and robustness in size spectra.

The short list above is not intended to suggest any supposed superiority of size spectra (simpler and more consistent) over trophic pyramids (often based on complex models, higher uncertainty and variability). Both approaches have provided valuable insights into the structure and functioning of food webs across the planet over the past few decades, and remain valid and useful methods for summarizing highly diverse ecosystems. Rather than offering a “better vs worse” comparison, the list above aims to clarify the intrinsic differences between these two approaches and databases. Some of these differences have hitherto been overlooked (e.g., in Woodson et al., 2018). For example, this comparison has helped to identify possible origins of “inverted” trophic pyramids, such as the increase in size bin width with trophic level (Topic VI above), a mathematical artifact that has hitherto been largely ignored.

Combining these two data sources - the numerous size-spectra databases (e.g., the TRIATLAS size-spectra database) and Ecopath with Ecosim models compiled in ECOBASE - is far from trivial. However, if successfully achieved, such integration could lead to substantially improved global ecosystem models with significant value for climate prediction. Understanding the intrinsic differences between these two approaches is a first, necessary step in the direction of their integration into a unified, general ecosystem theory, such as PETS.

*Paradox B. Phyto- and zooplankton are described by different models - but have identical size spectra: the "**phyto- zooplankton size spectra paradox"***

The "phyto- zooplankton size spectra paradox" can be summarized by the contradiction between nutrient flux and metabolism being the main topic for phytoplankton *size distribution* literature, while trophic interactions are the basis for zooplankton size spectra models although both communities evidently follow the same size spectrum shape, slope and intercept, being absolutely perfectly aligned (Dugenne et al., 2024, Schwamborn et al., submitted).

The PETS Framework and the predictions listed above have simple straightforward and unquestionable explanations for this apparent paradox. The most simple explanation is that when two such communities are so perfectly aligned there is an intrinsic trophic coupling between them. This indicates that the phytoplankton community is not limited by nutrients but by zooplankton grazing, and that the top down cascades as shown above in SOFT scenario 1, will invariably lead to a continuation of the linear zooplankton spectrum deep into the phytoplankton spectrum, as long as grazing is the dominant process in shaping the phytoplankton biomass. Interestingly, the situation where all micro and nanophytoplankton are subject to control through grazing would lead to a situation where these two communities are well aligned.

In a hypothetical scenario, where grazing control through zooplankton is limited only to the larger-sized phytoplankton (for example, when zooplankton is feeding only microphytoplankton), this top-down control horizon would shift towards the micro-nanoplankton boundary and lead to a peak in the nanoplankton. Thus, from the comparison of nutrient-based models and top-down control models, we can predict that a discrete "top-down control size horizon" (TopDoSH) must exist within the phytoplankton. Both models (nutrient-related and trophic) are correct, and they shape the phytoplankton size spectrum in different sections of the size spectrum and under different nutrient flux scenarios.

*Paradox C. Different growth and metabolism, but similar size spectrum slope in plankton and fish: the "**CritGill paradox"***

The observation that large-sized nektonic organisms, such as mesopelagic fish and macroinvertebrates (Fock et al., 2026, Schwamborn et al., submitted), and large marine mammals, actually follow the global plankton size spectrum (Hatton et al., 2021), in spite of completely different growth dynamics, physiology etc., may be called the "CritGill paradox", which still requires more data and theory to be quantitatively confirmed. We may regard this apparent paradox as a further proof of the existence of a universal law (represented by the PETS equilibrium equations and mechanisms) that shapes and stabilizes the log-linear size spectrum.

The clear definition of the three paradoxes above and detailed description of how they were solved by PETS provided important clues regarding the trophic status of marine ecosystems and possible nutrient limitation and food-web regulation mechanisms. These mechanisms and regulation processes, as described above, may help in the interpretation of size spectra datasets and in the elaboration of future size-based trophic models that account for flexible functional response of predators within SOFT-PATER and RISE-TOPS mechanisms.

## 11. From mass to abundance, turnover and production spectra

While there is no doubt regarding a ubiquitous value of "b", the actual value depends on the units used. Several units have been proposed and are currently used for size spectra analysis. For example, Boudreau and Dickie (1992) used energy units per square meter, ($kCal/m^2$), instead of mass per volume.

The abundance-based size spectrum i.e., NNSS (normalized numbers size spectrum, Vandromme et al., 2012, Figueiredo et al., 2025) is considerably steeper than the NBSS (normalized biomass size spectrum, or normalized biovolume size spectrum, Dos Santos et al. 2017, Figueiredo et al., 2025). For example, Figueiredo et al. (2025) reported slopes of -1 to -0.8 for the NBSS (total biovolume *vs* indiv. biovolume size spectrum) and -2 to -1.7 for the abundance *vs* individual biovolume size spectrum (a form of NNSS). It is generally assumed that the carbon biomass vs individual carbon mass size spectrum has a universal slope of -1, although the vast majority of *in situ* studies use other, more practical units, such as biovolume, ESD (equivalent spherical diameter), linear size, or wet weight (dos Santos et al. 2017).

The ubiquitous -1 slope of the weight-biomass spectrum can be used to make predictions for three other relevant derived spectra (Fig. 8): the abundance-weight, production-weight (the total mass produced in unit time and space, in each weight bin), and the turnover-weight spectrum (the turnover rate in each weight bin), using allometric conversions:

- ***Weight-Biomass spectrum*** (i.e., body-mass-Biomass spectrum) slope: $\beta = b_{weight\text{-}biomass}$ = **-1**
- ***Weight-Abundance spectrum*** (i.e., weight-numbers spectrum), obtained by Abundance = Biomass / M: $b_{weight\text{-}Abundance}$ = -1 -1 = **-2** (Figueiredo et al., 2025; Landreth et al., 2025)

- ***Size-Biomass spectrum*** $b_{size\text{-}biomass}$ = -1 -2 = **-3** (Lombard et al., 2019; Dugenne et al., 2024)
- ***Size-Abundance spectrum*** (linear size): $b_{size\text{-}Abundance}$ = -2 -2 = **-4** (Behrenfeld et al., 2021)

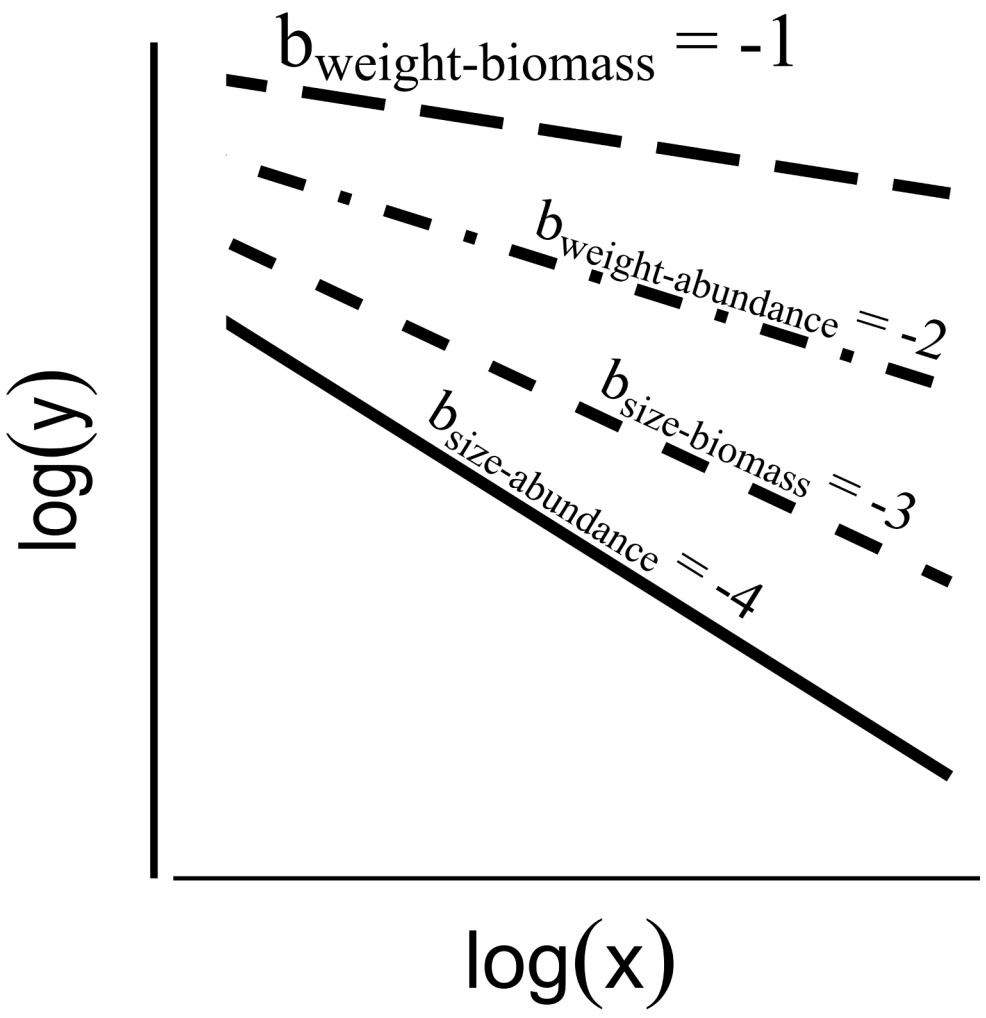

**Figure 8. The pelagic ecosystem size spectrum in different units.** Note that only the "weight-biomass spectrum" (i.e., the plot of log Body Mass vs log Biomass) has a slope of -1.

Thus, it is incorrect to state that the "size spectrum slope has slope of -1" as often seen in the literature. Only the weight-biomass spectrum (i.e., the plot of log Body Mass vs log Biomass, Fig. 8), has a mean slope of -1 (see also Sprules and Barth, 2026). Interestingly, variability-reducing binning artifacts in weight-biomass linear models, revealed in a recent study (Schwamborn, submitted), could lead, in the future, to a widespread preference of alternative models, such as the kernel density estimation (KDE) of weight-abundance spectra (i.e., the normalized numbers size spectrum NNSS, Vandromme et al., 2012, Figueiredo et al., 2025). So, in the future, size spectra theory studies may possibly analyse and discuss the "ubiquity of the $\beta_{weight\text{-}Abundance}$ = -2 weight-abundance spectrum slope". Since nowadays, most size spectra studies use units of weight and biomass, we will herein continue to keep using the weight-biomass log-log-linear model. This is based on the preliminary assumption that the same basic phenomena and processes affect weight-biomass and weight-abundance spectra (Figueiredo et al., 2026, Schwamborn, submitted).

Yet, none of these units, whether biomass, abundance, weight or individual size, compare in importance to a central parameter in ecology, which determines the magnitude of key dynamic processes: Production. The conversion from mass (i.e., biomass, when looking at living organisms only) into Production (generation of mass per unit time and space), is far from simple.

One big problem, when reviewing the literature on production, is an astonishing amount of confusion in units, terms (e.g., productivity vs production), and dimensions. Many authors (e.g., in zooplankton research, such as Huntley and Lopez, 1992, Zhou, 2006) look at mass-specific growth g (which is equivalent to P/B, i.e., turnover), others, especially fisheries scientists, report instantaneous growth rate (dM/dt) as function of asymptotic weight ($W_\infty$), within von Bertalanffy growth models (von Bertalanffy, 1938). Production, turnover, population growth, and body growth are often confused or ill-defined.

For phytoplankton, there are countless studies that focus on population growth, body growth, gross primary production or net production (obtained with different methods such as the oxygen method, or the $^{14}C$ method), leading to an immense fragmentation of units of P of P/B, maximum growth rates, and maximum nutrient uptake rates (Marañón et al., 2013), and *in situ* growth rates. Such rates are often related to cell size (dM/dt / dL) instead of mass-specific growth g ($dM/dt\ dM^{-1}$). For phytoplankton, previous studies (see review in Marañón et al., 2013) have reported unimodal distributions of maximum growth rate with a discrete constant temperature independent maximum at a discrete size class, based on the considerations regarding nutrient uptake, allometry, and metabolism.

Here, for the sake of simplicity, and standardization, the biomass production rate, in each size bin (and for the whole ecosystem), is defined as P = dB/dt. For living organisms, P can be regarded as the sum of all processes that (if unchecked by loss terms, such as mortality and export), would lead to an increase in total biomass of a size bin, i.e., the sum of processes, such as body growth (dM/dt), population growth (dN / dt) and reproduction ($dB_{eggs}$ /dt). In copepods, usually reproduction (egg production) dominates P, and is used to assess secondary production. In larger organisms, such as fish and macroinvertebrates, the assessment of body growth (dM/dt) is used to estimate P.

For any given type of non-organismic particles, an analogous term could be the mass of particles added per unit time (e.g., the mass of copepod faecal pellets added to the ecosystem per unit space and time), that is related to P.

There are numerous equations that relate production with individual mass for primary producers, zooplankton, and fish.

For small-sized, non-gill-bearing organisms, such as phyto- and zooplankton, we may define the mass-specific growth rate as:

$$g = dM/dt\ M^{-1} = a * M^{\mu} \qquad (29)$$

There is an immensely vast literature available on body growth of aquatic animals, but generally the instantaneous growth (dM/dt) of fishes and macroinvertebrates is a function of the asymptotic weight ($W_\infty$) for each species. For large-sized, gill-bearing organisms, such as fish and macroinvertebrates, instantaneous growth “g’ = dM / dt” usually follows the von Bertalanffy Growth Function (VBGF, von Bertalanffy, 1938). Fenchel (1974) found that for phytoplankton, mass-specific growth rates (i.e., growth turnover) scale as $\mu = -0.25$. The Metabolic Theory of Ecology (MTE) is intended as a general theory for all living organisms, including plankton and fish, where individual biomass production (i.e., body growth) is described as: $g \sim M^{0.75}$ and growth turnover is described as $P/B = \tau = g = M^{-0.25}$ (Fenchel, 1974, Brown et al., 2004, Kjørboe and Hirst, 2014).

Thus, we can assume that the slopes of marine weight-production and weight-turnover spectra are:

$$b_{weight\text{-}production} = -1 / 0.75 = -1.33$$

$$b_{weight\text{-}turnover} = -1 / -0.25 = 4$$

Thus, we can preliminarily conclude that the weight-production spectrum of marine ecosystems should have a mean slope of $\rho$ = -1.33, based on the (rather simplistic) MTE equations.

Although it is obvious that most vital services and processes in the oceans (e.g., fisheries and carbon pumps) are related to production (not biomass, or abundance), the weight-production slope $\rho$ has received surprisingly little attention within the marine science community, biogeochemists, and global ecosystem modellers.

## 12. Comparison with previous trophic equilibrium size spectra models

This short thought exercise builds upon the work on the relationship between PPMR and predator size by Barnes et al., (2010) and many other earlier works (Borgmann, 1982, Borgmann, 1987, Andersen et al., 2009) on size spectra theory. The most important argument for a trophic equilibrium theory, such as PETS, is the ubiquitous constant size spectrum slope of generally b = -1. The equations above add further robustness and detail to the understanding of size spectra dynamics, additionally to solving the key problems and paradoxes stated at the beginning and explaining key phenomena related to the ten predictions stated above.

Barnes et al. (2010), proposed that $\beta$ could be described by:

$$\beta = ( \log(SPTE) / \log(PPMR) ) - m \qquad (30)$$

, where SPTE is “the ratio of the production of a trophic level or mass category to that of its prey” (Barnes et al., 2010).

The PETS equilibrium equation $\beta = ( E * S ) - (m_i - m_0)$ is conceptually similar to the (Barnes et al. (2010) equilibrium equation, as the terms S and 1/log(PPMR are identical. However, log(SPTE) and the term “E” are completely different (Independently of whether SPTE is based on trophic level or weight categories, which the authors did not define), since SPTE is clearly related to production, not total biomass, as in E. Thus, when compared to PETS, the Barnes et al. (2010) equation proposes a completely different form of equilibrium, which is actually related to trophic pyramids (TL-based), rather than to size spectra (weight-bin-based).

Also, Barnes et al. (2010), included “m” into their equilibrium equation, which as we showed above, is unnecessary due to the “Kleiber-scaling” of trophic efficiency (see above). Similarity, Brown et al. (2004) also concluded that trophic efficiency is already Kleiber-scaled and thus, they did also not include a term for “m” into their basic equilibrium equation.

O spite of these differences, the PETS equilibrium approach is conceptually similar to that of Barnes et al. (2010), where β is also proportional to log(“some form of efficiency”) and to 1 /log(PPMR).

It is important to note that the simple PETS theoretical framework builds upon the previous work by (Borgman 1982, Borgman,1987, Andersen et al., 2009, and Barnes et al., 2010), and that this study does not ignore the plethora of theoretical work that has been published by numerous authors, with a large diversity of approaches, (e.g. Platt and Denman, 1977, Zhou and Huntley, 1997, Brown et al. 2004, Zhou, 2006, and Dalaut et al., 2025). The comparatively simple PETS approach does replace or conflict with these approaches, rather, it may be regarded as a simplification and extension of these previous modelling efforts.

There have been many attempts to derive the exact value of size spectra slope, from first principles and equations. For example, Brown et al. (2004) dedicated an extensive section to this topic, within their extensive explanations of MTE (West et al., 1997) equations in various fields, highlighting the importance of mass and temperature for a large list biological and ecological parameters, such as from metabolic rates, to biodiversity, growth and mortality. Within this extensive description of MTE, they tried to derive a convincing, unique solution that would lead to the explanation of the universal size spectrum slope, of -1, from an efficiency/PPMR ratio, that is conceptually similar to the equations proposed by Barnes et al., 2010, and to the present study. Yet, maybe because they were based on the erroneous assumption that the abundance-weight size spectrum should have a slope of -1 ( it actually has a slope close to -2), they developed a set of equations that led to an abundance-weight slope of-1 and a completely “flat” (!) weight-biomass spectrum with a slope of zero. Also there were other misconceptions regarding the marine size spectrum theory in this part of the description of the MTE, maybe because the focus of authors was on terrestrial plants and other terrestrial and mammal examples. As the MTE is designed to be a “theory of everything”, explaining biodiversity, life span, and many other phenomena, the seminal Brown et al. (2004) paper, devotes only a minuscule part of its attention to size spectra. This may explain why their approach to this area of research was not conclusive, though an important step towards our understanding of the trophic equilibrium. In their description of MTE, they did not mention, for example, the obvious trophic pyramid constant-size spectrum Paradox, which is fundamental for the understanding of the functioning of ecosystems. This probably explains why only a small section is devoted to this topic and why they produced absurdly erroneous concepts (e.g. a completely flat weight-biomass spectrum, with b = 0). There are many fundamental difference, also between the MTE approach of Brown et al. (2004) and the present PETS framework, for example that Brown et al. (2004) assumed a constant Lindeman efficiency of 10% Lindeman, 1942) and a variable PPMR, and did not recognize the equilibrium of c = E * S, with variable values of E and S. The treatment of size spectra in their work was, however, consistent with knowledge and concepts at that time, more than two decades ago. No dos resent any possible mechanism within MTE that would plausibly explain the equilibrium between trophic efficiency and PPMR, and variation in PPMR, as recognized by the authors "We do not yet have a mechanistic theory to explain… ". Also, MTE predicts that "energy flux or productivity per unit area of an ecosystem is therefore predicted to be independent of body size" (Brown et al., 2004), which is obviously erroneous (see above).

Thus, the PETS equations, MTE equations (e.g., "log ("Lindeman efficiency") / (log PPMR) = -¼ ", Brown et al., 2004) and the ones presented by Barnes et al. (2010) are conceptually similar, as they represent the idea of an equilibrium between some measure of trophic efficiency and 1/log(PPMR), but with completely different assumptions, terms, units and scalings, which probably represent the knowledge at the time when they were elaborated. Most importantly, these two previous studies (e.g. Brown et al., 2004 and Barnes et al. 2010), did not present any possible mechanism that would explain the equilibrium between trophic efficiency and PPMR, based on size-specific optimal foraging (SOFT), as proposed here. In summary, the PETS equations and theoretical framework are completely different from MTE, and although, similarly to MTE and GOLT, PETS is also derived from the original allometric theory (Kleiber, 1932), in combination with equilibrium-related concepts and equations (Lindeman, 1942, Polovina, 1984), optimal foraging theory (MacArthur and Pianka, 1966; Werner and Hall, 1974; Krebs, 1977), and predator size - PPMR relationships (Barnes et al., 2010).

Rather than presenting a completely new concept *ex nihilo*, this study builds on a large body of previous work, and is intended as a revision, correction, generalization and extension of the equations presented by Borgmann (1987), Andersen et al. (2009), and Barnes et al. (2010).

## *13. Towards a "compleat" generalised size spectra model, including organisms and particles*

It is obvious that the oceans are filled with myriads of particles that are not living organisms, (Fig. 9), such as near-infinite different types of detritus, marine snow and microplastics (Silva et al., 2019, Lins-Silva et al., 2021, da Cruz et al., 2023, Lins-Silva et al., 2024). Although this fact is the base for many biogeochemical models, it has been completely ignored within the size spectra modeling community, leading to many misconceptions and considerable bias.

An example of such a misconception, is that the total mass or volume of a zooplankton sample caught with a plankton net is still widely considered an appropriate indicator of zooplankton biomass. However, a recent study (Silva et al., 2019) showed that zooplankton samples actually contain relevant amounts of non-organismic, discernible, robust particles (that are not destroyed by the towing of plankton net, and can be easily distinguished from living organisms), which can make up more than 50% of the mass or volume in such common, formaline-fixed samples (Silva et al., 2019). Furthermore, discernible robust particles are only a minuscule fraction of all non-living particles in pelagic ecosystems. Results from on-board staining experiments showed that dead copepod carcasses can make up more than 90% of the zooplankton (da Cruz et al., 2023). Such carcasses are counted as living biomass in routine plankton work, based on formaline-fixed samples, as long as they are morphologically intact.

Even more important than robust particles and carcasses are ephemerous liposomes, and numerous types of fragile, gelatinous particles such as “porous aggregates”, which may make up the vast majority of detectable particles in the oceans, being much more abundant than living organisms, as has been recently observed with large datasets obtained with the UVP *in situ* imaging device (Stemmann et al., 2008). Only few studies have attempted to quantify the trophic relevance of ubiquitous detritus for zooplankton feeding, with considerable ingestion rates, although fresh and living food items are generally selectively preferred (Schwamborn et al., 2006). The feeding selectivity behaviour of each single organism, and the abundance and composition of particles in the size spectrum will define the actual relevance of non-living particles in the food web. Yet, even the most simple information, the size spectrum of particles, is still largely unknown.

Considering the ubiquity and unquestionable dominance of non-organismic particles in pelagic environments, it is quite surprising and unexplainable that until now, no efforts have been undertaken to integrate particles into current size spectra theory. For example, there are no available estimates of mass, size, and turnover rate ($\tau$, $d^{-1}$) of non-living particles ever mentioned in the vast literature on size spectra.

The turnover rate $\tau$ has been well established for many living organisms, for example for fish populations that suffer from natural mortality “$Z_{natural}$” and Fisheries mortality “$Z_{fisheries}$” (e.g., Pauly and Christensen, 1991):

$$\tau = Z = Z_{fisheries} + Z_{natural} \qquad (31)$$

Equilibrium terms for sinking particles vary between types of particles, but they can be separated into positive (particle generating) and negative (particle removal) processes. For example, for nutritious, edible particles (e.g., carcasses, aggregates, plant detritus, etc.) it is possible to calculate $\tau$ from the weight-specific sinking rate s’ ($d_{-1}$) and weight-specific ingestion loss rate i’ ($d^{-1}$), i.e., the loss through ingestion by consumers:

$$\tau = s' + i' \quad (32)$$

Biogeochemical models often consider particle-related processes and variables, but current size spectra theory does not, leading to a fractionation of knowledge and models.

**The complete ecosystem mass spectrum**

(living organisms and particles)

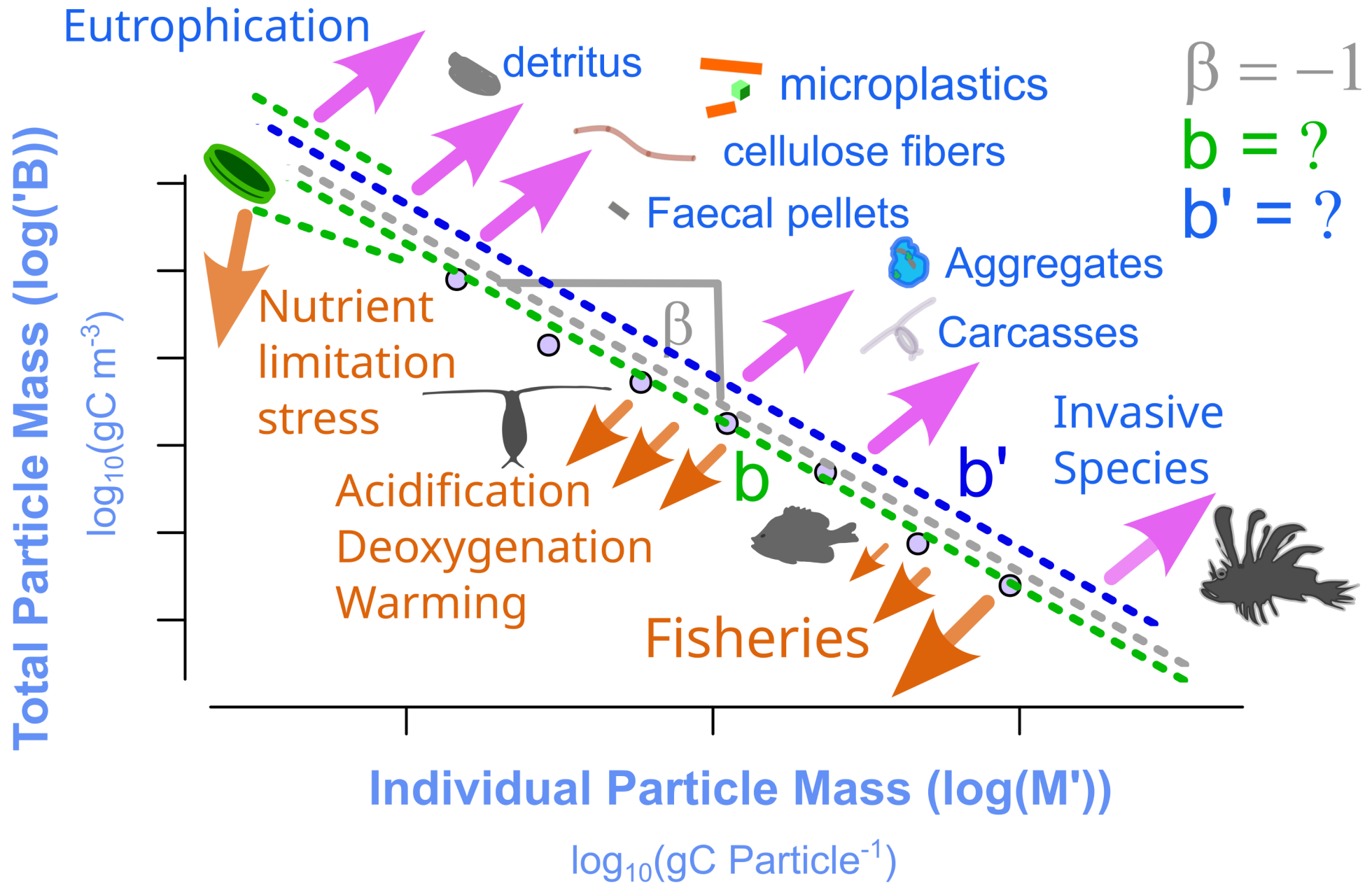

β: Equilibrium spectrum (β = -1)

b: Observed ecosystem biomass spectrum (living organisms)

b': Complete ecosystem mass spectrum (living organisms and particles)

**Figure 9. The complete ecosystem weight-mass spectrum.** Depicted are the main positive (purple arrows) and negative (orange arrows) processes and types of particles that act upon the original biomass equilibrium spectrum. Note, for example, the depicted decrease in phytoplankton biomass and change in phytoplankton size spectrum slope (affecting b and b') due to nutrient limitation stress (oligotrophic ecosystems), in the upper left corner. For many of these anthropogenic and natural processes, the impact on the size spectrum b' (organisms and particles) is still unknown.

There have been very few attempts to quantify the complete ecosystem size spectrum, including particles and organisms. Lins-Silva et al. (2024) showed that the size spectra of non-organismic particles, including biogenic particles and microplastics, are characteristic for each ecosystem, when comparing tropical estuarine coastal and shelf areas in northeastern Brazil. Biogenic particles, such as mangrove detritus, had a significant impact on the overall ecosystem spectrum, leading to higher mass per bin and flatter "complete size spectra" (i.e., weight-mass spectra), slopes, as compared to weight-Biomass spectra (living organisms only), in estuarine and marine waters (Lins-Silva et al., 2024).

This shows that ignoring non-organismic particles leads to an incomplete, misleading picture of marine ecosystems. One example of the fact that non-organismic particles have been generally neglected is the widespread use of non-imaging particle counters, such as the LOPC and OPC (e.g., Zhou, 2006), that supposedly quantify "biomass" spectra, while actually not biomass but total (organisms and particles) abundance and volume of living and non-living particles is quantified. A LOPC will count bubbles, detritus, marine snow (e.g., aggregates, crustacean exuviae, larvacean houses, animal (e.g., copepod) carcasses), liposomes, microplastics, and suspended mineral particles, leading to all kinds of "exotic" size spectra (Zhou, 2006).

Since the very beginning of marine size science, many seminal studies (e.g., Sheldon and Parsons, 1967, Sheldon et al., 1972; Sheldon and Sutcliffe, 1973, Sheldon et al., 1977) and recent publications (Zhou and Huntley, 1997, Zhou, 2006, Marcolin et al., 2013) have used datasets that pool the sums of nonliving particles and living organisms together (e.g. from particle counting devices, such as LOPC), for whatever reason neglecting the non-organismic particles and assuming that such data can represent the zooplankton biomass. However, their theory and equations refer to biomass (living organisms only). Yet, the present study is the first to highlight the importance of explicitly including particles in the size spectrum analysis, based on recent field observations (e.g., Lins-Silva et al., 2024).

*Considering non-living particles in size-spectra data analysis*

Note that size spectra obtained from optical instruments that do not distinguish particles and organisms (such as the LOPC in Zhou and Huntley, 1997, Zhou 2006, and Da Rocha Marcolin et al., 2013), also count structures such as bubbles, marine snow, and suspended mineral particles. This highlights the need of future field studies and theoretical work (i.e., a generalized theory) to describe and explain and predict the complete size spectrum through dynamic processes.

A preliminary comprehensive trophic equilibrium equation for the observed complete spectrum b' (organisms and particles *sensu* Lins-Silva et al., 2024) can be formulated as:

$$b' = b + \rho' + \varepsilon' \quad (33)$$

, where

b': slope of the complete weight-mass spectrum, i.e., the complete size spectrum obtained *in situ*, including all living organisms and all nonliving particles (e.g. for data from LOPC and UVP), where b' = dlog(B') / dlog(w'), and B' = total particle mass (living and nonliving), and M' = mean individual mass of all particles (living and nonliving).

b: slope of the weight-Biomass spectrum (living organisms only).

ρ': slope-changing effect ($\rho' = dlog(B') * (dlog(w'))^{-2}$ ) of non-living particles ε' : stochastic error due to the addition of non-organismic particles.

The relationship between the size spectrum slope of non-living particles ($b_p$) and the size spectrum slope of living organisms, in establishing the complete size spectrum b', may be described by a simple weighted mean:

$$b' = \text{weighted mean ( mean}(b), \text{mean}(b_p) \text{ )} \qquad (34)$$

, while

$$\tau_p = \text{weighted mean ( mean}(b_{p1}), \text{mean}(b_{p2}), \text{mean}(b_{p3}), \dots , \text{mean}(b_{pi}) \text{ )} \qquad (35)$$

where:

$\tau_p$ = mean turnover rate of non-living particles.

$b_p$: size spectrum slope of all non-living particles.

$b_{pi}$: size spectrum slope of non-living particles of a specific type (e.g., plant detritus), where $b_{pi}$ is proportional to $\tau_i$

Note that $\tau_p$ is not equal to $b_p$. This conception is in analogy to the observation that for each population of living organisms, $b_i$ is proportional to $\tau_i$, but overall ecosystem b is determined by the log-linear carrying capacity spectrum of all populations, not by the turnover and $b_i$ of each population (governed by g/z in each population).

Whether there is a (log-linear) carrying capacity spectrum of non-living particles (controlled by density-dependent grazing) is yet unknown (subject to future studies), also there may be exist a carrying capacity spectrum of living+non-living particles (as in Silva et al., 2024), which may contain a “Nutritiousness Factor” considering that not all particles are equally nutritious and palatable (analogous to the vulnerability term for prey-predator interactions). While the existence of a log-linear carrying capacity spectrum of living organisms has been recently observed (Schwamborn et al., submitted), the existence of such a **carrying capacity spectrum for non-living, or for all (living+non-living) particles** has yet to be investigated *in situ*. Far from being a merely academic question, this hypothesis (i.e., the “carrying capacity spectrum of particles” - hypothesis) has far-reaching consequences for our understanding of ecosystems, for example regarding the effects of the **size spectrum of microplastics** on living beings.

Numerous processes affect nonliving particles, such as, for example:

$r_P$ Particle mass production rate (dB/dt) (e.g. the mass of fecal pellets produced per day per unit volume),

$r_S$ linear sinking loss rate (dB/dt),

$r_A$ (dB/dt) linear aggregation gain rate (from smaller particles),

$r_D$ (dB/dt) linear destruction loss rate (decomposition and into smaller particles),

$r_I$ (dB/dt) linear ingestion loss rate (loss through ingestion), and

$r_d$ (dB/dt) linear dissolution loss rate (loss through dissolution into DOM).

Thus, a preliminary, simple mass balance equation may look like this:

$$r_P + r_A = r_D - r_I - r_d \quad (36)$$

Analogously to the trophic bottom-up and top down-processes that transfer Biomass between weight bins, destruction and aggregation rates will analogously transfer mass between weight bins, generating smaller and larger particles from each other.

A log-linearly declining particle size spectrum may be obtained, for example by faster sinking of larger particles (according Stokes’ law). This process alone may be sufficient to create a log-log linearly declining spectrum with a power law distribution. Additionally, the balance between aggregation and destruction loss rates which will lead to exchanges between size bins (analogously to trophic processes between trophic levels) may also lead to a power law distribution of particles. Both processes together will most likely shape the particle size spectrum together with the dynamics of the diverse particle generating processes, such as carcasses, exuviae, and fecal pellets.

Considering that exuviae, carcasses and fecal pellets probably have a similar size distribution to living organisms, they probably also have a power law distribution, but with a slope that is steeper than -1, given that Stokes's law will remove larger carcasses (and larger fecal pellets) much faster than smaller ones, by sinking. Conversely, Lins-Silva et al. (2024) showed that natural biogenic particles in a tropical estuary had a size spectrum slope that was much flatter than the size spectra slope of living zooplankton. Many data compilations and modeling efforts are still necessary for a proper understanding of the relationships between nonliving particles and living organisms in the oceans (Stemmann and Boss, 2012). Although previous authors have already suggested such an integrated approach (e.g., Stemmann and Boss, 2012), until now very little has been done to integrate size spectra of nonliving particles and living organisms.

## 14. Alternative approaches, viewpoints and models

Alternatively, several other approaches and models, based on processes other than trophic equilibria, may be considered to analyse and explain size spectra in marine environments.

*Growth and mortality*

One might be tempted to explain the ecosystem size spectrum through population-level processes, such as body growth and mortality. For example, for any given (adult) population, the number of individuals from birth, in any cohort, starts with very large numbers, which decrease exponentially. This decrease is described by the mortality parameter "Z" (d-1). Combining mortality and weight-specific body growth g (dM/dt $dM^{-1}$), will produce the exact size distribution of the population (Abundance or mass as a function of size or weight). Thus, it is possible to use the size distribution to estimate mortality Z, when body growth (and thus, individual age) is known, into a length-converted numbers-age distribution (the "length converted catch curve", Baranov, 1918, Pauly, 1983). The slope of the numbers-age distribution can indeed be used to study the mortality of natural populations. It makes perfect sense to investigate ecosystem size spectra under the aspect of growth and mortality. The observed mortalities and growth rates are highly variable. However, species with higher Z tend to have faster growth, which may lead to a population level equilibrium of mortality / growth ratios. Numerous studies have been dedicated to the monumental challenges involved in the assessment in body growth (e.g., Schwamborn et al., 2019; Schwamborn and Schwamborn, 2021; Schwamborn et al., 2023; Wilhelm, et al., 2025) and mortality / growth ratios. (e.g., Schwamborn, 2018; De Barros et al., 2024). Actually, it is not absurd to imagine that a growth mortality / equilibrium qualifies for a reasonable explanation of the universally observed constant size spectrum slope. Yet, the processes at population level have limited effects, within a discrete size range, while size spectra analysis covers a large number of orders of magnitude, species, life history stages and taxonomical groups. In conclusion, it seems that growth / mortality equilibrium processes have a limited scale (few orders of magnitude, single-species only) within the overall ecosystem size spectrum, although g/z has been indeed included in PETS equation 26 (to consider the possibility of such z/g - driven variations in size spectra), it does not qualify as the main determinant for the size spectrum slope of a whole ecosystem, within PETS.

One particularly important argument that speaks against the importance of the growth / mortality /ratio as a key determinant of ecosystem size spectra is that the observed species- and life history-stage-specific size spectra are generally not perfectly log-linear and rarely exhibit a slope of -1 (Lira et al., 2024; Schwamborn et al., 2025). However, it is important to highlight that E can also be regarded as a direct representation of g/z. Higher mass-specific trophic efficiency "E" reflects a lower g/z ratio. Conversely to PETS, such diverse processes as growth, mortality, recruitment, and food-web-related processes were indeed included as key factors within another approach, that has been widely cited: the highly complex equations of Zhou (2006).

The comprehensive set of equations of Zhou (2006) include many different parameters, such as birth and mortality rates, net mortality including birth, death and predation, the recycles of biomass between different trophic levels and the (mass-specific?) "growth of all individuals", "assimilation efficiency of the community", with a focus on "size-dependent growth". Zhou (2006) presents several time-independent state solutions and a first-order wave equation. Different ecosystems with different size spectra slopes are analyzed by assessing the "numbers of internal biomass recycles", assuming a constant invariable "community assimilation efficiency of 0.7" which may be questionable in view of the highly variable efficiency values reported in the literature. The author himself notes that "Note that there is no justification for choosing the assimilation efficiency of 70% ….". This extremely complex approach is certainly capable of producing a number of different outputs, for example the author lists scenarios with biomass size spectra that range from -2 to –0.5. In summary, the seminal paper by Zhou, 2006, can be seen as an important contribution to the discussion regarding size spectra but it fails to recognize a universally invariant slope of b = -1, probably because it uses numerous LOPC-derived size spectra, with variable results that are interpreted with regard to a fixed assimilation efficiency of 70%. Then, using this (questionable) fixed assimilation efficiency value, the size spectrum slope is used to estimate the food chain length (i.e., the "number of recycles"). Most importantly, Zhou (2006) does not present testable predictions, and thus, it is extremely difficult to test the premises and accuracy of this approach. Maybe, future studies, by quantitatively analyzing food chain length, size spectra slopes and assimilation efficiency, will be able to verify the accuracy of the Zhou (2006) equations (although no exact units and values are given). This is probably out of reach for contemporary science, given the immense uncertainties regarding food chain length and PPMR across complete ecosystems (e.g., from stable isotopes, as in Figueiredo et al., 2020).

*Oxygen supply, body size, and respiratory surfaces*

The Gill-Oxygen Limitation Theory (GOLT; Pauly, 1981; Pauly et al., 2010; Pauly, 2019, 2021; Cheung & Pauly, 2016) is based on the allometric rationale that body mass, body surface, and respiratory surfaces change in different scalings with body size, and that oxygen uptake is a key challenge for aquatic organisms. Thus, GOLT could lead to the assumption that size-spectrum slopes should vary among small, gill-less organisms that respire through their body surfaces, larger, gill-bearing taxa, and air-breathing organisms (e.g., reptiles and mammals). Conversely, numerous studies have documented a uniform biomass-weight spectrum with a slope of -1 across a broad range of body sizes and taxonomic groups (Boudreau & Dickie, 1992; Hatton et al., 2021; Dugenne et al., 2024; Figueiredo et al., 2025; Fock et al., 2026; Schwamborn et al., submitted). This observation does not negate the fundamental principles of GOLT; instead, PETS indicates that respiration-related physiological changes in growth are compensated by the universal, mass-specific equilibrium mechanisms described above, thus leading to the invariant $\beta$ size spectrum.

*Space and time*

The most simple way to look at the size distribution of living beings in nature is to verify the relationship between size and the minimum space necessary for each organism (i.e., the carrying capacity of the system, for each organism). For large sized-apex predators, indeed the carrying capacity has been often defined by the space (area, territory, or volume) necessary to maintain their numbers. Many studies have reported increasing stress, disease and agonistic intra-specific interactions, when apex predators exceed their carrying capacity. Extrapolating from this simplistic approach from apex predators to the whole food web, one may propose that a consistent global maximum-carrying-capacity-body-size relationship is universal throughout the food web, for all living organisms. Convertino et al. (2013) modeled the size spectra of different physical and biological systems, including food webs, with such a simple geometrical approach, and were surprisingly successful. Thus, it is not absurd to imagine that there are physical (spatial and geometrical) laws in place, that define a carrying capacity size spectrum and that natural populations will grow in numbers until reaching this exact carrying capacity maximum. Interestingly, such models are relatively simple, based on contact surface areas of individual spaces, and do not actually need to include any terms for growth, mortality, metabolism, or trophic interactions. Yet, such simplistic geometric models (albeit compelling), may not be able to explain certain phenomena, such as the extremely high biomass of zooplankton in the oceans, above the global “$\beta$” linear model (Fock et al., 2026, Schwamborn et al., submitted), and the extremely high biomass of bacteria.

Another way to look at size spectra, which interestingly was one of the earliest (Platt and Denman, 1977), is to convert size and mass into units of time. Platt and Denman (1977) considered time (e.g., turnover time of body mass and the time scale of system energy loss) as key to understanding and modeling the marine size spectrum.

Time scales of basically all physiological and population processes are related to size (or weight). Obviously, very small organisms, such as bacteria, have population processes such as the recovery from a catastrophic event, in terms of minutes and hours, zooplankton may have population processes in the scale of weeks or months, most fish have population processes in the scale of years and decades, and large sharks and mammals may have population processes in the scales of decades to centuries. Thus, it is possible to look at the x-axis of a size spectrum (e. size or individual weight) as a representation of time. Similarly, the y-axis for example abundance or biomass can be seen as a representation of the space (i.e. volume) used by each individual. Thus the size spectrum would be a representation of the log-log linear relationship between space and time. This is analogous to the common representation of longevity and size in animals which has a similar shape.

In spite of being very interesting and inspiring thought exercises, the practical usefulness of these approaches, for example for modeling and predicting changes, in the context of global warming, may be rather limited.

*Universally constant PPMR and E*

The PETS theory described above assumes a constant ideal size spectrum slope of $\beta = -1$ that is maintained by highly variable PPR and E, which are in equilibrium (i.e., “E ~ log(PPMR)”). Alternatively, it is also possible to imagine a universe where PPMR and E are constant within and across ecosystems and thus, there is no need for any equilibrium mechanisms at all. So far, there have been no attempts made to quantify E *in situ*, and actually there is no well-established method yet available to achieve this. Conversely, there have been already two published attempts to quantify PPMR in marine zooplankton food webs, encompassing many trophic levels, orders of magnitude of size and biomass, and thousands of species, *in situ* (Hunt et al., 2015, Figueiredo et al., 2020). Both studies showed a considerable amount of uncertainty regarding their numerical estimates of PPMR, but both studies showed a set of consistent linear relationships that indicate a constant PPMR throughout the marine zooplankton, in spite of many different taxonomic groups and biochemical compositions. It may initially seem absurd to imagine a universally constant PPMR, considering extreme singleton PPMR examples in nature, such as, for example baleen whales x krill, *Oithona* spp. x dinoflagellates, white sharks x seals, brachyuran crab larvae x copepods, and the huge variation in singleton PPMR values reported in the literature (e.g., Barnes et al., 2010). Yet, it is actually possible to imagine, in theory, that mean log(PPMR), and thus mean “E”, are relatively constant within and across complex and highly diver marine food webs with innumerable species and life history stages. Considering that the only two previous studies Hunt et al., 2015, Figueiredo et al., 2020) focused on zooplankton food webs, it is also possible to imagine that mean log(PPMR) is highly variable in less diverse higher trophic levels but near-constant within the highly diverse zooplankton.

Studies on single-species singleton PPMR values (e.g., Barnes et al., 2010) are not adequate to assess the mean food web PPMR, given the high complexity, diversity of prey and predators, ontogenetic changes, and numerous possible different prey-predator functional responses (e.g., Type I, II, and III). Further studies are urgently needed to quantify PPMR precisely in different food webs and ecosystems, as to assess the global variability in mean food-web-scale PPMR values.

*Variable values of "b" explained by local phenomena*

Throughout this paper, a constant, ubiquitous size spectrum slope of β = -1 has been assumed. Conversely, several authors have observed sized spectra slopes that are considerably different from -1 and have explained these variations by local phenomena, such as the "Island Mass Effect" and larval release from oceanic islands (Lira et al., 2024), or the effect of estuarine plumes (Santana et al., 2025). Other authors have tried to to explain such deviations from -1 by variations in trophic efficiency, PPMR, or food chain length (Zhou, 2006), but which may actually be due to methodological issues (e.g., the LOPC counting non-living particles). Thus, it is important to remember that PETS does not assume a constant value of b = -1 in all ecosystems and in all communities within a given food web. Many variations are of course possible, many have been described (see examples above), and certainly, many more will be found in the future. However, it is an undisputed fact that marine pelagic ecosystems have a general food-web-wide size spectrum slope of -1, especially in the zooplankton (Dugenne et al., 2024, Soviadan et al., 2024).

*Dynamic size niche interactions and gap-filling evolutionary strategies*

Another approach to understanding regulation and equilibrium mechanisms in size spectra is to look at the taxon-by-taxon, bin-by-bin dynamic interactions in real time. This detailed "size-niche interactions" approach has been successfully implemented in a recent study (Schwamborn et al., 2025). The authors described how gaps in the size spectrum were immediately filled by well-specialized species and life history stages, specifically decapod crustacean larvae. Also, they found that the insertion of new, additional individuals within a given size bin into the system, such as the release of meroplankton in immense numbers in tropical estuaries, leads to adaptations in the size spectrum, namely the functional replacement of species within specific size bins, until the general linear size spectrum is restored. They found that the addition of meroplanktonic brachyuran larvae did not lead to bumps in the spectrum, rather these added larvae replaced previously existing organisms within any given size bin, such as similarly-sized copepods. The authors hypothesize that copepods that are eliminated from the size spectrum by similarly sized crab larvae, are most likely predated upon by similar-sized brachyuran crab larvae. Alternatively, copepods may be evading predatory crab larvae by migrating (e.g., by vertically escaping into deeper, less resource-rich layers). Similarly to the present study, they found that size spectra in pelagic ecosystems are regulated mainly by top-down control mechanisms and a carrying capacity spectrum.

Obviously, the mean spectrum reflects the average across numerous temporal scales that vary according to size from seconds to months. It is important to remember that predator-prey interactions are highly stochastic and dynamic in time and space. When assuming the existence of Lotka-Volterra dynamics between predators and prey, even in a constant resource-rich environment, there will be regular oscillations in time due to the intrinsic dynamics of predator-prey interactions. Such dynamics may be one of the reasons for the observed peaks and bumps in marine ecosystem size spectra. Yet the observation of such dynamic behavior and bumps and domes does not invalidate the assumption of a constant invariant size spectrum with a constant slope of -1 between and across ecosystems. Our ability to actually observe such a constant size spectrum depends on sampling the adequate numbers of organisms and repeated sampling in space and time which is often not feasible within limited resources for marine sciences, e.g., in many developing countries. Imaging devices, such as the FlowCam, ZooScan, and the UVP are promising tools to develop large-scale databases of size spectra, although their costs are prohibitive for many institutions.

Interestingly, the comparison of samples obtained *in situ* with different concentrations of non-living particles and their effect on the size spectra slopes can be used to verify whether nonliving particles (that don't have growth and mortality) fit into the expected size niches and whether such size niches and their carrying capacities ($B_{max}$) are actually constant, and independent of metabolic and trophic processes. These size-niche interactions, gap-filling processes, and functional replacement may be summarized as the **RISE** Hypothesis (**R**eplacement **i**n **S**ize-structured **E**cosystems, see above).

Further studies are necessary to investigate these relationships, especially considering that *in situ* observations as opposed to balanced laboratory experiments can not be confidently used to analyze causal effects and rapid dynamic processes in size niches (such as functional replacement and competition avoidance, see Schwamborn et al., 2025).

*Size-niche interactions within an evolutionary perspective*

In contrast to theories for innate systems (e.g., in physics and geology), any theory that intends to explain biological systems has to consider not only the "how" but also the "why", i.e., the evolutionary advantage of a given phenomenon or pattern over other possible alternative configurations. Clearly, the existence of a universal β spectrum in three-dimensional ecosystems indicates that there is an evolutionary advantage for natural food webs to distribute biomass and abundance according to the perfectly log-log linear (power-law-shaped) β spectrum. The hypothetical alternatives of flexible (variable), inverted, dome- or hourglass-shaped size spectra are obviously less sustainable in the long term, as compared to the observed power-law shape, considering its ubiquity in the marine pelagos. Interestingly, the possible existence of inverted (or hourglass-shaped) biomass-trophic-level pyramids does not seem to hamper the sustainability of natural ecosystems. This further supports the WETBIO hypothesis, that the relationships between volume, weight and mass are governed by universal laws (e.g., $V_{min} \sim M^2$), while trophic levels, production, feeding selectivity, consumption, and mass transfer up the food chain are dynamic and variable, as is the size- specific predation strategy of predators (SOFT and PATER), that can be highly variable and flexible, according to (type I, II or III) size-specific responses.

Within an evolutionary perspective, it seems likely that the power-law- (log-log-linearly-) shaped β spectrum is more stable than any "exotic" spectra, for pelagic ecosystems.

Accordingly, all living organisms in these ecosystems have evolved towards an optimized niche-occupation strategy (within competition avoidance and functional placement strategies) within this size spectrum, i.e. towards successfully filling a specific “prey size niche”, “predator size niche”, and “body size niche” (Schwamborn et al., 2025), that fits perfectly into the surrounding background $\beta$ spectrum (see the *RISE* mechanism described above, within PETS theory).

## 15. Estuarine and coastal size spectra - why so steep?

A remarkable exception to the b = -1 paradigm is the observation of extremely steep size spectra in tropical, subtropical and temperate estuaries, lagoons, and estuarine plumes, with values often steeper than b = -2, for weight-biomass or weight-biovolume spectra (e.g., Tao et al., 2008; Ke et al., 2018; Mallick et al., 2024; Schwamborn et al., 2025; Santana et al., 2025)

Several factors and processes may be considered, that are probably contributing to this intriguing phenomenon Fig. 10). First of all, coastal lagoons and estuaries are shallow environments. If we assume that diel vertical migration (DVM) is a key factor in shaping zooplankton size spectra, with larger organisms needing a considerable water column depth to conduct this migration, we may conclude that shallow estuaries and lagoons have less large-sized organisms because they are not able to conduct DVMs, which is necessary, as to hide from visual predators at night. Thus, organisms are more prone to visual predators such as anchovies, sardines and mojarras, which are generally extremely abundant in coastal lagoons, estuaries, and estuarine plumes. Since large-sized organisms conduct a higher amplitude DVM than smaller ones, the impossibility to conduct DVM will affect mostly larger-sized organisms such as large-sized copepods and euphausiids, and would less affect smaller-sized organisms, such as copepod nauplii, which are extremely abundant in estuaries and lagoons. This hypothesis would also mean that shallower shelf systems should have steeper slopes then deeper systems, such as observed in tropical (Figueiredo et al., 2025) and temperate seas (Sourisseau and Carlotti, 2006).

Additionally, estuaries are dynamic outflushing systems, that are continuously exporting organisms towards the sea, which is a form of pseudo- mortality, reducing the numbers of older, large-sized organisms in relation to small-sized, younger ones. Also, estuaries are often important nursery areas, being characterized by considerable abundances of larval organisms, such as decapod crustacean larvae (Schwamborn et al., 2025). Their fast growth and larval export ("pseudo-mortality") contribute to the steep estuarine size spectrum.

Another central characteristics of estuaries are the extremely high non-predatory mortality rate due to variable and often lethal conditions, including low oxygen concentrations, high ammonia concentrations, and extremely variable salinities (with salinity shock, up to complete freshwater, and complete marine conditions, varying within a very short time span). Night-time anoxia is especially dangerous for larger-sized organisms (GOLT theory, Pauly, 2021). The same possibly applies to daytime oxygen hyper-saturation and oxidative stress. A further important aspect is high predation mortality in estuaries due to the high densities of preys and predators (i.e., eutrophic systems, such estuaries and lagoons, are “food heaven, but predation hell”, Bakun, 2006).

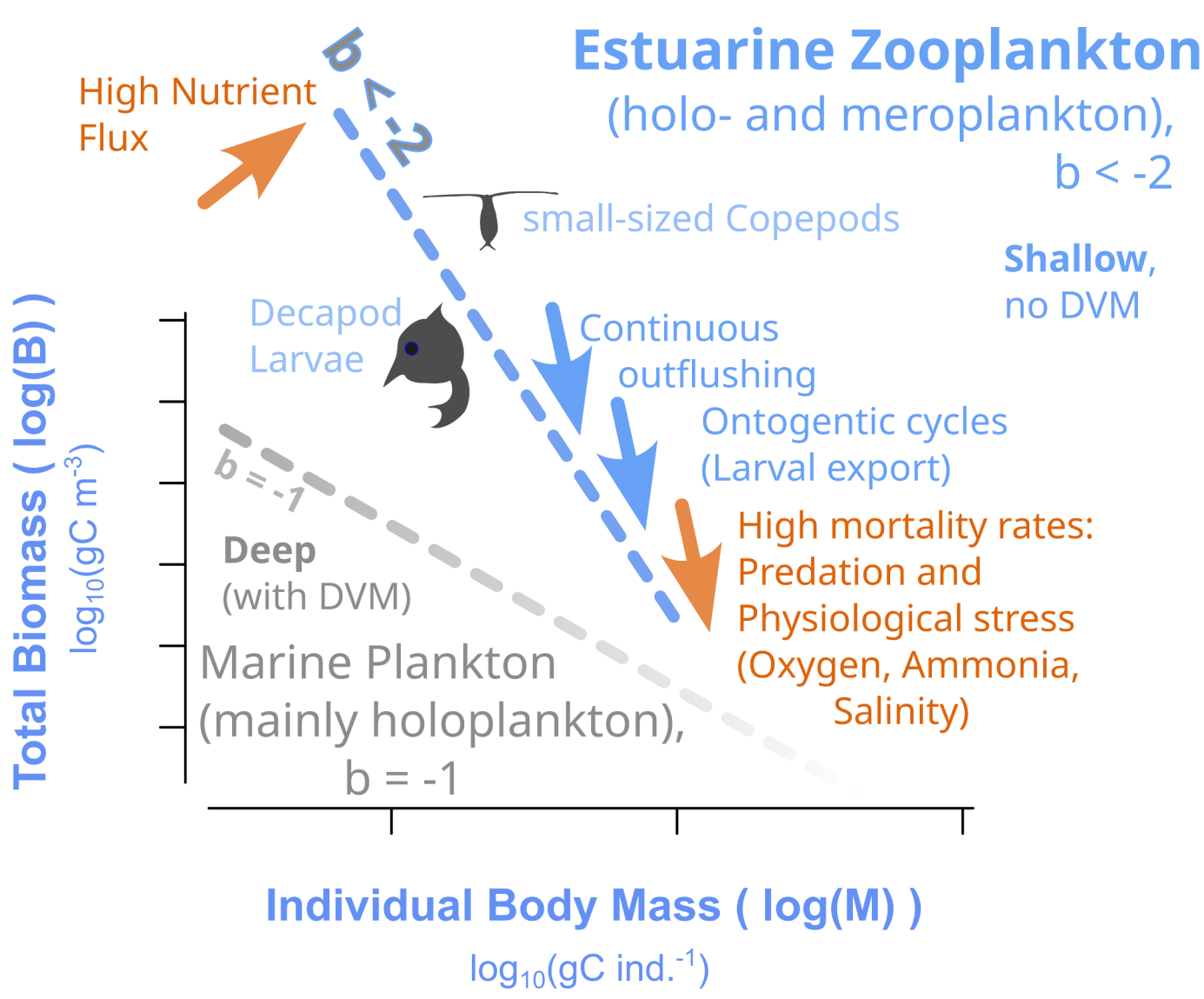

**Figure 10.** Factors and processes affecting zooplankton size spectra in estuaries, as compared to marine ecosystems. DVM: diel vertical migration.

Also, estuaries receive immense amounts of nutrients. This nutrient flux will lead to high primary and secondary productivity. The equilibrium between high primary production, high mortality (by predation, parasites, and physiological stress), and outflushing (pseudo-mortality), together with the effect of shallow depth (no DVM), may explain the observed steep spectra.

Considering that upwelling ecosystems are also very eutrophic and have high production, and high mortalities, yet often present a β slope of approximately -1, it is still an open question what actually leads to the observed extremely steep size spectrum in estuaries. This is still an intriguing phenomenon, and will be the subject of numerous future studies. Probably, the “pseudo- mortality” (i.e. the effect of continuous outflushing of organisms) has a central role in this phenomenon, but also the regular (e.g., tidal) intrusion of planktivorous and piscivorous fish from adjacent marine areas.

## 16. Do trophic cascade waves propagate within or below the carrying capacity spectrum ?

It would be incorrect to state that PETS implies that all ecosystems on this planet are trophically regulated and have a continuous size spectrum, with a constant slope of b = -1 . Many deviations from b = -1 are predicted from PETS, such as extremely flat systems (e.g. in hyper-oligotrophic seas), as well as extremely steep spectra (e.g., for picoplankton and bacteria). Non-linearly shaped (with bumps, gaps, and domes) spectra are expected for ecosystems with low diversity (when few species and size classes exist, as in many temperate and polar lakes and seas, and in many important fishing grounds), or in non-linear, non-equilibrium dynamic situations, (e.g. in non-equilibrium systems, that are disturbed by varying fisheries moralities). In such systems, population-level processes (recruitment, growth, and mortality) will appear as characteristic biomass domes or bumps in the spectrum, for each species. Such domes or bumps may vary according to trophic interactions, taxonomic diversity, size diversity, functional diversity, population processes, and fisheries, and may be easily confounded with top-down or bottom-up traveling cascade waves, and even more probably, with possible stationary (predicted, but not yet proven *in situ*) trophic cascade waves. It will be an immense challenge for size spectra research and future studies to distinguish such diversity-related bumps and domes from dynamic trophic cascade waves, thus resolving possibly ongoing misinterpretations (Rossberg et al., 2019). A perfectly continuous linear size spectrum only forms, when there is a myriad of species, life history stages and populations (as in zooplankton, including holo-, mero-, and ichthyoplankton) with strong size distribution overlap, as often observed in tropical marine plankton (e.g., Figueiredo et al., 2020, Lira et al., 2024, Schwamborn et al., 2025).

Bixed, bottom-up, top-down and resource-limit-stress controls are expected to co-occur in natural systems. Rather than expect that all ecosystems have “b = -1”, it can be expected that the carrying capacity (i.e., the observed maximum biomass values, in large datasets) in each bin follows the ideal, non-resource-limited ceiling of β = -1.

Conversely, earlier attempts to define a carrying capacity were based on the assumption that carrying capacity is defined by an ecosystem’s realized primary production and respiration rates (e.g., Christensen and Pauly, 1998), not by a theoretical maximum biomass based on the carrying capacity spectrum, i.e., the theoretical, ideal “β spectrum”, as in this study. Thus, PETS suggests a new set of concepts and paradigms regarding carrying capacity in natural ecosystems, while defining the carrying capacity for each size bin, based on the β spectrum model. The intercept ($\alpha$) of the ideal carrying capacity model β, is probably defined by several factors, such as the maximum possible primary producer biomass at TopDoSH ($B_{max}$ at $M_{TopDoSH}$), trophic efficiency “E”, the maximum possible $B_{max}$ at $M_{max}$, large-sized apex predator biomass, and the dynamic, size–specific interactions between top-down and bottom up cascade waves, in the context of SOFT and PETS.

An ecosystem with stable maximum biomass, but with variable primary and secondary production, means that production P and turnover $\tau$ may increase with primary production, but not biomass. Higher biomass (i.e., a higher intercept) is only possible if there is an increase in E (i.e., more efficient food webs, such as in densely compacted ecosystems in upwelling regions). The carrying capacity at the primary producer (bottom) end of the food web (i.e., at TopDoSH) is limited by the maximum possible biomass in each size bin, which may be due to several space use, resource use and physiological ("m-scaled") constraints (e.g., night-time anoxia and ammonia toxicity at extreme cell densities), and top-down regulation (grazing), but not by maximum primary production or realized primary production (as in Christensen and Pauly, 1998).

Many resource-limited (i.e., oligotrophic) systems will show mean and median slopes with biomass below predicted from β, and flat or irregular (non-linear) shapes. So, PETS actually does not predict that b = -1. It predicts that observed mean values of "b" should be between -1 and -0.375, and that the slope of the *carrying capacity spectrum* is equal to β.

In summary, PETS predicts that most common marine pelagic size spectra (that are trophically regulated, not resource-stress–regulated) can be expected to be shaped by trophic cascade waves, that propagate up and down between apex predators and phytoplankton, like waves propagating along a stretched string, including the mean position of the string.

Thus, for common trophically interacting food webs, one can expect, for any given sample, to obtain varying shapes, with bumps, waves and domes, above and below b = -1, but when considering many samples, there should be a mean spectrum that follows a linear, β-like shape (e.g., as in Figueiredo et al., 2020).

For resource-stress–regulated (e.g., hyperoligotrophic) systems, one may predict that the biomass in each bin is constrained between between a "ceiling" described by the ideal carrying capacity "β spectrum" and a "floor" described by a "m spectrum" (for stress-limited, infinitely sparsely distributed, 100% non-interacting populations). Considering the ubiquitous size spectrum of b = -1, especially for the zooplankton and larger-sized phytoplankton, it seems likely that today, most natural ecosystems (except for extreme settings, such as in deserts, hyperoligotrophic marine ecosystems, and physiologically extreme environments) are shaped by trophic interactions, not by resource-limit stress.

Within PETS, ecosystems with higher primary productivity will also produce higher secondary productivity (but not higher biomass per bin), while retaining the overall biomass size spectrum shape. Any increase in primary productivity will be passed up the food web through the complex zooplankton food webs, without increasing the biomass in any size bin.

Such bottom-up-cascade waves travel faster in the small-sized bins (rapid turnover, fast body growth and population growth.), than in communities with large-sized animals, which may explain the increasing amplitude. This may lead to a pile-up of such a bottom-up cascade wave signal (e.g. from a phytoplankton bloom), similarly to physical ocean swell waves piling up and increasing in amplitude at the beach. This is in accordance with mizer models and many *in situ* observations, where stable size spectra are found in marine zooplankton vs bump-shaped size spectra in large-sized marine animals (Scott et al., 2014). However, such a bottom-up cascade wave (e.g.. due to a phytoplankton bloom) can hardly become stationary, and will probably be limited to seasonal or stochastic bloom events that are propagated up the food web, within dynamic cascade trophic waves (Fig. 11). Similarly, one may propose that top-down propagating dynamic trophic cascade waves are initially generated at high amplitude (for example, forced due to changes in fisheries, or by intrinsic

Lotka-Volterra-dynamics) and then are subsequently stabilized (damped) when reaching high-turnover food webs in the plankton (as in Andersen and Pedersen, 2010).

From an allometric and energy dissipation perspective, one may easily explain the decrease in amplitude with decreasing body size (Fig. 11) when considering that the energy dissipation Φ is Kreiber-scaled with body mass, where $\Phi \sim M^{-m}$. Simply put, decaying wave amplitude with decreasing size may be explained by organism size and metabolism, if we assume that energy use per unit mass is much higher in smaller organisms. Thus, wave damping and energy dissipation is much more intensive in small-sized organisms, similarly to the damping of physical ocean waves, which are also subject to energy dissipation.

Such wavelike dynamics have been extensively described for predator-prey interactions since the early work of Lotka and Volterra (Lotka, 1925, Volterra, 1927). Interestingly, such waves may erupt spontaneously in highly dynamic predator-prey relationships, even when there is no external forcing, such as fisheries. However, given their highly dynamic nature in time, such Lotka-Volterra cycles can hardly lead to stationary (permanent) wave patterns and bumps.

The larger amplitude within higher trophic levels, or inversely formulated, the more linear spectra found in marine zooplankton, may be due to several interacting factors, such as the 1.) top-down cascade-wave pile-up in stationary and dynamic trophic cascade waves,

2.) different levels of overlap in taxonomy and function, with much more species and life history stages in the plankton (all fish and macroinvertebrate larvae) than in larger size classes,

3.) fisheries affecting largest-sized animal only (not plankton), and

4.) faster growth and turnover in plankton than in large animals.

Interestingly, the size of morphological transition (metamorphosis) from exponential growth to slow growth, is defined by the "size at critical surface area / volume ratio", where gills become necessary, according to GOLT theory (Pauly, 2010). At this size, growth changes completely from very fast exponential weight-specific growth, to a much slower asymptotic VBGF growth. This discrete "critical gill-bearing size" (***CritGill***), or zooplankton-nekton-boundary, coincides with the limit of linear *vs* bump-shaped spectra reported in the literature (e.g., Scott et al., 2014).

Thus, from the fundamental principles described above, PETS predicts a perfectly linear size spectrum in natural pelagic ecosystems to occur within the size range between TopDoSH and CritGill. This size range coincides approximately with the nanophytoplankton, microphytoplankton, meso- and macrozooplankton, where perfectly linear size spectra have indeed been observed. This also explains why size spectra theory is a central topic in plankton research, but has been widely ignored in fisheries research and traditional trophic modeling (e.g. in EwE models).

Conversely, modern size-based trophic models such as APECOSM and mizer, do represent a stable fixed β "background resource" spectrum for plankton and a variable, cascade-shaped spectrum for higher trophic levels, which is in accordance with PETS (Fig. 11).

However, whether changes in larger size classes can affect the size spectrum intercept within the plankton, is still subject to debate and has not yet been properly modeled and verified *in situ,* in large numbers of marine ecosystems. Also, it is not clear whether such stationary trophic cascade waves (if their actual existence is further confirmed by future large *in situ* data sets and ongoing modeling efforts, such as Scott et al., 2014) are ancient natural phenomena, or whether they have been created recently by human action, such as whaling and fisheries.

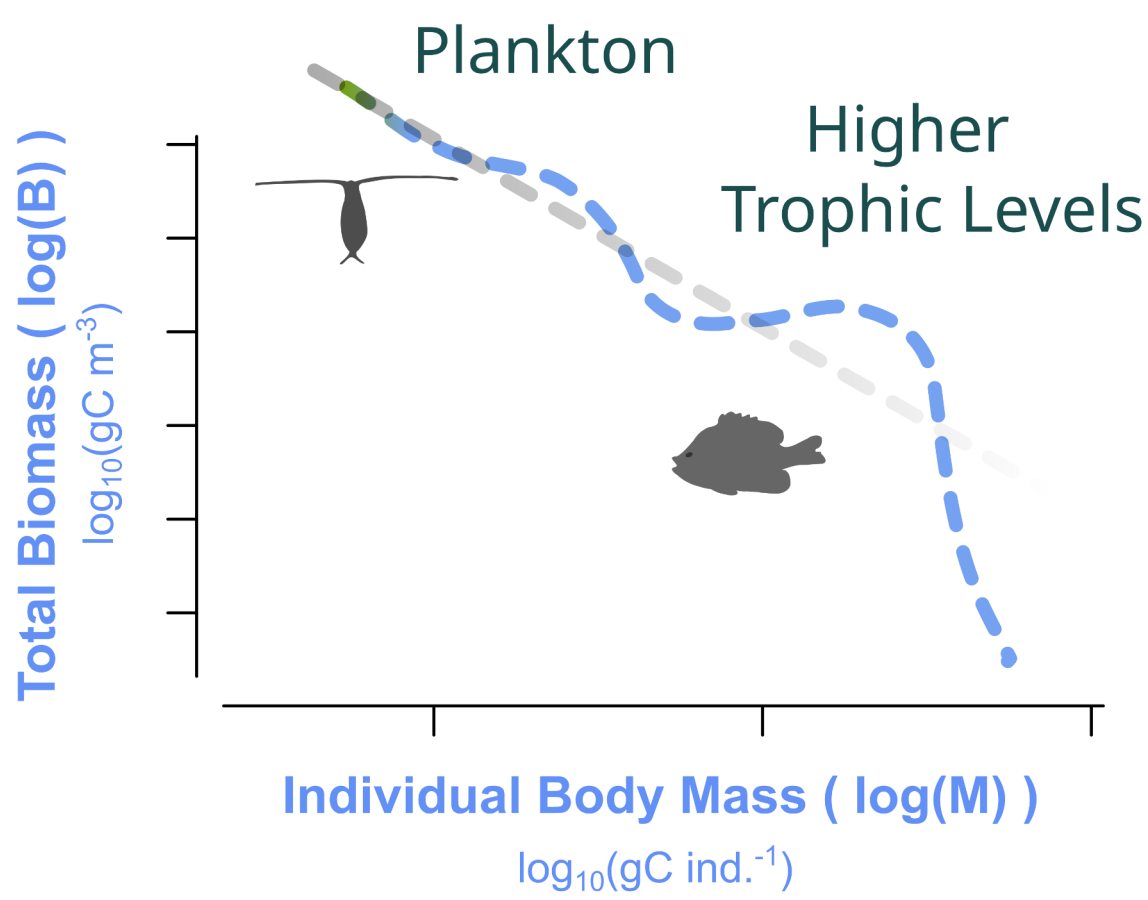

**Figure 11. Hypothetical example of a stationary trophic cascade wave in the weight-biomass spectrum** (similar to Andersen and Pedersen. 2010, Scott et al., 2014, modified). Note the larger amplitude within higher trophic levels.

## 17. Are pelagic size spectra shaped by unicellular organisms or by large animals, or both?

For the largest size bins ($M_{max}$), it is possible to imagine that their carrying capacity is limited by their maximum possible biomass only (i.e, by the minimum space unit, surface or volume, such as $V_{min}$ needed per individual, considering agonistic interactions between individuals and predation with the multi-species size bin $M_{max}$). Alternatively, large-sized animals may be limited by food abundance, i.e., by the maximum possible biomass at the small–sized extreme end of the food web (i.e., $B_{max,i}$ at Mi = $M_{TopDoSH}$), combined with the mass-specific biomass change efficiency β = -1. Most likely, both limiting factors ($V_{min}$ and food abundance) are important in defining the carrying capacity of large-sized animals in nature. When food is abundant, and large-sized ($M_{max}$) biomass $B_{Mmax}$ is at maximum, agonistic processes, predation, cannibalism, and within size-class functional replacement processes related to $V_{min}$ take over. Under prolonged food scarcity, $B_{Mmax}$ declines and will be far lower than $M_{\beta}$, or even completely collapse, as observed for marine mammals off Peru during strong El Niño events.

Similarly, instead of imagining that primary producers are limited by their physiological, space and resource (i.e., nutrient) needs only, they may also be limited by zooplankton grazing and top-down cascade waves, i.e., by the maximum possible biomass for the largest biomass in the system (Bmax at Mmax) which is mediated though β. Most likely, the extremes of the size spectrum (Mmax and MTopDoSH size bins) interact through the food web, in a "tug-of-war"-like mechanism, thus establishing the equilibrium β spectrum. This is in accordance with several previous authors, who have emphasized the special "key" controlling roles of primary producers, and especially of apex predators, in natural food webs (Hairston et al., 1960, Power, 1990, Pauly et al., 1998, Pauly and Palomares, 2005).

Given the importance of the theoretical, non-trophically regulated biomass $B_{max}$ in the extreme bins ($M_{max}$ and $M_{TopDoSH}$ size bins), it may be useful to analyze which kind of processes may actually regulate population biomasses in these bins.

The shapes of phytoplankton size spectra in nutrient-limited scenarios have been detailed above. In nutrient-rich situations, e.g., where there is a constant nutrient flux from upwelling or terrestrial sources, hyper-eutrophic food webs are quickly established, where there are considerable densities of zooplankton and thus, effective zooplankton grazing (Schwamborn et al., 2004). Yet, in a theoretical, non-nutrient-stressed scenario with phytoplankton that is not limited by grazing, huge phytoplankton biomasses can build up. Under these "green-water" conditions, the main limiting factors, that lead to mass mortality, are the nighttime oxygen depletion and toxic ammonia concentrations, which both occur at extreme phytoplankton densities, as in hyper-eutrophic estuaries and bays (Schwamborn et al., 2004). However, such hyper-eutrophic situations, with Secchi depths of a few centimeters only, do not usually occur in the oceans. Thus, they can hardly be considered a reference for the carrying capacity in the open ocean. From this thought experiment we may conclude that phytoplankton is generally not limited in the open ocean by predation-independent processes, but rather by zooplankton grazing, as in most marine ecosystems, which exhibit a β spectrum (and by extreme nutrient limitation stress, in hyper-oligotrophic systems).

A different picture is seen when looking at the largest size bin ($M_{max}$), where there are usually many apex predators (e.g., large-sized sharks, giant squid, and odontocetes), but also large-sized planktivores (such as filter-feeding sharks and baleen whales). Since the proportions of these groups (apex predators *vs* filter-feeders) are highly variable among ecosystem, it is not an easy task (or may not make sense at all) to determine the trophic level of the largest size bin, within a global ocean perspective. The largest size bin in the "global ocean size spectrum", if one would try to compile it (as in Hatton et al., 2021), would have a relatively low trophic level, of approximately 3 to 4 (baleen whales). Actually, the largest size bin of our planetary “global size spectrum” has a trophic level of 1, being occupied by giant sequoia trees (Tekwa et al., 2023). This indicates that a “global size spectra model”, for the whole planet may be implausible, possibly invalidating the idea of an “average global size spectrum” (e.g., Hatton et al. 2021, Tekwa et al., 2023), and highlighting the need for ecosystem-specific size spectra models (as in Dugenne et al., 2024, Fock et al., 2026, Schwamborn et al., submitted).

Many authors have suggested a regulation mechanism for the largest sized animals ($M_{max}$) that is independent of food supply. Actually, cannibalism and self-regulation is a common phenomenon and well-described in apex carnivores (Polis, 1981, van den Bosch et al., 1988). Carnivores are often limited through cannibalism instead of food supply. Thus, it is possible to argue in favor of a hypothesis that apex carnivores and other large-sized animals, in a pristine pre-anthropocene setting, were limited in their biomass through intrinsic, density-dependent processes that were not related to food supply in most natural ecosystems. Indeed, most field studies and food-web models support the notion that apex predators exert strong top-down control on natural food webs. Thus, we may preliminarily accept this plausible hypothesis: the “top predators define the structure of food webs through top-down control” - hypothesis or analogously, from a PETS perspective, we may state the “$B_{max}$ at $M_{max}$ affects the size spectrum intercept through top-down control”- hypothesis Obviously, this assumes abundant food supply. Enduring food shortage will obviously lead to a decrease or even collapse of apex predator biomass (as observed in strong El Niño events off Peru).

Considering that in laboratory cultures, much denser phytoplankton cultures are theoretically possible than ever observed in the open ocean, and the linear shape of microphytoplankton size spectra, it is likely that the system carrying capacity is not defined by the maximum possible phytoplankton biomass, but rater controlled through top-down regulation (i.e., grazing) processes by the zooplankton community, which itself is most likely controlled, in its biomass, by higher trophic levels. Thus, PETS may predict that overfishing of zooplanktivores (sardines and anchovies) may lead to zooplankton biomass above predicted from the global linear model. This highlights the ecosystem-wide impact of fisheries (and of the introduction of exotic apex predators, e.g., lionfish) as demonstrated in several case studies (e.g., Jackson et al., 2001). Thus, the shape (i.e., the intercept) of the size spectrum and the total system biomass in a pristine, trophically coupled ecosystem, could be defined by the carrying capacity of the largest size bins (i.e, the maximum possible biomass of the largest organisms, $B_{max}$ at $M_{max}$), together with E and PPMR (Fig. 12).

Considering a detailed analysis (Fig. 12) of how taxon-specific spectra compose the total ecosystem spectrum, based on Lira et al. (2024), Dalaut et al. (2025), Schwamborn et al. (2025), and Schwamborn et al. (submitted), we may note that there are numerous differently-shaped taxon-specific spectra. However, their biomass spectra near maximum body size (i.e., the biomass of the largest individuals) are well aligned below the *Maximum Carrying Capacity Spectrum* (MCCS). Interestingly, the taxon-specific spectra and their alignment to the ecosystem spectrum varies considerably across taxa and sizes. At larger fish sizes, the alignment of the biomass of the oldest organisms (large-sized adults) within each fish population is mathematically defining (by the sum) the total ecosystem size spectrum (Fig. 12). This may also highlight the size-spectrum-modifying impact of the removal of largest-sized fish from the ecosystem (as generally done by fisheries).

In contrast to fish (which have a wide size range across many orders of magnitude in body size and mass, from eggs, larvae, juveniles, to adults), in phyto- and zooplankton (fast growth, narrow size range), the shape of the total ecosystem-level size spectrum is not defined by the largest individuals. Instead, the main taxon-specific peak within each phyto- and zooplankton assemblage mathematically determines the total plankton ecosystem spectrum (Fig. 12).

The alignment of the plankton taxa biomass peaks and the biomass of large-sized adult fish below the MCCS occurs regardless of the numerous shapes (i.e., varying growth / mortality ratios) of the size spectra of the individual taxa and populations. This further underlines that *trophic processes* (e.g., top-down control) shape the pelagic biomass spectrum, rather than population-level variations (growth/mortality ratios).

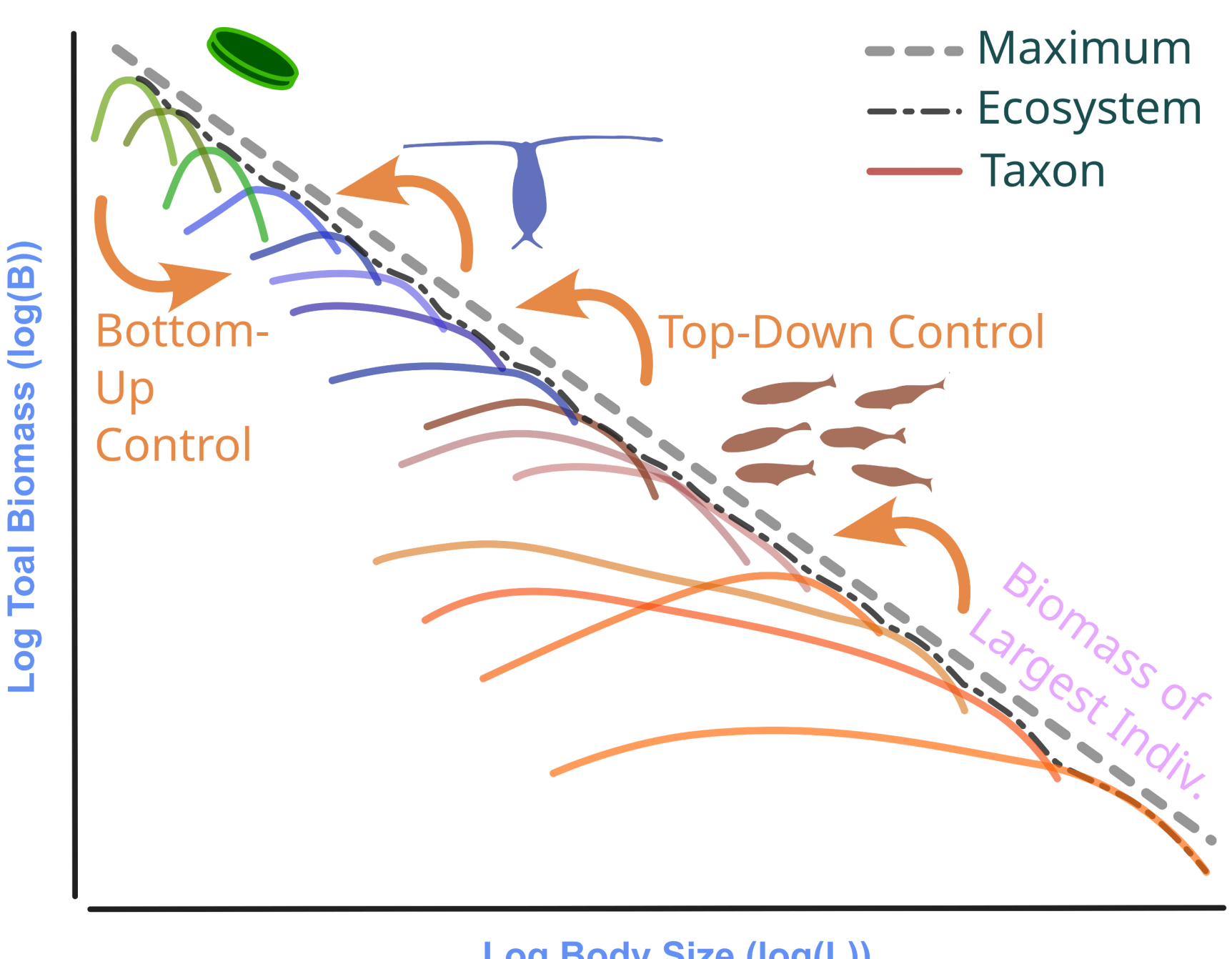

**Figure 12. Hypothetical example of a size-biomass spectrum composed of taxon-specific spectra (fish species and phyto- and zooplankton taxa), maximum carrying capacity spectrum, and total ecosystem spectrum** (based on Andersen et al., 2016, Lira et al., 2024, Dalaut et al., 2025, Schwamborn et al., 2025, Schwamborn et al., submitted, *modified*). Note the numerous differently-shaped taxon-specific size spectra. Yet, their biomass peaks (e.g., peaks of adult fish biomass) are well aligned below the maximum carrying capacity spectrum (MCCS). The alignment of these peaks composes the shape of the total ecosystem-level size spectrum, regardless of the shapes (i.e., growth / mortality ratios) of the size spectra of the individual taxa. Note that large-sized organisms (e.g., fish) have a wider life-history size span than smaller ones (e.g., phytoplankton). Orange: Large-sized fish species; Brown: small coastal epipelagic fish species; Blue: macro- and mesozooplankton; Green: microphytoplankton.

The question whether global marine ecosystems are limited by current apex predators, maximum theoretical carrying capacity at $M_{max}$ or by the maximum possible biomass at $M_{TopDoSH}$, is still subject of debate, and cannot be conclusively settled with the current knowledge and data. The idea that top predators control the whole food web, including herbivores, and thus the removal of apex predators leads to a greener world, is also called the Hairston-Smith-Slobodkin-theory (or HSS theory), which has gained immense popularity in the past decades.

Thus, it may be interesting to consider, from a WETBIO and PETS perspective, the theoretical possibility that the pristine carrying capacity of large animals, which includes the pre-whaling whale biomass and pre-fisheries fish stock biomass, could be a determinant of present size spectra shapes in small-sized animals. If this was confirmed, it would actually explain why extant large-sized organisms (whales and fish stocks) display biomass well below predicted from plankton size spectra (Scott et al., 2014). Also, if this preliminary hypothesis (i.e. the “pristine biomass - carrying capacity size spectrum” - hypothesis) is confirmed, it would mean that we can estimate the pristine biomass of large-sized organisms just by looking at the predicted biomass from linear models that were adjusted to the invariant net-caught zooplankton size spectrum. This could be called the “Constant Biomass At Lower Trophic levels” (COBALT) hypothesis.

Alternatively, it is also possible that current higher trophic level (HTL) biomass controls and linearizes (trough top-down cascades) the biomass of lower trophic level size spectra, down to the TopDoSH, except when there is extreme nutrient limitation stress (then, there is bottom-up control). This could be called the CATCH hypothesis (“Current And Top-down Control by Higher trophic level biomass & carrying capacity size spectra” - hypothesis. While CATCH is in full concordance with HSS theory, they are not identical. While HSS considers trophic levels only (not size), CATCH is size-specific and focuses on the effect of the removal of large-sized animals on the size spectrum.

Both COBALT (little or no top-down control on plankton) and CATCH (top-down control by higher trophic levels) are in accordance with minimum-volume considerations (i.e., both can coexist with the WETBIO hypothesis), but with important differences regarding the dominant mechanisms (top-down or bottom-up), and the effects of whaling and fisheries. In both scenarios (limitation by the minimum volume needed by each individual, or alternatively, by the current biomass of higher trophic levels) the linear size spectrum could be explained by a carrying capacity spectrum that is attained by natural populations until reaching the WETBIO carrying capacity.

A global ocean according to the CATCH hypothesis assumes natural pelagic size spectra and ecosystems with high predation vulnerability, except for phytoplankton under extreme nutrient limitations stress. Interestingly, short-term nutrient limitations stress would affect only the phytoplankton, while zooplankton, small pelagic fishes, and higher trophic levels would contain the same biomass but at lower production (feeding mostly on alternative, less nutritious food sources, such as protozoans and abundant non-living particles). Although counter-intuitive at first, the idea that there is a constant biomass of top-down controlled food webs (independently of minor variations in primary production and food supply) is conceptually similar to the GOLT theory, where limitations in oxygen uptake determine biomass and production, not food supply.

The extent of the bottom-up control through nutrient limitation stress depends on the intensity and especially on the duration of such stress events. Short-term stress events in the scale of a few days may have only limited effects on the biomass of higher trophic levels. However, prolonged and intensive nutrient supply failure events may lead to catastrophic biomass collapse throughout the whole food web, as regularly observed in strong El Niño events off Peru (bottom-up control). The resilience and capability to survive long periods of food shortage and starvation depends mainly on body size (e.g., many whale species can live for months without feeding, while some zooplankters may experience "point of no return" starvation effects after a few hours). Thus, the extent of the nutrient limitation stress (bottom-up control) signal up the food web and into larger size classes depends specifically on the body size and the duration of such resource limitation stress events. Conversely, oxygen-stress events will mostly affect large-sized, gill-bearing animals (GOLT theory).

When looking at the importance of top-down cascade waves induced by fisheries, It is important to remember that the bulk of fish biomass on our planet is concentrated in mesopelagic fishes, most of which are not subject to direct fisheries effects, and that fish stocks actually represent only a small fraction of fish biomass in the oceans. Interestingly, it is possible to imagine a scenario of functional replacement, where fisheries may have led to the replacement of epipelagic fish stocks (e.g., sardines, mackerels and tuna) by increasing biomass of unfished animals such as jellyfish and mesopelagic fish. Functional replacement may be one of the most deleterious, and often irreversible, consequences of apex predator removal (e.g., Jackson et al., 2001, Estes et al., 2011, Hobbs, et al., 2024). Such functional replacement may be facilitated by recent anthropogenic global warming (e.g., Li and Convertino, 2021).

Dynamic functional replacement processes, leading to the maintenance and stabilization of the smooth, continuous size spectrum, have been recently demonstrated by Schwamborn et al. (2025). Thus, it is possible to hypothesize that the global ocean size spectrum is actually not permanently modified by the removal of fish stocks, as long as there is a functional replacement by other, unfished groups occurs (e.g., jellyfish). Such a hypothetical size-niche-specific (*sensu* Schwamborn et al., 2025) functional replacement mechanism may be able to re-stabilize the size spectrum, in a completely different, stable state, with completely different species composition, trophic dynamics and ecosystem services (e.g., fisheries and carbon pumps). Simply put, warming, fisheries, and whaling may have led us onto a completely different planet, with more jellyfish blooms (e.g., Pauly et al., 1998, Daskalov et al., 2007), but with a constant, immutable size spectrum, governed by geometric laws (WETBIO) and equilibrium mechanisms (PETS).

A "compleat" size spectra model should be able to accurately predict the (living + non-living) mass in each size or weight bin from first principles and simple equations, and not only the overall value of "b". Defining the exact value and key processes for carrying capacity for a given population, species, community, and size bin, is far from simple and is a task still to be solved by future field sampling and modeling efforts.

Historical reconstructions of past ecosystems however, show that previous ecosystems were much more biomass-rich than present (especially for higher trophic levels, Pauly et al., 1998), indicating that present ecosystems may not be representative of historical carrying capacity, due to persistent, long-term effects of drastic cross-ecosystem changes, especially whaling and fisheries (Jackson et al., 2001, Estes et al., 2011). Many populations of large-sized whale species are still at a minuscule fraction of their pre-whaling biomass, many decades after whaling was interrupted. For example, Antarctic blue whales remain at about 1% of their pristine population (Thomas et al., 2016).

## *18. Key conclusions - and 10 open questions*

PETS is certainly not the only possible theory for natural ecosystem size spectra (see chapter 14, and especially the widely cited model of Zhou, 2006). Far from being conclusive, in the future, PETS may be extended, amended, improved, corrected, and replaced by another theory that is better in explaining and predicting natural phenomena. Interestingly, both MTE and PETS are theories that are derived from the observation of consistent scaling in nature. MTE is derived from the observation that the weight-metabolism relationship in many eukaryotes scales with m = 0.75, while PETS is derived from the observation that the weight-biomass relationship in three-dimensional ecosystems scales with $\beta$ = -1. Both theories have in common that they are not made up out of thin air, but derived from many observations in real nature, and their analysis, conclusions, and equations are subsequently applied to numerous other situations and phenomena. Interestingly, MTE is in stark contradiction with PETS, since MTE fails to predict the -1 slope of the pelagic size spectrum (Brown et al., 2004). None of the extant volume-, mass-, weight-, or abundance-related relationships discussed herein scale with 0.75, or can otherwise be derived directly from MTE. Yet, the inclusion of the term "- ($m_i$ - $m_0$)" into the PETS equilibrium equation "$\beta$ = ( E * S ) - ( $m_i$ – $m_0$ )" shows that metabolism is an intrinsic part of PETS.

There are numerous theories, models and hypotheses that tried to explain the weight-metabolism scaling of 0.75 (among which one of the most popular is still the heavily debated, much contested, highly speculative WBE fractal network model, West, Brown & Enquist, 1997). There is still an ongoing vivid debate, and there is no single conclusive explanation or model that can explain the m = 0.75 scaling in many eukaryotes. Notwithstanding, MTE, that builds upon this observation, has become highly popular, since it could be used to explain and predict numerous natural phenomena (see Brown et al., 2004), with varying success. Similarly, the processes and equations that were herein used to explain the ubiquitous phenomenon of $\beta$ = -1 within PETS, may not be absolutely conclusive, but, most importantly, the formal concepts, terminology, applications and predictions of PETS may be useful for further studies on ecosystem size spectra and beyond. PETS intends to explain the $\beta$ = -1 slope (by a combination of the minimum-volume weight scaling of $V_{min}$ / M ~ M, the concept of a size-specific carrying capacity, and size-specific food-web processes, i.e., WETBIO, SOFT, and PATER). Based on the insights and equations that helped us to understand the key processes that shape a size spectrum, it was possible to explain and predict a series of key phenomena in pelagic, benthic and terrestrial ecosystems.

Far from being conclusive, this study highlights the importance of 10 key questions for our understanding of ecosystem size spectra:

1.) Is there a constant **equilibrium between trophic efficiency and PPMR**? If so, how can it be confidently quantified?

2.) What is the relevance and **relative importance of the growth / mortality ratio and E / log(PPMR) ratio** in shaping size spectra?

3.) How are **growth / mortality ratios and E related**? The relationship between g/z and E will certainly be the subject of many future theoretical considerations, proposals and numerical modeling efforts in the future.

4.) What are the precise **numerical values** for E, PPR and “c”? Given the high uncertainty and low number of available estimates for PPMR and E in real ecosystems, we are still far from obtaining a reliable numerical estimate for the proposed trophic equilibrium constant “c”. The exact value of “c” obviously depends on the units used for E and PPR. The proposed PETS theory is intended to serve as an inspiration for further investigation and highlights the need for quantitative estimates of such key ecosystem descriptors. The huge number of available size spectra studies has revealed a constant, temperature-invariant value for β (e.g., dos Santos et al., 2017), which favors the proposed concept of a universal value for “c”, with an exact numerical value that still has to be precisely determined in the future. PPMR can be approximately assessed from nitrogen stable isotopes (Figueiredo et al., 2020). By using amino-acid specific stable isotopes (GC-IR-MS), a very precise method, reliable estimates of FCL and PPMR can be obtained for pelagic ecosystems (and thus, precise numeric estimates of “E” across the ecosystem), a promising path for future studies.

5.) How important is **non-predation mortality**? Recent studies (da Cruz et al., 2023) have shown that non-predatory mortality is extremely relevant in pelagic ecosystems and that dead copepod carcasses are ubiquitous and abundant. However, explanations for the main causes of non-predatory mortality and an adequate modeling that can explain the factors causing non-predatory mortality and the effects of $z_0$ on size spectrum models (i.e., the effect of $z_0$ on E) are still far down the road.

6.) How are **biological carbon pumps** (e.g., sinking phytoplankton blooms, carcasses, exuviae, fecal pellets, aggregates, gelatinous appendicularian houses, vertical migration, etc.) related to the size spectrum? Several numerical models are already available that relate plankton size spectra to the vertical carbon export from the ocean surface (e.g., Serra-Pompei et al., 2022). However, this area of research and modeling is still incipient and lacks integration to other size-based models and approaches, such as for higher trophic levels and fisheries.

7.) How does the **size spectrum of non-living particles** (nutritious particles and inert microplastics) affect the structure of marine food webs and weight-biomass size spectra?

8.) Is there a (log-linear?) **carrying capacity spectrum of non-living particles** (controlled by density-dependent grazing? By the size-specific production of particles?)? Is there a (log-linear?) **carrying capacity spectrum for all (living+non-living) particles** (controlled by nutritiousness, vulnerability, and predator-and-prey-density-dependent trophic processes)? What is the effect of **size spectrum of microplastics** on aquatic ecosystems?

9.) **How far down the food web** does the direct and indirect effect of top-down cascade waves reach? Are there indirect effects below the TopDoSH? If so, ecosystems with strong metazoan grazing pressure should present high nanoplankton biomass (just beyond the TopDoSH), with a discernible peak, above predicted from the $\beta$ spectrum. Analogously, how far up the food web and in which timescales do bottom-up disturbances propagate?

10.) How does the **size distribution of fisheries mortality** affect marine food webs and ecosystem services? This question is probably the oldest and most intensively investigated, in fisheries science, since more than 100 years, and has led to an extremely vast theoretical framework, numerous modeling software, and intensive debates (see Pope et al., 2006), for example regarding "Balanced Harvesting" (Kolding et al., 2016) and the preservation of large-sized "super spawners" (Froese, 2004). However, as far as the recent literature has been reviewed, there are no published attempts to understand fisheries mortality $Z_f$ explicitly in the context of size spectra equilibrium theory (which has mostly focused on plankton).

Within a preliminary and superficial evaluation, it may be assumed that variable removal of large size organisms may lead to a non-equilibrium state of the system, leading to bumps and stationary waves (Scott et al., 2014). If the CATCH hypothesis is correct, the constant removal of large-sized animals by fisheries (i.e., "*Fishing Down the Size Spectrum*" analogously to Pauly et al., 1998) has possibly already lead to a decrease in overall ecosystem biomass through SOFT-mediated top-down trophic cascades throughout the food web (e.g., Frank et al., 2005), while maintaining a constant -1 spectrum slope and an increased biomass of extremely small-sized primary producers (e.g., picoplankton) beyond the top-down-control size horizon (TopDoSH).

A proper understanding and modeling of the processes discussed herein, and their relative importance in shaping aquatic ecosystems, is fundamental to our ability to predict the future of key ecosystem services, such as fisheries and the biological carbon pump in the context of climate change.

Rather than presenting a conclusive, complete (or “compleat” *sensu* Gayanilo et al., 1988) theory, this study is intended as a contribution to our understanding of the functioning of marine ecosystems and as a contribution to the ongoing discussions and modeling efforts.

## Acknowledgements

The author is greatly indebted to many collaborators and colleagues. To Denise F. M. C. Schwamborn and Wesley de Oliveira Neves for revising the text and equations. Many thanks to Gaby Gorsky and Marc Picheral for introducing me to the ZooScan method more than two decades ago, which opened new paths towards the analysis of zooplankton size spectra. Many thanks to all students, postdocs, and colleagues in UFPE's zooplankton lab for companionship, unity, friendship, and intensive work on zooplankton size spectra with ZooScan, especially to Lúcia Gusmão (*in memoriam*) and Sigrid Neumann Leitão. Many thanks to Nathália Lins-Silva, Felipe Kessler, and Catarina Marcolin, for co-developing our common zooplankton, biogenic particles and microplastics research stream, which helped me understand the importance of explicitly including particles in size spectrum analysis. Many thanks to M. L. “Deng” Palomares for encouraging me and for pushing me deeper into the work on sized-based methods, more than 17 years ago. Many thanks to Daniel Pauly for inspiring me, along many years (since 1991), to do in-depth theoretical work on this subject, although he in 2018 repeatedly discouraged any effort to establish a general size-based theory for particles, plankton and fisheries, which he considered impractical and out of reach at that time, but which actually motivated me even more to work intensively on the development of this theory. The author is especially indebted to the TRIATLAS size spectra working group for attending many size spectra workshops and numerous inspiring discussions. Many thanks to Cristina González-García and Emilio Marañón for important discussions regarding the size spectra of phytoplankton, specifically the nutrient flux and mixed layer depth effects. Many thanks to Matthias Wolff, Werner Ekau an Ulrich Saint-Paul for early discussions on size spectra, many decades ago. Many thanks to Henrike Andresen, Tim Dudek, and Heino Fock for inspiring comments and lively discussions in Bremen and during the TRIATLAS conferences. Many thanks to Arnaud Bertrand for inspiring comments, continuous support, and friendship. Many thanks to Simone A. Lira, Gabriela Figueiredo, Claudeilton de Santana, and Iurick Saraiva Costa for important comments.

Many thanks to Olivier Maury and Laureline Dalaut for introducing me to the APECOSM model and for inviting me to participate in their recent work. To Mathilde Dugenne, Dodji Soviadan, and Lars Stemmann for in-depth discussions on UVP size spectra and their relationship to the complete ecosystem size spectrum. Many thanks especially to Laurent Dragorn for specifically encouraging the elaboration of such a general theory, that is intended to help us better understand the complete size spectrum of marine ecosystems.

## Artificial intelligence declaration

During the preparation of this manuscript, the author used artificial intelligence (AI) tools (Grammarly, Google Scholar and ChatGPT) for text editing (e.g., grammar and spelling corrections) and literature searches. Generative AI was used in the elaboration of the Graphical Abstract, of four schematic illustrations of example organisms in Fig. 2, and for the lawnmower sketch in Fig. 3, that were partially elaborated with the aid of ChatGPT (v.5.5, OpenAI). No AI tools were used to generate ideas, scientific content, analyses, interpretations, tables. All content is the original work of the author.